\documentclass[prd,preprint,superscriptaddress,preprintnumbers,eqsecnum,showpacs,nofootinbib,
nobibnotes,noeprint,aps]{revtex4-1}
\pdfoutput=1
\usepackage{amsfonts,bm}
\usepackage{amsfonts,amssymb,amsmath}
\usepackage{graphicx}
\usepackage{paralist}
\usepackage[english]{babel}
\usepackage{slashed}
\usepackage[usenames]{xcolor}
\usepackage{mathtools}
\usepackage[makeroom]{cancel}
\usepackage{color}

\usepackage[linkcolor=blue,citecolor=blue,urlcolor=blue,colorlinks=true]{hyperref} 
 \usepackage{breakurl}

\newcommand{\be}{\begin{equation}}
\newcommand{\bea}{\begin{eqnarray}}
\newcommand{\ee}{\end{equation}}
\newcommand{\eea}{\end{eqnarray}}
\newcommand{\Dr}{\mathcal Z}

\def\1eq#1{Eq.~(\ref{#1})}

\def\2eqs#1#2{Eqs.~(\ref{#1}) and~(\ref{#2})}
\def\3eqs#1#2#3{Eqs.~(\ref{#1}),~(\ref{#2}) and~(\ref{#3})}

\def\fig#1{Fig.~\ref{#1}}

\def\ie{{\it i.e.}, }
\def\eg{{\it e.g.}, }

\newcommand{\Ls}{ \mathit{L}_{{sg}}}   
\def\g{\Gamma}
\def\gz{\Gamma_{\!0}}

\def\s#1{{\scriptscriptstyle #1}}

\newcommand{\fatg}{{\rm{I}}\!\Gamma}
\newcommand{\fatgt}{\widetilde{{\rm{I}}\!\Gamma}}

\newcommand{\w}{{\cal W}}

\newcommand{\Cfat}{{\mathbb C}}

\newcommand{\C}{{\mathcal C}}
\newcommand{\Ctilde}{{\widetilde{\mathcal C}}}

\newcommand{\CB}{{\mathbb C}_{\star}}
\newcommand{\Cc}{{\mathcal C}_{\star}}

\newcommand{\Cgh}{C}

\begin{document}

\title{Emergence of mass  in the gauge sector of QCD}

\author{J. Papavassiliou}
%\email{Joannis.Papavassiliou@uv.es}
\affiliation{\mbox{Department of Theoretical Physics and IFIC, 
University of Valencia and CSIC},
E-46100, Valencia, Spain}

\begin{abstract}

It is widely accepted nowadays that gluons, while massless at the level of the fundamental QCD Lagrangian, 
acquire an effective mass through the non-Abelian implementation of the  
classic Schwinger mechanism. The key dynamical ingredient that triggers the onset of this mechanism 
is the formation of composite massless poles inside the fundamental vertices of the theory. 
These poles enter in the evolution equation of the gluon propagator, and affect   
nontrivially the way the Slavnov-Taylor identities of the vertices are resolved,
inducing a smoking-gun displacement in the corresponding Ward identities.   
In this article we present a comprehensive review of the pivotal concepts associated with  
this dynamical scenario, emphasizing the synergy between functional methods and lattice simulations, 
and highlighting recent advances that corroborate the action of the 
Schwinger mechanism in QCD.

\end{abstract}

\pacs{
12.38.Aw,  % General properties of QCD (dynamics, confinement, etc)
12.38.Lg, % Other nonperturbative calculations
14.70.Dj %Gluons
}

\maketitle

\section{Introduction}
\label{sec:intro}

Gluons are massless at the level of the fundamental Lagrangian that describes pure Yang-Mills theories
or the gauge sector of Quantum Chromodynamics (QCD)~\cite{Marciano:1977su}, and the use of symmetry-preserving regularization
schemes, such as dimensional regularization~\cite{Collins:1984xc}, 
enforces their masslessness at any finite order in perturbation theory.
Nonetheless, mounting evidence indicates~\cite{Cloet:2013jya,Aguilar:2015bud,Roberts:2020hiw,Roberts:2021xnz,Horak:2022aqx} that the nonperturbative 
gluon self-interactions give rise to a dynamical gluon mass, or mass gap, 
as originally asserted four decades ago in a series of seminal works~\cite{Cornwall:1979hz,Parisi:1980jy,Cornwall:1981zr,Bernard:1981pg,Bernard:1982my,Donoghue:1983fy}, and subsequently explored in a variety of contexts~\cite{Mandula:1987rh,Cornwall:1989gv,Wilson:1994fk,Lavelle:1991ve,Halzen:1992vd,Mihara:2000wf,Philipsen:2001ip,Kondo:2001nq,Aguilar:2002tc}.
In principle, this mass sets the scale for dimensionful quantities such as glueball masses~\cite{Athenodorou:2020ani,Athenodorou:2021qvs} 
and ``chiral limit'' trace anomaly~\cite{Collins:1976yq}, and cures the instabilities (\eg Landau pole) stemming from 
the infrared divergences of the perturbative expansion. The gluon mass gap underlies the concept of 
a ``maximum gluon wavelength'' above which an effective decoupling (screening) of the gluonic modes occurs~\cite{Brodsky:2008be},
and is intimately connected to confinement, fragmentation, and the
suppression of the Gribov copies, see, \eg\cite{Braun:2007bx,Binosi:2014aea,Gao:2017uox} and references therein.

In a strict sense, 
the term {\it mass gap} is understood to mean a {\it physical} scale, which is
independent of the gauge-fixing procedure used to quantize the theory,
and invariant under changes of the renormalization scale $\mu$. 
Of course, when the emergence of such a mass is exhibited by  
the off-shell $n$-point correlation (Green) functions of the theory,
the resulting effects  
are both gauge- and $\mu$-dependent. Nevertheless, 
the distinctive patterns induced by the gluon mass to the infrared behavior  
of two- and three-point functions admit a pristine physical interpretation, 
providing invaluable information on the nature and operation of the
underlying dynamical mechanisms. Moreover, a special combination of these 
correlation functions, denominated {\it process-independent QCD effective charge}~\cite{Cornwall:1981zr,Binosi:2002vk,Aguilar:2009nf}, allows the definition of  
a renormalization-group-invariant gluonic scale of about half the proton mass~\cite{Binosi:2016nme,Cui:2019dwv}.

Particularly conclusive 
in this context is the characteristic feature of {\it infrared saturation} displayed by the gluon propagator,
which has been observed in numerous large-volume lattice simulations~\cite{Cucchieri:2007md,Cucchieri:2007rg,Bogolubsky:2007ud,Bogolubsky:2009dc,Oliveira:2009eh,Oliveira:2010xc,Cucchieri:2009zt} 
and explored within
a variety of continuum approaches~\cite{Aguilar:2008xm,Fischer:2008uz,Binosi:2012sj,Serreau:2012cg,Tissier:2010ts, Aguilar:2016vin, Pelaez:2014mxa,Dudal:2008sp,RodriguezQuintero:2010wy,Pennington:2011xs,Meyers:2014iwa,Siringo:2015wtx,Cyrol:2018xeq}. 
This special attribute is rather general, manifesting itself 
in the Landau gauge,  when other gauge-fixing choices are implemented~\cite{Cucchieri:2009kk,Cucchieri:2011pp,Bicudo:2015rma,Epple:2007ut,Campagnari:2010wc,Aguilar:2016ock,Glazek:2017rwe},  and in the presence of dynamical quarks~\cite{Bowman:2007du,Kamleh:2007ud,Ayala:2012pb,Aguilar:2013hoa,Aguilar:2019uob}. 
In all these cases, as the scalar form factor, $\Delta(q^2)$, of the gluon propagator 
reaches a finite nonvanishing value in the deep infrared, a gluon mass may be defined through the simple identification
$\Delta^{-1}(0) = m^2$, as happens in the case of ordinary massive fields.
However, as we will see in detail in what follows, the field-theoretic circumstances that account for this exceptional
behavior are far from ordinary, involving a
subtle interplay between nonperturbative dynamics and symmetry.

The way to reconcile local gauge invariance with a gauge boson mass was elucidated long ago
by Schwinger~\cite{Schwinger:1962tn,Schwinger:1962tp}: 
a gauge boson may acquire a mass, even if the gauge symmetry forbids
a mass term at the level of the fundamental Lagrangian, provided that,
at zero momentum transfer ($q^2 = 0$), its vacuum polarization develops a pole with positive residue.
In what follows we will refer to this fundamental idea as the ``{\it Schwinger mechanism}''.
As we will demonstrate, this special mechanism is indeed operational in the gauge sector of QCD. 

The precise implementation of the Schwinger mechanism in the case of Yang-Mills theories
is commonly explored with continuum Schwinger function methods, such as 
the Schwinger-Dyson equations (SDEs)~\cite{Roberts:1994dr,Alkofer:2000wg,Fischer:2006ub,Roberts:2007ji,Binosi:2009qm,Cloet:2013jya,Huber:2018ned}
and the functional renormalization group~\cite{Pawlowski:2003hq,Pawlowski:2005xe,Cyrol:2017ewj,Corell:2018yil,Blaizot:2021ikl},
which describe the momentum evolution of correlation functions.
A crucial ingredient in all such studies is  
the incorporation of {\it longitudinally coupled massless poles}~\cite{Aguilar:2011xe,Ibanez:2012zk,Binosi:2017rwj,Aguilar:2017dco,Binosi:2017rwj,Eichmann:2021zuv} in the fundamental interaction vertices of the theory. 
These poles{ \it carry color} and correspond to massless bound state excitations,
whose formation is governed by appropriate Bethe-Salpeter equations (BSEs)~\cite{Aguilar:2011xe,Ibanez:2012zk,Aguilar:2017dco,Binosi:2017rwj}.
The inclusion of these poles in the diagrammatic expansion of the SDE that determines the function $\Delta(q^2)$, or,
equivalently, the gluon vacuum polarization, 
triggers the Schwinger mechanism, giving rise to a dynamically generated
gluon mass~\mbox{\cite{Aguilar:2011xe,Binosi:2012sj,Ibanez:2012zk,Aguilar:2016ock,Aguilar:2017dco,Binosi:2017rwj}.} 
It is important to emphasize that these massless poles do not produce divergences in physical observables (see Sec.~\ref{sec:sdec}),
and are intimately connected with the vortex picture of confinement, 
see, \eg Ch.7 of~\cite{Cornwall:2010upa} and references therein.

Since the massless poles are longitudinally coupled, they drop out from the {\it transversely projected} vertices  
employed in lattice simulations~\cite{Cucchieri:2006tf,Cucchieri:2008qm,Athenodorou:2016oyh,Duarte:2016ieu,Boucaud:2018xup,Aguilar:2019uob},  and only the {\it pole-free} parts contribute to the lattice results.
Nonetheless, the information on the existence of the massless poles is unequivocally encoded in the  
pole-free parts. 
Indeed, the additional key role of the massless poles is their participation in the 
realization of the Slavnov-Taylor identities (STIs)~\cite{Taylor:1971ff,Slavnov:1972fg} satisfied by the
vertices: the form of the STIs remains intact, but they are 
resolved through the crucial participation of the massless poles~\mbox{\cite{Eichten:1974et,Poggio:1974qs,Smit:1974je,Cornwall:1981zr,Papavassiliou:1989zd,Aguilar:2008xm,Binosi:2012sj,Aguilar:2016vin}}. Thus, when the gluon momentum of the STIs is taken to vanish, 
the  Ward identities (WIs) 
satisfied by the pole-free parts are displaced by a characteristic amount, dubbed the {\it displacement function}~\cite{Aguilar:2021uwa};
quite remarkably,  it is exactly identical to the {\it BS amplitude} for the pole formation found when solving
the corresponding dynamical equations.
The WI displacement function serves as a smoking-gun signal, whose precise measurement furnishes a highly nontrivial confirmation
of the action of the Schwinger mechanism in QCD~\cite{Aguilar:2016vin,Aguilar:2021uwa}.

In the present work we review the central concepts and techniques that are instrumental for the aforementioned framework,
focusing especially on recent developments that have enabled the systematic scrutiny and preliminary substantiation of 
this entire approach~\cite{Aguilar:2021uwa}. Furthermore, 
we emphasize the creative synergy between continuum methods and gauge-fixed lattice simulations of Schwinger functions, 
and elaborate on the tight connection between dynamics and symmetry,
as expressed through the SDEs and BSEs on the one hand, and the special displacement of the WIs on the other.

The article is organized as follows. 
In Sec.~\ref{sec:prel} we introduce the notation and the basic SDEs that govern the relevant two- and three-point correlation functions.
Then, in Sec.~\ref{sec:smg} we discuss in detail the salient features of the Schwinger mechanism in the context of the pure Yang-Mills theory. 
Sec.~\ref{sec:bse} is dedicated to the derivation and solution of the BSE that controls the emergence of poles in the three-gluon and
ghost-gluon vertices. 
In Sec.~\ref{sec:mass} we explain in detail how the gluon mass gets induced  at the level of the gluon SDE, once the
massless poles have been formed.
In Sec.~\ref{sec:widis} we introduce the concept of the WI displacement, and discuss its origin and implications, while 
in Sec.~\ref{sec:widis3g} we derive the  WI displacement function of the three-gluon vertex.
Then, in Sec.~\ref{sec:wilat} we determine this particular function from the
judicious combinations of ingredients obtained from lattice QCD.   
In Sec.~\ref{sec:int} we present a deep connection between the WI displacement and an important identity
that enforces the nonperturbative masslessness of the gluon in the absence of the Schwinger mechanism.
In Sec.~\ref{sec:theI} we analyze in detail the structure of the transition amplitude that connects an off-shell gluon with the composite excitation, 
and derive a compact formula that relates its value at the origin with the gluon mass. 
In continuation, in Sec.~\ref{sec:sdec} we demonstrate with an explicit example the mechanism that leads to the cancellation of the massless poles from the $S$-matrix.
Finally, our concluding remarks are presented in Sec.~\ref{sec:conc}. 

\section{ Notation and general framework}
\label{sec:prel}

In this section we establish the necessary notation and comment on the basic functional equations that determine the
dynamics of the correlation functions that we consider in this work. 

The Lagrangian density of the SU(N) Yang-Mills theory with covariant gauge-fixing is given by 
\begin{align}
 {\cal L}_{\mathrm{YM}} = -\frac14F^a_{\mu\nu}F^{a \mu\nu} + \frac{1}{2\xi} (\partial^\mu A^a_\mu)^2  - {\overline c}^a\partial^\mu D_\mu^{ab}c^b\,,
	\label{lagden}
\end{align}
where $A^a_\mu(x)$, $c^a(x)$, and ${\overline c}^a(x)$ denote the gauge, ghost, and antighost fields, respectively, 
$F^a_{\mu\nu}=\partial_\mu A^a_\nu-\partial_\nu A^a_\mu+gf^{abc}A^b_\mu A^c_\nu$ is the antisymmetric field tensor, 
$D_\mu^{ab} = \partial_\mu \delta^{ac} + g f^{amb} A^m_\mu$ is the covariant derivative in the adjoint representation,
and $\xi$ represents the gauge-fixing parameter. For the color indices we have
$a=1,\dots,N^2-1$, and $f^{abc}$ are the totally antisymmetric structure constants of SU($N$).
Obviously, the transition to QCD is implemented by adding to ${\cal L}_{\mathrm{YM}}$  
the appropriate kinetic and interaction terms for the quark fields. In what follows we will
work exclusively with \1eq{lagden}, corresponding to pure Yang-Mills.

Throughout the article we carry out calculations employing Feynman rules derived in the Minkowski space; then, 
the final expressions are passed to the Euclidean space, where their numerical evaluation is carried out\footnote{Strictly speaking, the rigorous definition of Quantum Field Theory takes place in Euclidean space, while its
 Minkowski formulation is largely a matter of convenience.}.  
Note that all derivations are valid for space-like momenta only; this allows 
the use of inputs taken from lattice simulations, and facilitates the comparison of the functional results to those of the lattice. 

In the {\it Landau gauge} that we employ,
the gluon propagator, \mbox{$\Delta^{ab}_{\mu\nu}(q)=-i\delta^{ab}\Delta_{\mu\nu}(q)$},
assumes the completely transverse form
\begin{align}
\Delta_{\mu\nu}(q) = \Delta(q^2) {P}_{\mu\nu}(q)\,, \qquad {P}_{\mu\nu}(q) := g_{\mu\nu} - q_\mu q_\nu/{q^2}\,, \quad\quad\Delta(q^2)= \Dr(q^2)/q^2\,,
\label{defgl}
\end{align}
where, for latter convenience, we have introduced the gluon dressing function,  $\Dr(q^2)$.

\begin{figure}[t]
  \includegraphics[width=1.0\textwidth]{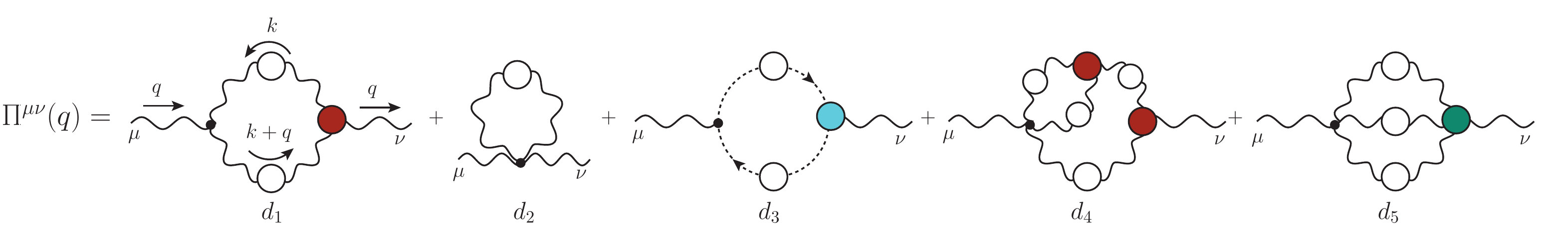}\\
\vspace{0.5cm} 
\includegraphics[width=0.9\textwidth]{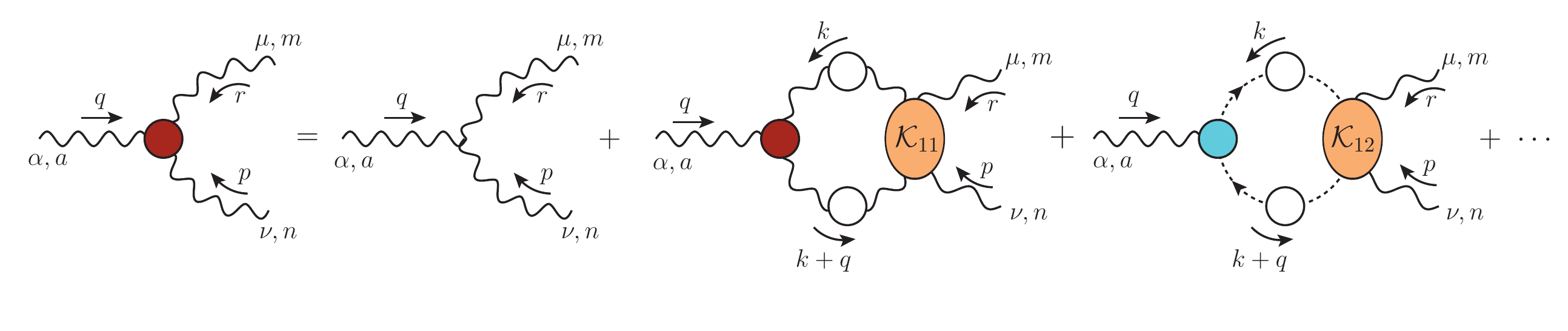}\\
\includegraphics[width=0.9\textwidth]{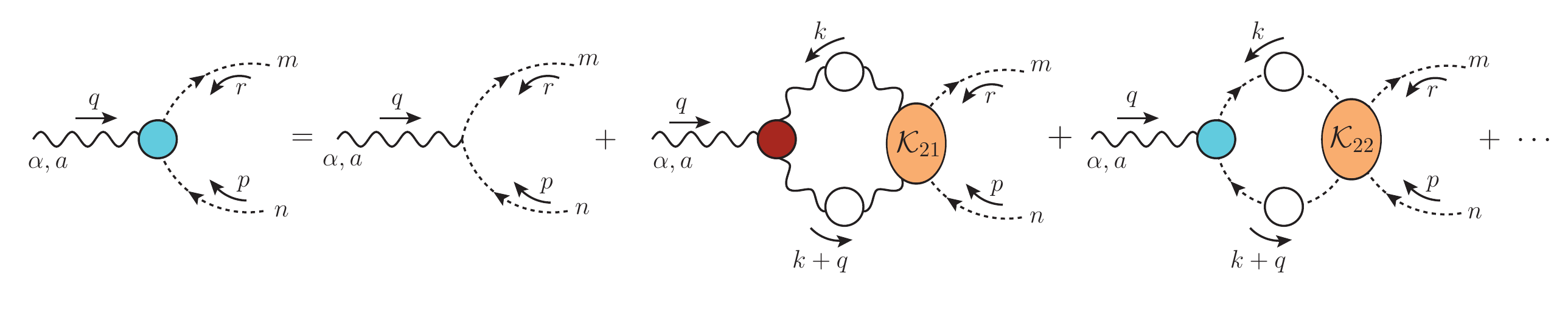}  
\caption{{\it First row}: The diagrammatic representation of the gluon self-energy. {\it Second row}: The SDE for the three-gluon vertex.
  {\it Third row}: The SDE for the  ghost-gluon vertex. White (colored) circles denote fully dressed propagators (vertices), while
the orange ellipses denote four-particle kernels.}
\label{fig:SDEs}
\end{figure}

The SDE for the gluon propagator is given by
\be
\Delta^{-1}(q^2)P_{\mu\nu}(q) = q^2P_{\mu\nu}(q)  + i {\Pi}_{\mu\nu}(q) \,,
\label{glSDE}
\ee
where  $\Pi_{\mu\nu}(q)$ is the gluon self-energy,  shown diagrammatically in the first row of \fig{fig:SDEs}.
Since $\Pi_{\mu\nu}(q)$ is transverse, 
\be
q^{\mu}\Pi_{\mu\nu}(q) = 0\,\,\,\,\Longrightarrow \,\,\,\, \Pi_{\mu\nu}(q) = \Pi (q^2) P_{\mu\nu}(q)\,,
\label{pitr}
\ee
from \1eq{glSDE} follows that  
\be
\Delta^{-1}(q^2) = q^2 + i \Pi (q^2).
\label{DeltaPi}
\ee
Note that, in the case of the infrared finite solutions for $\Delta^{-1}(q^2)$, found in lattice simulations
and numerous SDE analyses, the square of the gluon mass is identified with the finite
nonvanishing value of $\Delta^{-1}(q^2)$ at the origin~\cite{Aguilar:2006gr,Aguilar:2008xm}, namely 
\be
m^2 = \Delta^{-1}(0) \,.
\ee

In addition, we introduce the ghost propagator, denoted by $D^{ab}(q^2) = i \delta^{ab} D(q^2)$, 
and the corresponding dressing function, $F(q^2)$, defined as 
\begin{align}
    D(q^2) =  \frac{F(q^2)}{q^2}.
\label{defF}
\end{align}
According to numerous lattice simulations and studies in the
continuum, at $q=0$ the dressing function reaches a finite
nonvanishing value, see, \eg\cite{Aguilar:2008xm,Dudal:2008sp,Kondo:2009gc,Dudal:2012zx,Aguilar:2013xqa,Cyrol:2016tym,Boucaud:2018xup,Huber:2018ned,Aguilar:2021okw}.

%%%%%%%%%%%%%%%%%%%%%%%%%%%%%%%%%%
%Fig. 1
%%%%%%%%%%%%%%%%%%%%%%%%%%%%%%%%%%
%%%%%%%%%%%%%%%%%%%%%%%%%%%%%%

We next turn to the three-point sector of the theory, which, in the absence of dynamical quarks, contains 
the three-gluon and the ghost-gluon vertices, denoted by
\be
\fatg^{abc}_{\alpha\mu \nu}(q,r,p) =gf^{abc}\,\fatg_{\alpha\mu \nu}(q,r,p)\,,\qquad\qquad \fatg_\mu^{mna}(r,p,q) = -gf^{mna} \,\fatg_\mu(r,p,q) \,,
\label{twovert}
\ee
where all momenta are incoming, $q+p+r=0$. 
The corresponding SDEs that govern the evolution of $\fatg_{\alpha\mu \nu}(q,r,p)$ and  $\fatg_\mu(r,p,q)$ 
are shown in the second and third row of \fig{fig:SDEs}, respectively~\cite{Schleifenbaum:2004id,Huber:2012kd,Aguilar:2013xqa,Huber:2012zj,Blum:2014gna,Eichmann:2014xya,Williams:2015cvx,Aguilar:2017dco,Huber:2018ned,Papavassiliou:2022umz}. 
The omitted terms, indicated by ellipses, 
contain the fully-dressed four-gluon vertex (with incoming momentum $q$). In general,  
these latter contributions are technically harder to compute; nonetheless,
related studies seem to suggest that their impact on our analysis  
is likely to be rather small~\cite{Williams:2015cvx,Huber:2018ned}.

Note that in \fig{fig:SDEs} we show the Bethe-Salpeter version of the vertex SDE, whose   
main difference is that, inside the loops, the 
tree-level vertices (with incoming momentum $q$) have been replaced by their fully-dressed counterparts.
This substitution may be carried out provided that 
the corresponding four-particle kernels ${\cal K}_{ij}$ have been modified accordingly, in order to avoid overcounting.
For example, ladder graphs (straight boxes) must be omitted, while cross-ladder (crossed boxes) are retained 
(see \eg Fig.~7 of~\cite{Aguilar:2011xe}). This particular formulation of the SDE offers an important technical advantage: 
vertex renormalization constants, which otherwise
would appear explicitly multiplying individual diagrams, are fully absorbed by the additional dressed vertices
(see Sec.~\ref{sec:mass}).

The algebraic manipulations of potentially divergent integrals require the use of symmetry-preserving
regularization schemes. This is particularly important, because a flawed regularization procedure may introduce
artifacts that mimic the effects of a gluon mass. Dimensional regularization~\cite{Collins:1984xc} is especially well-suited for
this purpose, and will be adopted in what follows. Note, in fact, that its use is crucial for the demonstration of the
seagull identity~\cite{Aguilar:2009ke,Aguilar:2016vin} (see Sec.~\ref{sec:int}),
whose validity, in turn, guarantees the nonperturbative masslessness of the gluon,
when the Schwinger mechanism is not activated. 

For the loop integrals regularized with dimensional regularization we introduce the short-hand notation 
\be
\int_{k} :=\frac{\mu_{\s 0}^{\epsilon}}{(2\pi)^{d}}\!\int_{-\infty}^{+\infty}\!\!\mathrm{d}^d k\,,
\label{dqd}
\ee
where $d=4-\epsilon$ is the dimension of the space-time, and $\mu_{\s 0}$ denotes the 't Hooft mass.
It is understood that the regularization is employed until certain crucial cancellations
have taken place, and the procedure of renormalization has been duly carried out. Past that point, the
resulting equations are finite, and no regularization is needed.

\section{Schwinger mechanism in Yang-Mills theories}
\label{sec:smg}

Endowing gauge bosons with a mass in a field-theoretically consistent way is particularly subtle.
In this section we review the generation of a gluon mass
through the nonperturbative realization  
of the well-known Schwinger mechanism~\cite{Schwinger:1962tn,Schwinger:1962tp} in the context of a Yang-Mills theory
described by \1eq{lagden}.

The general idea of the mechanism
is best expressed in terms of the dimensionless vacuum polarization,  $\overline \Pi(q^2)$, defined
in terms of the gluon self-energy ${\Pi}(q^2)$ through ${\Pi}(q^2) = q^2 \overline\Pi(q^2)$, such that  
\be 
\Delta^{-1}({q^2})=q^2 [1 + i \overline \Pi(q^2)]\,.
\label{vacpol}
\ee
Schwinger's fundamental observation states that 
if the vacuum polarization $\overline \Pi(q^2)$
develops a pole at zero momentum transfer ($q^2=0$), then the 
vector meson (gluon) acquires a mass, even if the gauge symmetry 
prohibits the inclusion of a mass term at the level of the defining Lagrangian.
Thus, one has 
\be
\lim_{q \to 0} i\overline \Pi(q^2) = m^2/q^2 \,\,\Longrightarrow \,\,\lim_{q \to 0} \,\Delta^{-1}(q^2) = \lim_{q \to 0} \,(q^2 + m^2) \,\,\Longrightarrow \,\,\Delta^{-1}(0) = m^2\,,
\label{schmech}
\ee
and the vector meson picks up a mass, in the sense that its propagator at the origin saturates at a finite nonvanishing value,  
which is determined by the (positive) residue of the pole.

The argument described above is completely general, and its key conclusion does not depend on the dynamical details
that lead to the appearance of a massless pole in $\overline \Pi(q^2)$. Of course, in practice, depending on the characteristics of each theory, the circumstances that trigger the sequence described in \1eq{schmech} may be very distinct~\cite{Jackiw:1973tr,Jackiw:1973ha}. 
In the case of Yang-Mills theories, the origin of the pole is purely
dynamical, as first described in the classic work by Eichten and Feinberg~\cite{Eichten:1974et}.
In what follows we will present the modern implementation of
this scenario, as it has been developed in a series of articles during the past few years.

\begin{figure}[t]
\includegraphics[width=0.8\textwidth]{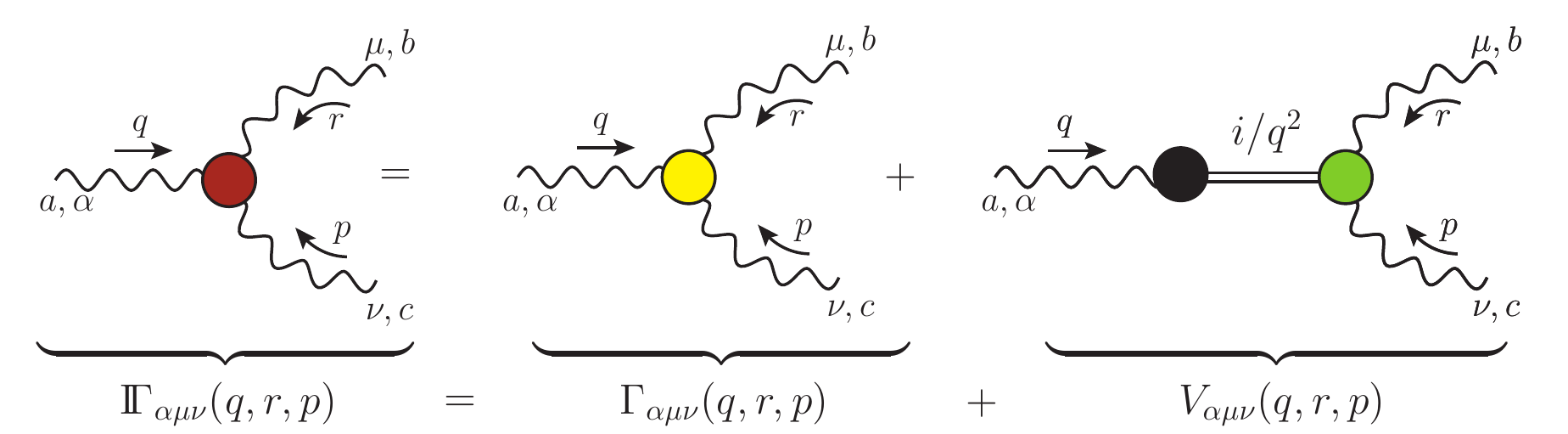} \\
\includegraphics[width=0.8\textwidth]{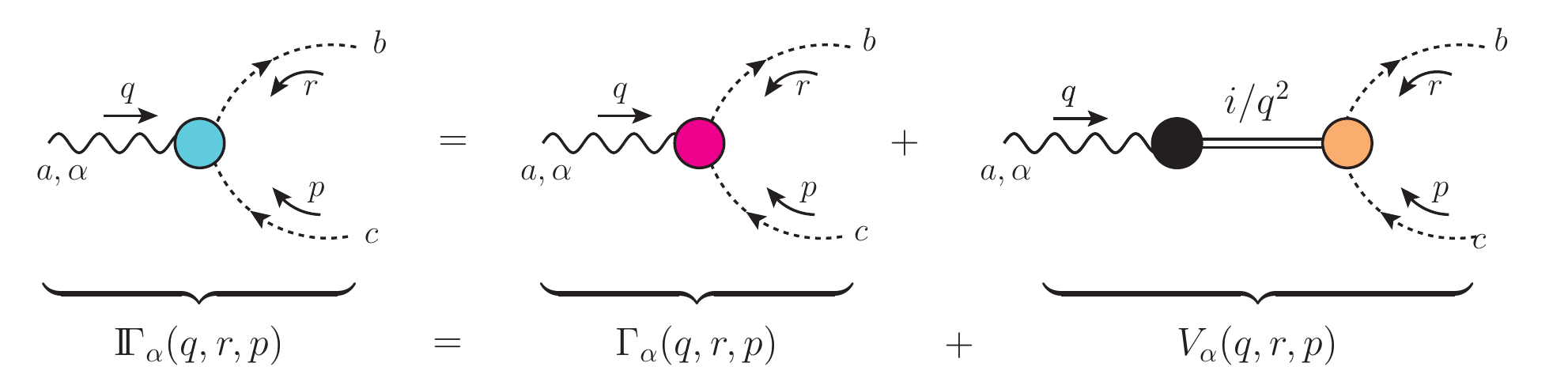}
\caption{The diagrammatic representation of the three-gluon and ghost-gluon vertices introduced in \1eq{fullgh}:
$\fatg_{\alpha\mu\nu}(q,r,p)$ (first row) and $\fatg_\alpha(q,r,p)$ (second row).
 The first term on the r.h.s. indicates the pole-free part, $\g_{\alpha\mu\nu}(q,r,p)$ or $\g_{\alpha}(q,r,p)$, 
while the second denotes the pole term  $V_{\alpha\mu\nu}(q,r,p)$ or $V_{\alpha}(q,r,p)$.}
\label{fig:poles}
\end{figure}

The general idea is that the nonperturbative vertices of the theory develop special 
massless composite excitations, which find their way into 
the gluon vacuum polarization through the SDE of \fig{fig:SDEs}~\cite{Aguilar:2011xe,Binosi:2012sj,Ibanez:2012zk,Binosi:2017rwj,Aguilar:2017dco}. 
In particular, the three-gluon and ghost-gluon vertices assume the general form (see \fig{fig:poles})
\bea 
\fatg_{\alpha\mu\nu}(q,r,p) &=& \g_{\alpha\mu\nu}(q,r,p) + V_{\alpha\mu\nu}(q,r,p) \,,\qquad
\nonumber\\
\fatg_\alpha(q,r,p) &=& \g_{\alpha}(q,r,p) + V_{\alpha}(q,r,p)\,,
\label{fullgh}
\eea
where $\g_{\alpha\mu\nu}(q,r,p)$ and $\g_\alpha(r,p,q)$ are the pole-free components of the two vertices, while 
$V_{\alpha\mu\nu}(q,r,p)$ and $V_{\alpha}(q,r,p)$ contain {\it longitudinally coupled} bound-state poles, \ie of the
special tensorial structure 
\bea 
V_{\alpha\mu\nu}(q,r,p) &=& \frac{q_\alpha}{q^2}  C_{\mu\nu}(q,r,p) + \frac{r_\mu}{r^2} A_{\alpha\nu}(q,r,p) + \frac{p_\nu}{p^2} B_{\alpha\mu}(q,r,p) \,,
\nonumber\\
V_\alpha(q,r,p) &=& \frac{q_\alpha}{q^2}\Cgh(q,r,p)\,.
\label{eq:Vgen}
\eea
We emphasize that the pole-free components are not ``regular'' functions, in the strict sense of the term,
because certain of their form factors diverge logarithmically in the infrared, see,\eg \cite{Aguilar:2013vaa}.
Note also that the corresponding tree-level expressions are given by  
\be 
\gz^{\alpha\mu\nu}(q,r,p) = (q - r)^\nu g^{\alpha\mu} + (r - p)^\alpha g^{\mu\nu} + (p - q)^\mu g^{\nu\alpha} \,, \qquad\qquad \gz^\alpha(q,r,p) = r^\alpha\,.
\label{bare3g}
\ee

%%%%%%%%%%%%%%%%%%%%%%%%%%%%%%%%%%
%Fig. 2
%%%%%%%%%%%%%%%%%%%%%%%%%%%%%%%%%%
%%%%%%%%%%%%%%%%%%%%%%%%%%%%%%%%%%

The longitudinal nature of the $V_{\alpha\mu\nu}(q,r,p)$ and $V_{\alpha}(q,r,p)$ is easily established
at the level of \fig{fig:poles}. Specifically,
the black circle denotes the transition amplitude, $I_\alpha(q)$, connecting a gluon with a (massless composite) scalar; 
since this amplitude depends on a single momentum, $q$, and a single Lorentz index, $\alpha$, its general form is simply
$I_\alpha(q) = q_\alpha I (q^2)$, where $I (q^2)$ is a scalar form factor~\cite{Aguilar:2011xe,Ibanez:2012zk}.
In the case of $V_{\alpha\mu\nu}(q,r,p)$, Bose symmetry enforces the same property in the remaining two channels,
thus finally accounting for the general form given in \1eq{eq:Vgen}. The form factor $I (q^2)$ is eventually absorbed into
$\Cgh(q,r,p)$ and $C_{\mu\nu}(q,r,p)$; additional details on the structure of $I_\alpha(q)$ will be given in Sec.~\ref{sec:theI}.

An immediate consequence of \1eq{eq:Vgen} is that $V_{\alpha\mu\nu}(q,r,p)$ and $V_{\alpha}(q,r,p)$ satisfy the crucial relations
\be
\label{eq:transvp}
{P}_{\alpha'}^{\alpha}(q){P}_{\mu'}^{\mu}(r){P}_{\nu'}^{\nu}(p) V_{\alpha\mu\nu}(q,r,p) = 0 \,,\qquad {P}_{\alpha'}^{\alpha}(q) V_\alpha(q,r,p) = 0\,;
\ee
therefore, they drop out from the typical quantities studied on the lattice,  
which involve the {\it transversely projected} vertices [see, \eg \1eq{asymlat}]. In fact, as we will see in detail in Sec.~\ref{sec:sdec}, \1eq{eq:transvp}
is instrumental for the cancellation of all pole divergences from physical observables, such as $S$-matrix elements.

Note that, even though $V_{\alpha\mu\nu}(q,r,p)$ possesses poles in all three of its channels,
only the one associated with the $q$-channel, \ie the channel that carries
the physical momentum entering in the gluon propagator, is actually relevant. 
In fact, the longitudinal structure of $V_{\alpha\mu\nu}(q,r,p)$, together with the fact that we work in the Landau gauge, and hence, 
gluon propagators inside Feynman diagrams are transverse, leads to the simplification 
\be
{P}_{\mu'}^{\mu}(r){P}_{\nu'}^{\nu}(p) V_{\alpha\mu\nu}(q,r,p) = \frac{q_\alpha}{q^2} {P}_{\mu'}^{\mu}(r){P}_{\nu'}^{\nu}(p) C_{\mu\nu}(q,r,p)\,.
\label{eq:PPG}
\ee
Thus, for our analysis we only require the tensorial decomposition of the term  $C_{\mu\nu}(q,r,p)$ in \1eq{eq:Vgen}, given by 
\be 
C_{\mu\nu}(q,r,p) = C_1 \, g_{\mu\nu}  + C_2\, r_\mu r_\nu  + C_3 \, p_\mu p_\nu  +  C_4 \, r_\mu p_\nu  + C_5 \,  p_\mu r_\nu  \,,
\label{eq:Cdec}
\ee
where \mbox{$C_j := C_j(q,r,p)$}. Now, when the $C_{\mu\nu}(q,r,p)$ of \1eq{eq:Cdec} is 
substituted into \1eq{eq:PPG}, and the relation $q+p+r =0$ is appropriately employed, only two form factors survive, 
\be
 {P}_{\mu'}^{\mu}(r){P}_{\nu'}^{\nu}(p) V_{\alpha\mu\nu}(q,r,p) = \frac{q_\alpha}{q^2} {P}_{\mu'}^{\mu}(r){P}_{\nu'}^{\nu}(p)
 \left[ C_1 \, g_{\mu\nu} + C_5 q_\mu q_\nu\right]\,.
\label{eq:PPG2}
\ee
Since we are mostly interested in the behavior of the gluon propagator at the origin, 
in what follows we will be expanding the relevant equations around
$q=0$, keeping terms at most linear in $q$. In such an expansion, the term proportional to $C_5$ in
\1eq{eq:PPG2} is subleading, being of order ${\cal O}(q^2)$. So, finally, one ends up with a single relevant form factor
per pole vertex, namely 
$C_1(q,r,p)$, related to $V_{\alpha\mu\nu}(q,r,p)$, and $\Cgh(q,r,p)$, the unique component of $V_\alpha(q,r,p)$.

In what follows we will repeatedly use the Taylor expansion 
of a function $f(q,r,p)$ around $q=0$ ($p=-r$), given by 
\bea
\lim_{q \to 0} f(q,r,p) &=& f(0,r,-r) + q^\alpha \left[\frac{\partial f(q,r,p)}{\partial q^\alpha}\right]_{q=0} + \cdots\,,
\nonumber\\
 &=& f(0,r,-r) + 2 (q\cdot r)\left[ \frac{\partial f(q,r,p)}{\partial p^2} \right]_{q = 0} + \cdots \,.
\label{Taylf}
\eea
where the ellipses denote terms of ${\cal O}(q^2)$ or higher. 

There are two important results relevant for the Taylor expansion of $C_1(q,r,p)$ and $\Cgh(q,r,p)$, namely  
\be
C_1(0,r,-r) = 0 \,,\qquad\qquad \Cgh(0,r,-r) = 0 \,.
\label{C1_0}
\ee
The first relation follows directly from the Bose symmetry of the three-gluon vertex, which implies that 
$C_1(q,r,p)=-C_1(q,p,r)$. The justification of the second relation in \1eq{C1_0} is less immediate, relying on special 
relations~\cite{Binosi:2002ez,Binosi:2009qm} linking $\fatg_\alpha(q,r,p)$ with the vertex $\widetilde\Gamma_\alpha(q,r,p)$
introduced in Sec.~\ref{sec:widis}.
As we will see in Sec.~\ref{sec:widis3g}, the first relation in \1eq{C1_0} will be derived 
in a completely independent way from the fundamental WIs satisfied by the three-gluon vertex.

In view of \1eq{C1_0}, the Taylor expansion of ${C}_1(q,r,p)$ and ${C}(q,r,p)$ around $q=0$ yields
\be
\label{eq:taylor_C}
\lim_{q \to 0} {C}_1(q,r,p) =  2 (q\cdot r) {\mathbb C}(r^2) \, + \cdots\,,  \qquad
\lim_{q \to 0} {C}(q,r,p) =  2 (q\cdot r){\cal C}(r^2) \, +   \cdots\,,
\ee
with
\be
{\mathbb C}(r^2) := \left[ \frac{\partial {C}_1(q,r,p)}{\partial p^2} \right]_{q = 0}\,,\qquad\qquad
{\cal C}(r^2) :=   \left[ \frac{\partial {C}(q,r,p)}{\partial p^2} \right]_{q = 0}\,.
\label{eq:theCs}
\ee
The functions ${\mathbb C}(r^2)$ and ${\cal C}(r^2)$ are central to the ensuing analysis.
In particular, there are three pivotal points related to them that will be elucidated in the next sections.
First, we will prove that nonvanishing ${\mathbb C}(r^2)$ and ${\cal C}(r^2)$ do indeed emerge 
from the corresponding dynamical equations for $\fatg_{\alpha\mu\nu}(q,r,p)$ and $\fatg_\alpha(q,r,p)$. In fact, 
as we will see in the next section, these two functions turn out to be the {\it BS amplitudes} describing the formation of a gluon-gluon
and a ghost-antighost {\it colored} composite bound state, respectively.
Second, we will derive the formula that expresses the gluon mass in terms of
${\mathbb C}(r^2)$ and ${\cal C}(r^2)$, and will demonstrate that it furnishes a result compatible with the lattice simulations;
that will be the subject of Sec.~\ref{sec:mass}.
Third, in Sec.~\ref{sec:widis}
we will elaborate on the notion of the WI displacement, and show that
${\mathbb C}(r^2)$ corresponds precisely to the
{\it displacement function} that quantifies the modification of the WIs satisfied by  $\Gamma_{\alpha\mu\nu}(q,r,p)$
in the presence of massless poles.

\section{Dynamical formation of massless poles}
\label{sec:bse}

We next turn to the study of the precise dynamics that leads to the formation of the poles that
comprise the components $V_{\alpha\mu\nu}(q,r,p)$ and $V_{\alpha}(q,r,p)$ entering in \1eq{eq:Vgen}.
The fundamental equations that control this process are the SDEs
for $\fatg_{\alpha\mu\nu}(q,r,p)$ and $\fatg_\alpha(q,r,p)$, 
which, in the limit $q\to 0$, provide a set of linear BSEs for the quantities 
${\mathbb C}(r^2)$ and ${\cal C}(r^2)$, defined in \2eqs{eq:taylor_C}{eq:theCs}.

In what follows we set $\lambda := i g^2 C_{\rm A}/2$, where 
$C_\mathrm{A}$ is the Casimir eigenvalue of the adjoint representation [$N$ for $SU(N)$].
Then, the system of SDEs shown in \fig{fig:SDEs} assumes the form~\cite{Aguilar:2021uwa}
\bea
\fatg^{\alpha\mu\nu} &=& \gz^{\alpha\mu\nu} - \lambda \int_k \fatg^{\alpha\beta\gamma}\Delta_{\beta\rho} \Delta_{\gamma\sigma} {\cal K}_{11}^{\mu\nu\sigma\rho}
+ 2 \lambda  \int_k \fatg^{\alpha} D D {\cal K}_{12}^{\mu\nu} \,, \nonumber\\
\fatg^{\alpha} &=& \gz^\alpha - \lambda \int_k \fatg^{\alpha\beta\gamma}\Delta_{\beta\rho} \Delta_{\gamma\sigma}{\cal K}_{21}^{\sigma\rho} 
- \lambda \int_k \fatg^{\alpha} D D {\cal K}_{22} \,,
\label{BSE_inhom}
\eea
where, for compactness, all momentum arguments, indicated explicitly on the diagrams of \fig{fig:SDEs}, have been suppressed.

Next, we substitute into \1eq{BSE_inhom} the expressions for the fully-dressed vertices given in \1eq{fullgh}.
In addition, in order to exploit \1eq{eq:PPG2}, the first of the two equations is multiplied by the factor ${P}_{\mu'\mu}(r){P}^{\mu'}_{\nu}(p)$.
Then, as the limit $q\to 0$ is taken, two tensorial structures emerge: one associated with the pole-free terms, which is 
proportional to $r^{\alpha}$, and one associated with the pole terms, being proportional to $q^{\alpha}$.
The matching of the terms proportional to $q^{\alpha}$ on both sides leads to the desired
BSEs, while the matching of the terms proportional to $r^{\alpha}$ furnishes a dynamical system for the so-called 
``soft-gluon'' form-factors of $\g_{\alpha\mu\nu}(q,r,p)$ and $\g_{\alpha}(q,r,p)$
[see, for example, \1eq{TGamma}]. Focusing on the BSEs, the limit $q\to 0$ 
activates \1eq{eq:taylor_C}, and the functions ${\mathbb C}(r^2)$ and ${\mathbb C}(r^2)$ make their appearance.

%%%%%%%%%%%%%%%%%%%%%%%%%%%%%%%%%%
%Fig. 3
%%%%%%%%%%%%%%%%%%%%%%%%%%%%%%%%%%
\begin{figure}[t]
\includegraphics[width=0.75\textwidth]{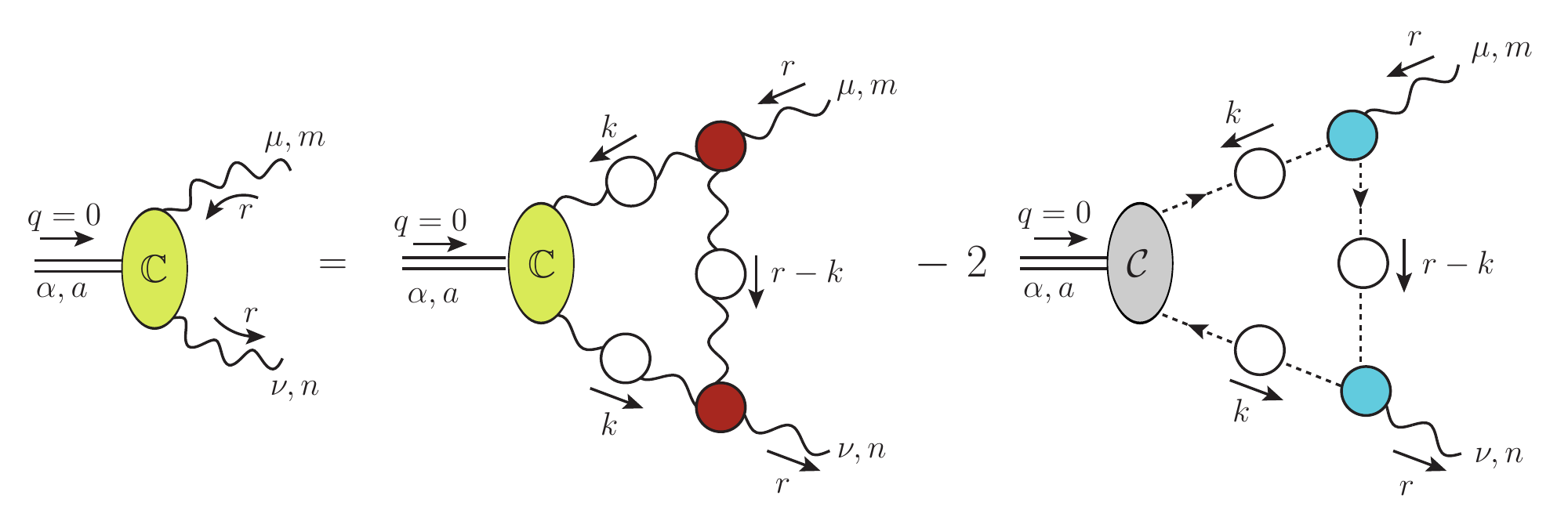}\\
\includegraphics[width=0.75\textwidth]{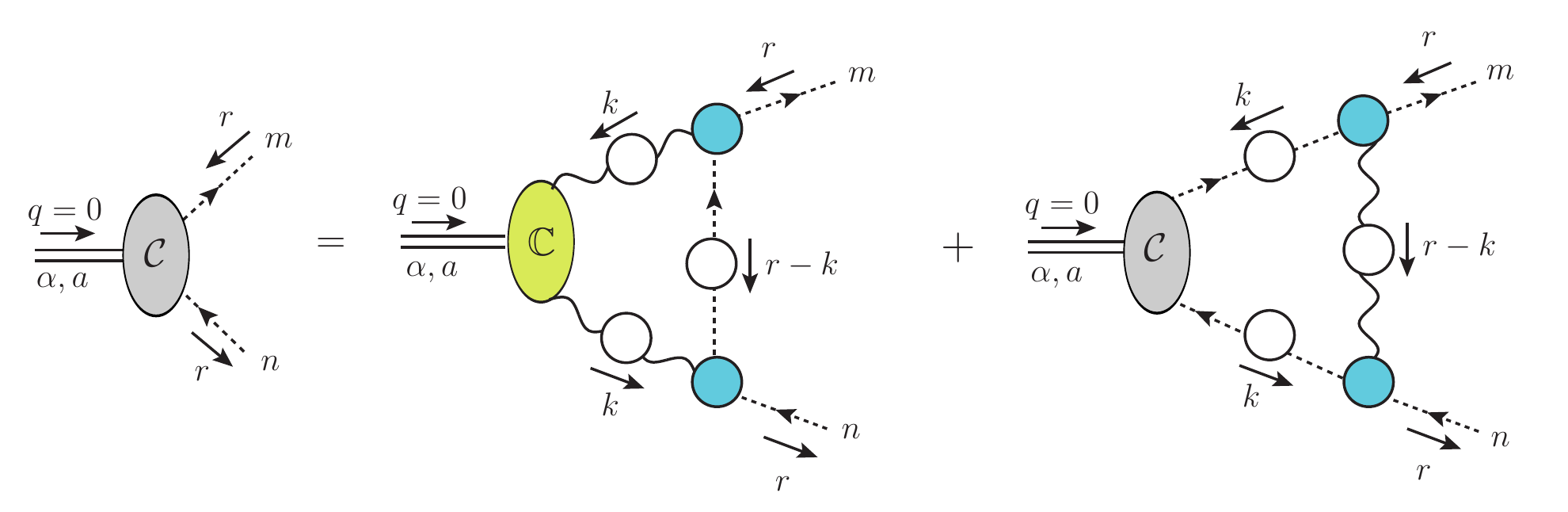}  
\caption{The diagrammatic representation of the coupled system of BSEs that governs the evolution of the 
functions ${\mathbb C}(r^2)$ and ${\cal C}(r^2)$. }
\label{fig:BSEs}
\end{figure}
%%%%%%%%%%%%%%%%%%%%%%%%%%%%%%%

Specifically, after employing the useful relation  
\be
\int_k (q\cdot k) \,f(k,r) = \frac{(q\cdot r)}{r^2} \int_k (r\cdot k) \,f(k,r)\,,
\label{auxrel}
\ee
with $f(k,r)$ denoting a generic kernel, we arrive at a system of homogeneous equations 
involving $\Cfat(r^2)$ and $\C(r^2)$, namely (see \fig{fig:BSEs})
\bea
\Cfat(r^2) &=& - \frac{\lambda}{3}\int_k \Cfat(k^2) \Delta^2(k^2)P_{\rho\sigma}(k)P_{\mu\nu}(r) \widetilde{\cal K}_{11}^{\mu\nu\sigma\rho} 
+ \frac{2 \lambda}{3} \int_k \C(k^2) D^2(k^2)P_{\mu\nu}(r) \widetilde{\cal K}_{12}^{\mu\nu}  \,, \nonumber\\
\C(r^2) &=& - \lambda \int_k \Cfat(k^2)\Delta^2(k^2)P_{\sigma\rho}(k) \widetilde{\cal K}_{21}^{\sigma\rho} 
- \lambda \int_k \C(k^2)D^2(k^2) \widetilde{\cal K}_{22} \,,
\label{BSE_hom}
\eea
where $\widetilde{\cal K}_{ij} := (r\cdot k /r^2)\, {\cal K}_{ij}(r,-r,k,-k)$. Note that the above
derivation has been carried out in Minkowski space, and hence the imaginary factor of $i$ in
the definition of $\lambda$. Before proceeding with the numerical analysis, the result must
be passed to the Euclidean space, following
standard conversion rules. Note, in particular, that the integral measure changes according to $d^4k \to i d^4k_{\s {\rm E}}$; this additional factor of $i$
combines with $\lambda$ to yield real expressions. 

The system of integral equations given in \1eq{BSE_hom} are the  
BSEs that govern the formation of massless colored bound states out of two gluons and a ghost-antighost pair; the
functions $\Cfat(r^2)$ and $\C(r^2)$ are the corresponding BS amplitudes. It is therefore of the utmost importance 
to find nontrivial solutions for these functions, even if certain simplifying assumptions will
be implemented at the level of the ingredients entering in \1eq{BSE_hom}. 

\begin{figure}[t!]
%\hspace{-1.5cm}
\begin{minipage}[b]{0.45\linewidth}
\centering
%\hspace{-1.5cm}
\includegraphics[scale=0.67]{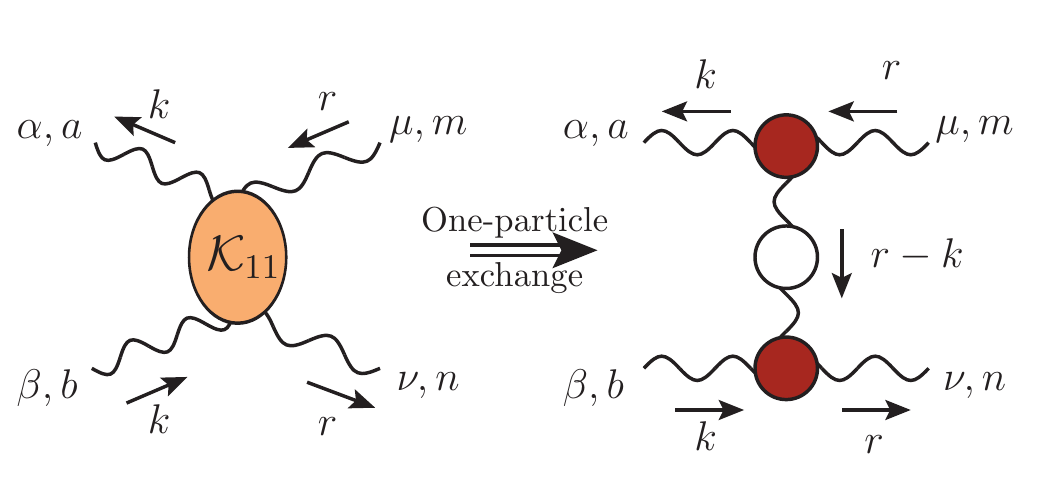} 
\end{minipage}
\hspace{0.5cm}
\begin{minipage}[b]{0.45\linewidth}
\includegraphics[scale=0.67]{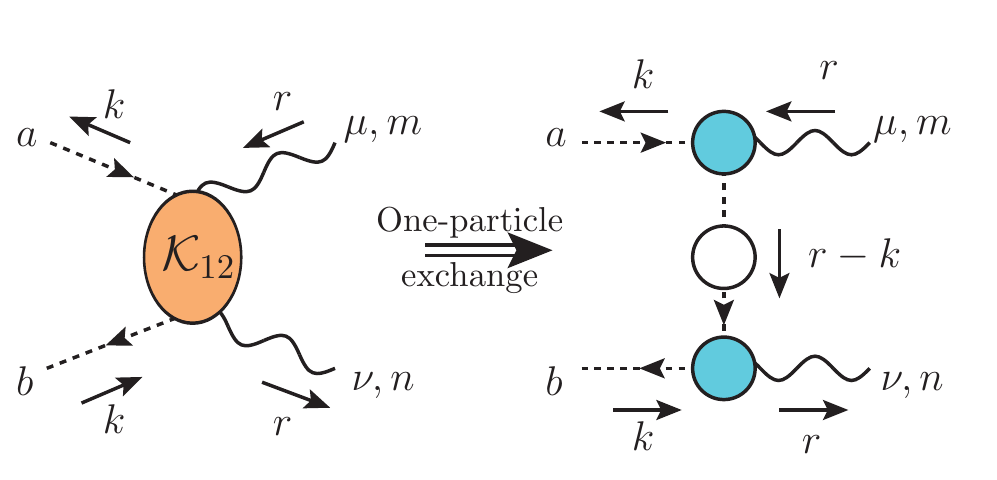}
\end{minipage}
\begin{minipage}[b]{0.45\linewidth}
\centering
%\hspace{-1.5cm}
\includegraphics[scale=0.67]{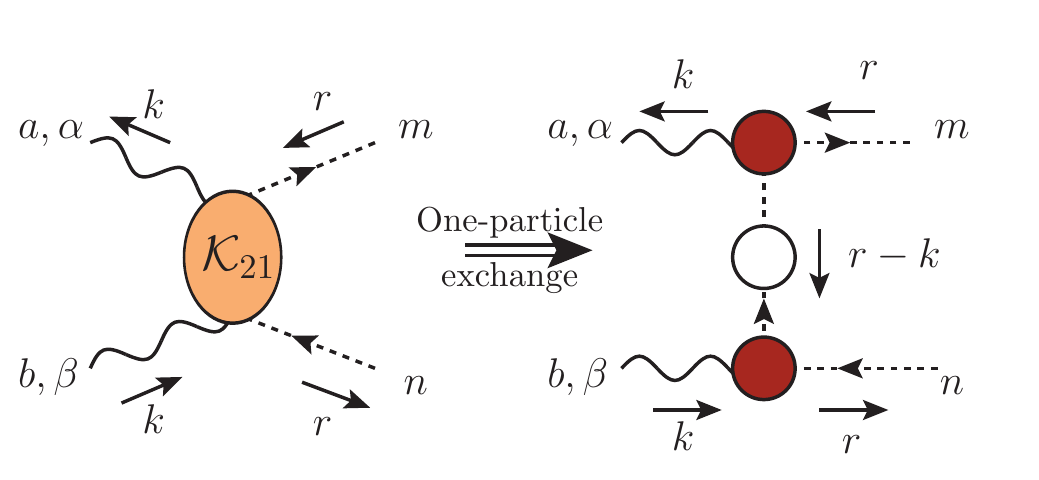} 
\end{minipage}
\hspace{0.5cm}
\begin{minipage}[b]{0.45\linewidth}
\includegraphics[scale=0.67]{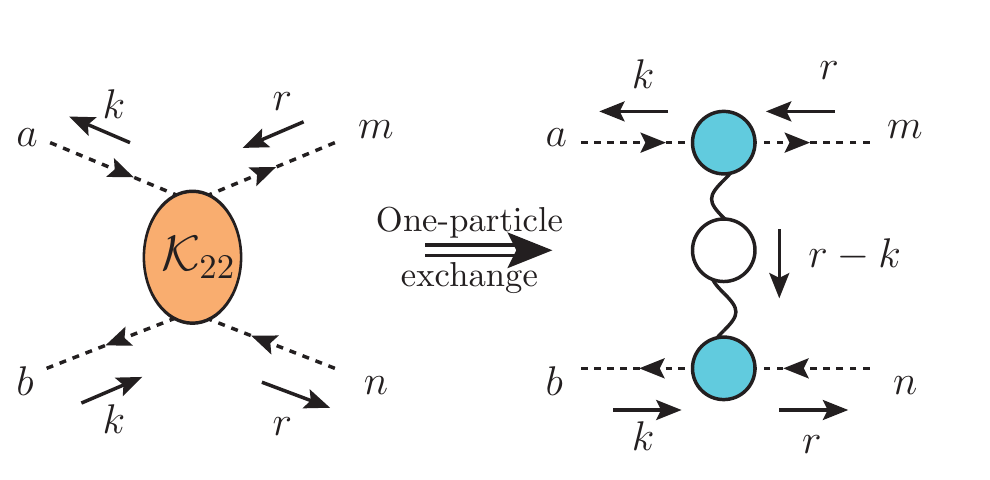}
\end{minipage}
\caption{The one-particle exchange approximations of the kernels ${\cal K}_{ij}$, and the associated kinematic conventions.}
\label{fig:kernels}
\end{figure}

To that end, we  employ the ``one-particle exchange'' approximation for the kernels  ${\cal K}_{ij}$, shown in \fig{fig:kernels};
the ingredients required for their evaluation 
are the fully dressed propagators and vertices.
Note that
only the pole-free parts $\g_{\alpha\mu\nu}$ and $\g_{\alpha}$ are relevant for the evaluation of the kernels ${\cal K}_{ij}$, because
the various projections implemented during the derivation of \1eq{BSE_hom} activate  \1eq{eq:transvp}; 
the reader is referred to~\cite{Aguilar:2021uwa}, and in particular Appendix A therein, for further details. 

%%%%%%%%%%%%%%%%%%%%%%%%%%%%%%%%%%
%Fig. 4
%%%%%%%%%%%%%%%%%%%%%%%%%%%%%%%%%%

\begin{figure}[t!]
%\hspace{-1.5cm}
\begin{minipage}[b]{0.45\linewidth}
\centering
\hspace{-1.5cm}
\includegraphics[scale=0.4]{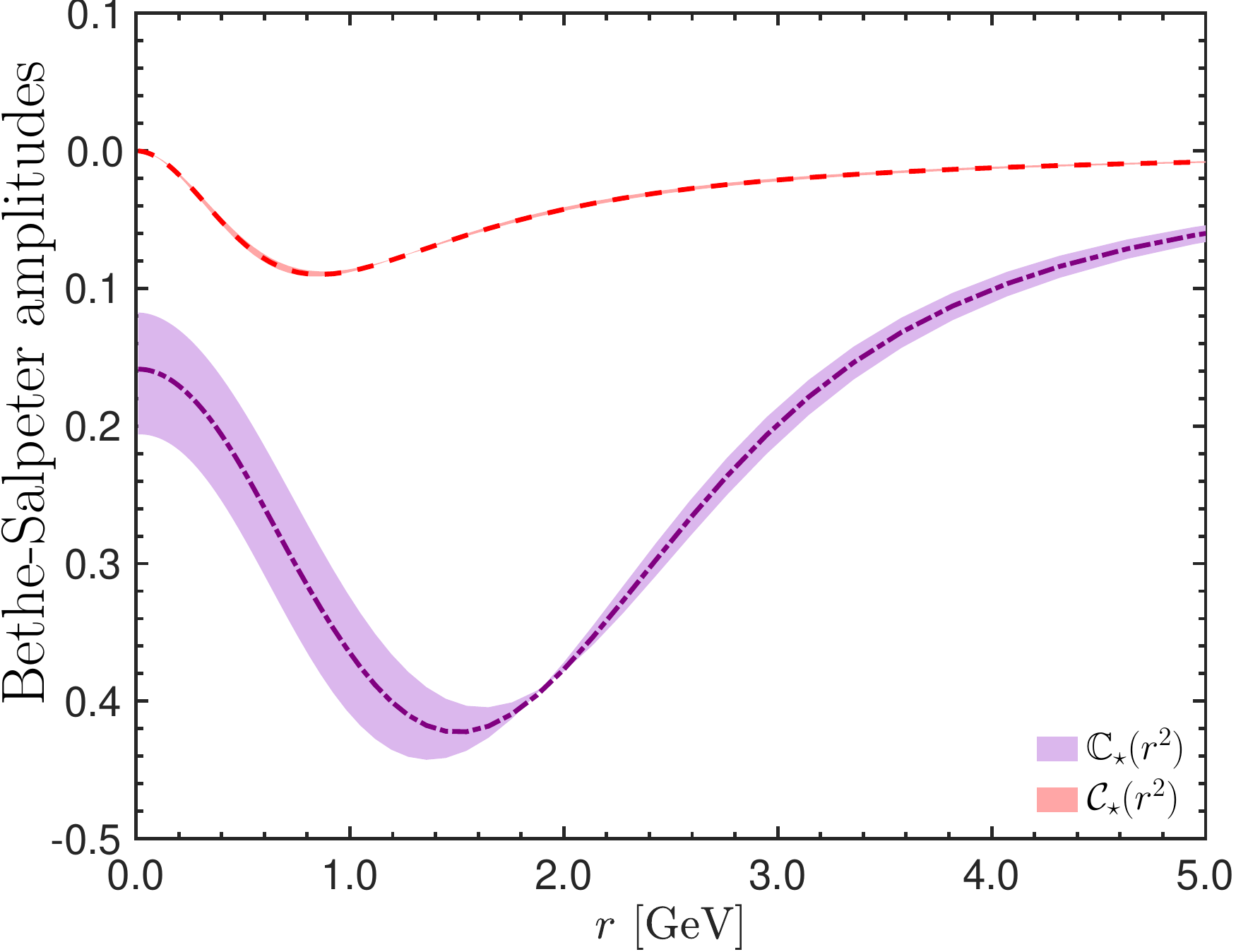}
\vspace{-1.1cm}
\hspace{-1.cm}
\end{minipage}
\begin{minipage}[b]{0.50\linewidth}
\includegraphics[scale=0.9]{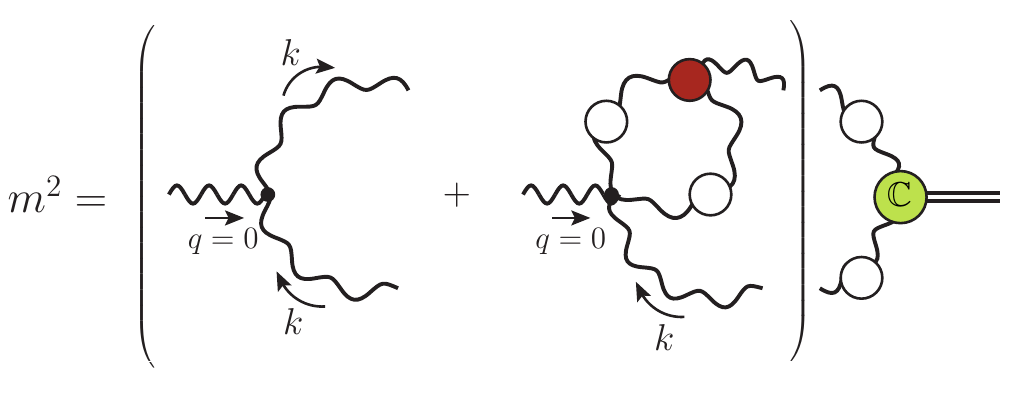} 
\end{minipage}
\vspace{0.65cm}
\caption{{\it Left panel}:
The solutions for $\CB(r^2)$ (purple dot-dashed) and $\Cc(r^2)$ (red dashed) obtained from the coupled BSE system of \1eq{BSE_hom}.
{\it Right panel}: Diagrammatic representation of the mass term that emerges from the insertion of the pole term
$V_{\nu\alpha'\beta'}$ into the diagrams
$d_1$ and $d_4$ of \fig{fig:SDEs}.}
\label{fig:C_gl_Cgh}
\end{figure}

The system of integral equations in \1eq{BSE_hom} is linear and homogeneous in the unknown functions, thus
corresponding to an eigenvalue problem, which finally singles out a special value for the strong coupling, 
$\alpha_s=g^2/4 \pi$. Specifically, we find that $\alpha_s=0.63$ when the renormalization point $\mu =4.3$ GeV.  
For this particular value of $\alpha_s$,  we find {\it nontrivial solutions} for $\Cfat(r^2)$ and $\C(r^2)$, which are 
shown in the left panel of \fig{fig:C_gl_Cgh}.
This value  of $\alpha_s$ is  to be contrasted with the corresponding value obtained within the
concrete renormalization scheme that we employ. Specifically, we work within the general framework of the  
momentum subtraction (MOM) scheme~\cite{Hasenfratz:1980kn}, where two-point functions acquire their tree-level
expressions at a given scale $\mu$, \ie \mbox{$\Delta^{-1}_{\s R}(\mu^2) = \mu^2$}. Within this scheme we adapt the so-called 
{\it asymmetric} version~\cite{Alles:1996ka,Davydychev:1997vh,Boucaud:1998bq,Boucaud:1998xi,Athenodorou:2016oyh,Aguilar:2019uob},
characterized by the condition $\Ls(\mu^2) =1$, where $\Ls(r^2)$ is the form factor of the three-gluon vertex
in the soft-gluon (``asymmetric'') configuration, see \1eq{asymlat}; the estimated value of $\alpha_s$ within this scheme
is $\alpha_s=0.27$~\cite{Athenodorou:2016oyh,Aguilar:2019uob}. 

It is of course natural to interpret  
this numerical discrepancy in the values of $\alpha_s$
as an artifact of the truncation employed, and especially the
approximation of the kernels ${\cal K}_{ij}$ by their one-particle exchange diagrams. It is worth mentioning that,  
according to a preliminary analysis,  moderate modifications
of the kernel affect the value of $\alpha_s$ considerably, leaving 
the form of the solutions found for $\CB(r^2)$ and $\Cc(r^2)$ essentially unmodified.
This observation implies that a more complete  
knowledge of the corresponding BSE kernels is required in order 
to decrease $\alpha_s$ towards its correct MOM value. On the other hand,   
the solutions obtained with the approximations described above should be considered as fairly reliable.

It is important to stress that, due to  the homogeneity and linearity of 
\1eq{BSE_hom}, the overall scale of the solutions is undetermined, since  
the multiplication of a given solution by an arbitrary real constant produces another solution.
In the case of the solutions shown in \fig{fig:C_gl_Cgh}, denoted by $\CB(r^2)$ and $\Cc(r^2)$, 
the scale has been fixed by requiring the best possible matching with the corresponding 
result obtained for $\Cfat(r^2)$ from the WI displacement in Sec.~\ref{sec:wilat}.
This scale ambiguity originates from considering only the leading order terms of the BSEs in the expansion around $q = 0$;
it may be resolved if further orders in $q$ are kept, because of the additional inhomogeneous terms that they induce, see, \eg\mbox{\cite{Nakanishi:1969ph,Maris:1997tm,Blank:2010bp}}.

Note finally that $\CB(r^2)$ is considerably larger in magnitude than $\Cc(r^2)$~\cite{Aguilar:2017dco}, indicating that the
three-gluon vertex accounts for the bulk of the gluon mass.

%%%%%%%%%%%%%%%%%%%%%
% Figure 5  - Bound state amplitudes obtained from BSE 
%%%%%%%%%%%%%%%%%%%%%%%%%%%%%
%%%%%%%%%%%%%%%%%%%%%%%%%%%%%

\section{Gluon mass via the Schwinger mechanism}
\label{sec:mass} 

In this section we elucidate in some detail on how the inclusion of vertices with massless poles
into the  gluon SDE triggers the Schwinger mechanism, leading to the generation of a gluon mass.

In order to fix the ideas, let us 
consider a specific example, namely the diagram $d_1^{\mu\nu}(q)$ shown in the first row of \fig{fig:SDEs}, corresponding
to the expression 
\be
i d_1^{\mu\nu}(q) =  \lambda \int_k {\Gamma}^{\mu\alpha\beta}_0(q,k,-k-q)
\Delta_{\alpha\alpha'}(k) \Delta_{\beta\beta'}(k+q) \fatg^{\nu\alpha'\beta'}(q,k,-k-q) \,.
\label{d1}
\ee
Let us next use for $\fatg^{\nu\alpha'\beta'}(q,k,-k-q)$  
a three-gluon vertex containing the type of massless poles described above (evidently with the appropriate relabelling of indices).
In particular, substitute  into \1eq{d1} the vertex given in the first equation of \1eq{fullgh},
with \mbox{$\{\alpha,\mu,\nu\} \to  \{\nu,\alpha',\beta'\}$}, and denote by ${\widehat d}_{1}^{\,\mu\nu}(q)$ 
the contribution originating exclusively from the term $V_{\nu\alpha'\beta'}(q,k,-k-q)$, namely
\be
  i {\widehat d}_{1}^{\,\mu\nu}(q) = \lambda \int_k {\Gamma}^{\mu\alpha\beta}_0(q,k,-k-q)
  \Delta_{\alpha\alpha'}(k) \Delta_{\beta\beta'}(k+q) V^{\nu\alpha'\beta'}(q,k,-k-q)\,.
\label{d1hat}
\ee
Now, from the general form of $V^{\alpha\mu\nu}$ given in \1eq{eq:Vgen} it is evident that, since we work in the
Landau gauge, the only term that survives in \1eq{d1hat} is proportional to $C^{\alpha'\beta'}(q,k,-k-q)$,
such that 
\be
 i {\widehat d}_{1}^{\,\mu\nu}(q) = \lambda \frac{q^{\nu}}{q^2} \int_k {\Gamma}^{\mu\alpha\beta}_0(q,k,-k-q)
 \Delta_{\alpha\alpha'}(k) \Delta_{\beta\beta'}(k+q) C^{\alpha'\beta'}(q,k,-k-q) \,.
\label{gen2}
\ee
Clearly, ${\widehat d}_{1}^{\,\mu\nu}(q)$ can only be proportional to $q^{\mu}q^{\nu}$; so, we set 
${\widehat d}_{1}^{\,\mu\nu}(q)= (q^{\mu}q^{\nu}/q^2) {\widehat d}_{1}(q^2)$, with
\footnote{Since $\Pi_{\mu\nu}(q) = (g_{\mu\nu} -q^{\mu}q^{\nu}/q^2 )\Pi(q^2)$, 
the term ${\widehat d}_{1}(q^2)$ contributes to the
value of $\Pi(q^2)$ with an additional minus sign.}
\be
  i {\widehat d}_{1}(q^2) = \lambda \frac{q_{\mu}}{q^2} \int_k {\Gamma}^{\mu\alpha\beta}_0(q,k,-k-q)
  \Delta_{\alpha\alpha'}(k) \Delta_{\beta\beta'}(k+q) C_{\alpha'\beta'}(q,k,-k-q)\,.
\label{pisca}
\ee
We next determine ${\widehat d}_{1}(0)$,
\be
i {\widehat d}_{1}(0)= \lambda \frac{q_{\mu}}{q^2} \,\lim_{ q \to 0 } \,\int_k {\Gamma}_0^{\mu\alpha\beta}(q,k,-k-q)
  \Delta_{\alpha\alpha'}(k) \Delta_{\beta\beta'}(k+q) C^{\alpha'\beta'}(q,k,-k-q)\,.
\label{lim2}
\ee
The expansion of the integrand around $q=0$ proceeds by inserting  \1eq{eq:taylor_C} 
and setting $q=0$ elsewhere, yielding  
\be
i {\widehat d}_{1}(0)  = 2 \lambda\,\frac{q_{\mu}q_{\nu}}{q^2} \int_k k^{\nu} {\Gamma}_0^{\mu\alpha\beta}(0,k,-k) P_{\alpha\beta}(k) \Delta^2(k^2) \Cfat (k^2) \,,
\label{mC}
\ee
with
\be
{\Gamma}_0^{\mu\alpha\beta}(0,k,-k) = 2 k^{\mu} g^{\alpha\beta} - k^{\alpha} g^{\mu\beta} -  k^{\beta} g^{\mu\alpha} \,.
\label{G0kk}
\ee
Since the integral is proportional to  $g^{\mu\nu}$, we find (using that $g_{\mu}^{\mu} =4$ and $P_{\mu}^{\mu}(q)=3$)  
\bea
i {\widehat d}_{1}(0) &=& \frac{\lambda}{2} \int_k   k_{\mu} {\Gamma}_0^{\mu\alpha\beta}(0,k,-k) P_{\alpha\beta}(k) \Delta^2(k^2) \Cfat (k^2)
\nonumber\\
&=&
3 \lambda \int_k k^2 \Delta^2(k^2) \Cfat (k^2)\,.
\label{mfin}
\eea
To establish explicit contact with the formulation by Schwinger described in Sec.~\ref{sec:smg},
notice that the contribution of ${\widehat d}_{1}^{\,\mu\nu}(q)$ 
to the gluon vacuum polarization, $\overline \Pi(q^2)$,
is simply given by ${\widehat d}_{1}(0)/q^2$; as advocated, it amounts to a massless pole, whose residue 
is precisely ${\widehat d}_{1}(0)$.

The full computation of the total gluon mass proceeds by including the effects of
diagrams $d_3$ and $d_4$, shown in the first line of \fig{fig:SDEs}. Specifically, diagram $d_4$ will
contribute to the mass for the same reason as $d_1$, namely due to the insertion of the
massless pole associated with the three-gluon vertex, proportional to $C_{\alpha'\beta'}$; eventually,
after the limit $q \to 0$ has been taken, $\Cfat(k^2)$ emerges once again. As a result,
the corresponding contributions from $d_1$ and $d_4$ may be naturally combined into a single expression,
whose diagrammatic representation is given in the right panel of \fig{fig:C_gl_Cgh}.
Note that the contribution from graph $d_4$ contains a function denoted by $Y(k^2)$, given by~\cite{Binosi:2012sj} 
\be
Y(k^2)= \frac{i\lambda}{2k^2}\, k^\rho\int_\ell\Delta_{\mu\rho}(\ell)\Delta_{\alpha\nu}(\ell+k) \Gamma^{\alpha\mu\nu}(k,\ell,-k-\ell)\,,
\label{Ydef}
\ee
whose origin is the one-loop subdiagram nested inside $d_4$.
In addition, the contribution of $d_3$ 
originates from the pole in the fully-dressed ghost-gluon vertex, $\fatg_\alpha$, in accordance with  
\2eqs{fullgh}{eq:Vgen}; it is proportional to $C(q,r,p)$, and once the limit  $q \to 0$ has been implemented,
to ${\cal C}(k^2)$. 

As with any SDE computation, multiplicative renormalization must be implemented following
the standard rules. In particular, we introduce the renormalized fields and coupling constant~\cite{Aguilar:2014tka}  
\be 
A^{a\,\mu}_{\s R}(x) = Z^{-1/2}_{\s A} A^{a\,\mu}(x)\,,\qquad
c_{\s R}^a(x) = Z^{-1/2}_{c} c^a(x)\,,\qquad 
g_{\s R} = Z_g^{-1} g\,,
\ee 
such that the associated two point functions are renormalized as 
\be
\Delta_{\s R}(q^2) =  Z^{-1}_{\s A} \Delta(q^2)\,,\qquad
D_{\s R}(q^2) = Z^{-1}_{c} D(q^2) \,.
\label{renprop}
\ee
Similarly, the renormalization constants of the three fundamental Yang-Mills vertices
(ghost-gluon, three-gluon, and four-gluon) are defined as 
\be
\fatg^{\mu}_{\!\!\s R} = \widetilde{Z}_1 \fatg^{\mu};\qquad  
\fatg^{\mu\alpha\beta}_{\!\!\s R} = Z_3 \fatg^{\mu\alpha\beta};\qquad
\fatg^{\mu\alpha\beta\nu}_{\!\!\s R} = Z_4 \fatg^{\mu\alpha\beta\nu}.
\label{renconst2}
\ee
In addition, we employ the following set of exact relations
\be
Z_g = \widetilde{Z}_1 Z_{\s A}^{-1/2} Z_c^{-1} = Z_3  Z_{\s A}^{-3/2} = Z_4^{1/2} Z_{\s A}^{-1}\,,
\label{STIrel}
\ee
which are enforced by the STIs of the theory.  
Once the renormalization procedure has been completed,
the subscript ``$\rm R$'' will be suppressed from all quantities, in order to avoid notational clutter.

Next we pass the answer to Euclidean space and make standard use of the hyperspherical coordinates,
carrying out the trivial angular integrations.
The final result reads
\be
m^2 = 3\hat\lambda\int_0^\infty\!\! dy \, \Dr^2(y) \left[6\pi\alpha_s C_\mathrm{A} Z_4 Y(y) -Z_3\right]{\mathbb C}(y)\, +\, \hat\lambda \,\widetilde{Z}_1\int_0^\infty\!\! dy \, F^2(y) \,{\cal C}(y) \,,
\label{eq:msm}
\ee
where we have employed the gluon and ghost dressing functions, $\Dr(q^2)$ and $F(q^2)$, introduced in \2eqs{defgl}{defF}, respectively,
and have set $\hat\lambda := C_\mathrm{A}\alpha_s/8\pi$. 

%%%%%%%%%%%%%%%%%%%%%%%%%%%%%%%%%%
%Fig. 6
%%%%%%%%%%%%%%%%%%%%%%%%%%%%%%%%%%
\begin{figure}[t]
\includegraphics[width=1.0\textwidth]{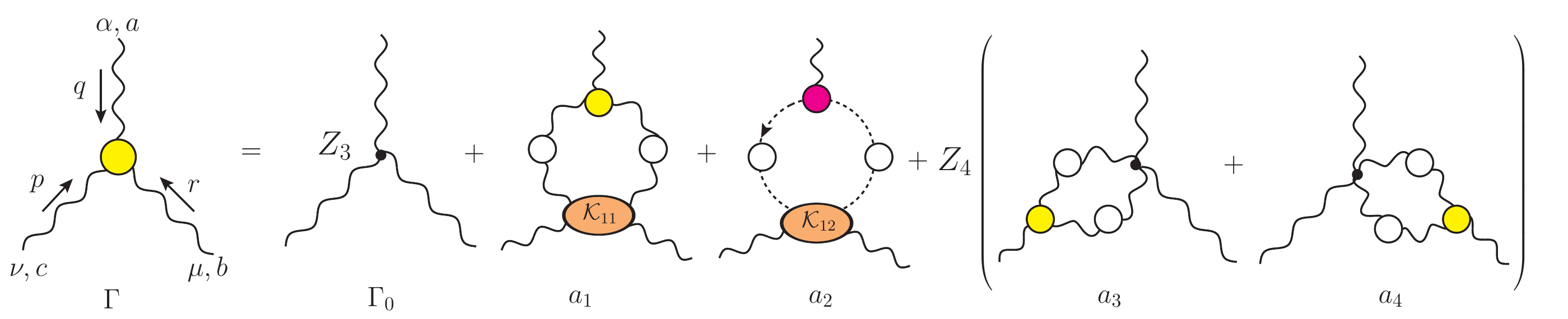} 
\caption{The SDE satisfied by the pole-free part of the renormalized three-gluon vertex, with the vertex renormalization constants
$Z_3$ and $Z_4$ explicitly indicated. The symmetry factor of diagrams $a_3$ and $a_4$ is $\frac{1}{2}$.}
\label{fig:vertwithZ}
\end{figure}
%%%%%%%%%%%%%%%%%%%%%%%%%%%%

Turning to the renormalization constants appearing in \1eq{eq:msm}, let us first point out that, in the Landau gauge,
$\widetilde{Z}_1$ has a finite, cutoff-independent value, by virtue of Taylor's theorem~\cite{Taylor:1971ff}; in fact,
in the so-called ``Taylor scheme''~\cite{Boucaud:2008gn,vonSmekal:2009ae,Boucaud:2011eh},
we have that $\widetilde{Z}_1=1$. However, in our analysis we will employ the ``asymmetric'' MOM scheme mentioned earlier, which yields a slightly
lower value, $\widetilde{Z}_1 \approx 0.95$~\cite{Aguilar:2021okw}. On the other hand, both $Z_3$ and $Z_4$ are cutoff-dependent, thus complicating considerably
the use of \1eq{eq:msm}.

To circumvent this difficulty, consider the renormalized SDE of the pole-free part, $\Gamma^{\alpha\mu\nu}$, 
shown in \fig{fig:vertwithZ}; the
renormalization constants that survive, after the relations in \1eq{STIrel} have been duly employed, are explicitly shown.
It is relatively straightforward to establish that the sum 
\be
{\cal G}^{\mu\alpha\beta}(q,r,p) := Z_3 {\Gamma}^{\mu\alpha\beta}_0(q,r,p) + Z_4 \left[ {a}^{\mu\alpha\beta}_3(q,r,p) + {a}^{\mu\alpha\beta}_4(q,r,p)\right]
\label{eq:theG}
\ee
is precisely the combination of vertex diagrams that appears inside the kernel of the mass equation (right panel of \fig{fig:C_gl_Cgh}),
in the special momentum configuration ${\cal G}^{\mu\alpha\beta}(0,k,-k)$~\cite{Aguilar:2014tka}.
Note that the graphs $a_3$ and $a_4$, after appropriate symmetrization, generate precisely the contribution associated with the 
function $Y(k^2)$; in fact, the symmetry factor of the diagrams $a_3$ and $a_4$ is $\frac{1}{2}$, exactly as needed for reaching \1eq{eq:theG}.  

Then, if we set 
\be
{\cal G}^{\mu\alpha\beta}(0,k,-k) = 2 {\cal G}(k^2)\,k^\mu g^{\alpha\beta} + \cdots \,,
\label{eq:theG2}
\ee
where the ellipsis indicates contributions proportional to 
$k^\alpha g^{\mu\beta}$,  $k^\beta g^{\mu\alpha}$, or $k^\mu k^\alpha k^\beta$, which get annihilated when contracted
by the projector $P^{\alpha\beta}(k)$, \1eq{eq:msm} may be written as 
\be
m^2 = 3\hat\lambda\int_0^\infty\!\! dy \, \Dr^2(y) {\cal G}(y)\, {\mathbb C}(y) \,+\, \hat\lambda\, \widetilde{Z}_1\int_0^\infty\!\! dy \, F^2(y) \,{\cal C}(y) \,.
\label{eq:msm2}
\ee
But, as is clear from the SDE, one may also set
\be
{\cal G}^{\mu\alpha\beta}(q,r,p) =  \Gamma^{\mu\alpha\beta}(q,r,p) - [{a}^{\mu\alpha\beta}_1(q,r,p) + {a}^{\mu\alpha\beta}_2(q,r,p)]\,.
\label{AZ3Z4}
\ee
For practical purposes, the main difference between \2eqs{eq:theG}{AZ3Z4} is the absence of renormalization
constants in the latter. In that sense, \1eq{AZ3Z4} is more reliable, and will be used for the actual determination of the
value of the gluon mass. 

In particular, we have that  
\be
\Gamma^{\mu\alpha\beta}(0,k,-k) = 2 \Ls(k^2) \,k^\mu g^{\alpha\beta} + \cdots,   \qquad {a}_i^{\mu\alpha\beta}(0,k,-k) = {a}_i(k^2) \,k^\mu g^{\alpha\beta}+ \cdots
(i=1,2)\,,
\label{Gtens1}
\ee
so that the form factor ${\cal G}(k^2)$, introduced in \1eq{eq:theG}, is now given by 
\be
{\cal G}(k^2) =   \Ls(k^2) - \frac{1}{2} [{a}_1(k^2) + {a}_2(k^2)]\,.
\label{Gdec}
\ee
This last form of ${\cal G}(k^2)$ will be used in \1eq{eq:msm2} for the numerical computation of $m^2$. 
The quantity $\Ls(k^2)$ is determined from large-volume lattice simulations~\cite{Athenodorou:2016oyh,Aguilar:2019uob,Aguilar:2021lke,Aguilar:2014tka},
while the form factors ${a}_1(k^2)$ and ${a}_2(k^2)$ must be computed from the graphs $a_1$ and $a_2$ in \fig{fig:vertwithZ}, where the one-particle
exchange approximation for the kernels ${\cal K}_{11}$ and ${\cal K}_{12}$,
shown in \fig{fig:kernels}, will be implemented. 

It is important to stress that the solutions for ${\mathbb C}(y)$ and ${\cal C}(y)$ obtained from the BSEs decrease 
sufficiently rapidly in the ultraviolet for the integrals of \1eq{eq:msm2} to be convergent.
In particular, for large values of $y$, we have that ${\mathbb C}(y) \sim y^{-1.45}$ and ${\cal C}(y)\sim y^{-1.12}$~\cite{Aguilar:2021uwa}.

The numerical evaluation of \1eq{eq:msm2} yields finally the value of \mbox{$m = 320 \pm 35$ MeV}~\cite{Horak:2022aqx}, where the error is estimated from the
uncertainties in the evaluation of the form factors ${a}_1(k^2)$ and ${a}_2(k^2)$.
The calculated value of $m$ is in very good agreement with the lattice value \mbox{$m_{\rm{\s {L}}} = 354 \pm 1$ MeV}, which is obtained
from the inverse of the saturation value of the gluon propagator,  
\mbox{$\Delta(0) = 7.99 \pm 0.05\, \rm GeV^{-2}$}, when the renormalization point is \mbox{$\mu=4.3$ GeV}; 
see \fig{fig:lQCD} and~\cite{Aguilar:2021okw}. Note that the lattice error reported is purely statistical. 

It is clear that the value of $m$ extracted in this manner depends on the choice of the renormalization point $\mu$, as already 
stated in the Introduction. Specifically, if a different point of renormalization, say $\nu$, had been chosen instead,  
the entire curve of the gluon propagator would be modified according to~\cite{Aguilar:2010gm}  
\be
\Delta(q^2,\nu^2) = \frac{\Delta(q^2,\mu^2)}{\nu^2 \Delta(\nu^2,\mu^2)}\,,
\label{mutonu}
\ee
which, for $q^2=0$, yields
\be
m^2(\nu^2) = m^2(\mu^2) \,\nu^2 \Delta(\nu^2,\mu^2)\,.
\label{mmutomnu}
\ee
Note finally that a renormalization-group-invariant gluon mass may be obtained by working with the process-independent
effective charge~\cite{Binosi:2016nme,Cui:2019dwv,Roberts:2020hiw}, which constitutes the QCD analogue
of the Gell-Mann--Low coupling known from QED~\cite{Gell-Mann:1954yli}.
The value of this mass turns out to be $m_{\rm{\s {RGI}}} = 430 \pm 10$ MeV.

\section{Ward identity displacement: general observations}
\label{sec:widis}

We will now turn to another central point of the entire approach, and elaborate on the displacement
that the Schwinger mechanism induces to the WIs satisfied by the pole-free parts of the vertices~\cite{Aguilar:2016vin}.  

In order to fix the ideas with a relatively simple example,  we consider the vertex $B^{a}_\alpha(q){\bar c}^{\,m}(r) c^{n}(p)$,
where $B^{a}_\alpha$ is ``background'' gluon, while  ${\bar c}^{\,m}$ ($c^{n}$) are the anti-ghost (ghost)
fields. This vertex has a reduced tensorial structure, and, due to the general properties of the Background Field Method (BFM)
~\cite{DeWitt:1967ub,Honerkamp:1972fd,tHooft:1971qjg,Kallosh:1974yh,Kluberg-Stern:1974nmx,Abbott:1980hw,Shore:1981mj,Abbott:1983zw} (see Sec.~\ref{sec:int}), 
it satisfies an Abelian STI. Specifically,  
after suppressing the gauge coupling $g$ and the color factor $f^{amn}$, the contraction of the remainder
of this vertex, to be denoted by  $\widetilde\Gamma_\alpha(q,r,p)$, yields  
\be 
q^\alpha \widetilde\Gamma_\alpha(q,r,p) = {D}^{-1}(p^2) - {D}^{-1}(r^2)\,,
\label{STI1}
\ee
where ${D}(q^2)$ is the ghost propagator defined in \1eq{defF}.

At this point we will assume that the form factors comprising $\widetilde\g_\alpha(q,r,p)$
do not contain poles, \ie the Schwinger mechanism is turned off. In that case, 
one may carry out the Taylor expansion of both sides of \1eq{STI1}, keeping terms at most linear in $q$: 
\be
[{\rm l.h.s}] = q^\alpha \widetilde\Gamma_\alpha(0,r,-r) \,,\qquad\qquad  [{\rm r.h.s}] = q^\alpha \frac{\partial {D}^{-1}(r^2)}{\partial r^\alpha} \,.
\label{lhs-rhs}
\ee
Equating the coefficients of the terms linear in $q^\alpha$ on both sides, one arrives at a simple QED-like WI
\be
\widetilde\Gamma_\alpha(0,r,-r) = \, \frac{\partial {D}^{-1}(r^2)}{\partial r^\alpha}\,.
\label{WInopole}
\ee

Since $\widetilde\Gamma_{\alpha}(0,r,-r)$ is described by a single form factor, namely 
\be
\widetilde\Gamma_\alpha(0,r,-r) = \widetilde{\cal A}(r^2)  r_\alpha \,,
\label{Gff}
\ee
we may cast \1eq{WInopole} into the equivalent form
\be
\widetilde{\cal A}(r^2) = 2 \frac{\partial {D}^{-1}(r^2)}{\partial r^2} \,.
\label{AwD}
\ee
Let us now activate the Schwinger mechanism,
and denote the resulting full vertex by $\widetilde\fatg_\alpha(q,r,p)$;
in complete analogy with \1eq{eq:Vgen}, it is comprised by a pole-free component and a pole term, according to  
\be
\fatgt_\alpha(q,r,p) = \widetilde\Gamma_{\alpha}(q,r,p) + \frac{q_\alpha}{q^2}{\widetilde C}(q,r,p) \,.
\label{ghsm}
\ee
As the Schwinger mechanism becomes operational,
the STIs satisfied by the elementary vertices retain their original form, but are now    
resolved through the nontrivial participation of the massless pole terms~\mbox{\cite{Eichten:1974et,Poggio:1974qs,Smit:1974je,Cornwall:1981zr,Papavassiliou:1989zd,Aguilar:2008xm,Binosi:2012sj,Aguilar:2016vin}}.
In particular, $\fatgt_\alpha(q,r,p)$ satisfies, as before, precisely \1eq{STI1}, namely
\bea 
q^\alpha \fatgt_\alpha(q,r,p) &=& q^\alpha \widetilde\Gamma_{\alpha}(q,r,p) + \widetilde{C}(q,r,p)
\nonumber\\
 &=& {D}^{-1}(p^2) - {D}^{-1}(r^2)\,.
\label{STI1sm}
\eea
Quite importantly, the contraction of $\fatgt_\alpha(q,r,p)$ by $q^\alpha$
cancels the massless pole in $q^2$, yielding a completely pole-free result.
Consequently, the WI obeyed by $\widetilde\Gamma_{\alpha}(q,r,p)$ may be derived as before,
by carrying out a Taylor expansion around $q=0$, keeping terms at most linear in $q$. In particular, we obtain 
\be
q^\alpha \widetilde\Gamma_{\alpha}(0,r,-r) = \widetilde{C}(0,r,-r) + q^\alpha \left\{ \frac{\partial {D}^{-1}(r^2)}{\partial r^\alpha}
- \left[\frac{\partial \widetilde{C}(q,r,p)}{\partial q^\alpha}\right]_{q=0}\right\}  \,.
\label{lhssm}
\ee
It is clear now that the only zeroth-order contribution present in \1eq{lhssm},
namely $\widetilde{C}(0,r,-r)$, must vanish: 
\be
\widetilde{C}(0,r,-r) = 0 \,. 
\label{Cant}
\ee
It is interesting to note that this last property is a direct consequence of the antisymmetry of  $\widetilde{C}(q,r,p)$ under $r \leftrightarrow p$, 
$\widetilde{C}(q,r,p) = - \widetilde{C}(q,p,r)$, which is imposed by the general ghost-antighost symmetry of the
$B(q){\bar c}(r) c(p)$ vertex. 
Let us now set
\be 
\left[\frac{\partial \widetilde{C}(q,r,p)}{\partial q^\alpha}\right]_{q = 0} \!\!\!\!\!= 
2 r_\alpha\,\Ctilde(r^2)\,, \qquad \Ctilde(r^2) := 
\left[ \frac{\partial \widetilde{C}(q,r,p)}{\partial p^2} \right]_{q = 0} \,,
\label{polegh}
\ee
and proceed with the matching of the terms linear in $q$, thus arriving at the WI 
\be
\widetilde\Gamma_{\alpha}(0,r,-r)  = \frac{\partial {D}^{-1}(r^2)}{\partial r^\alpha} - \underbrace{2 r_\alpha\,\Ctilde(r^2)}_{\rm WI\, displacement}\,.
\label{WIdis}
\ee
Evidently, the WI in \1eq{WIdis} is {\it displaced} with respect to that of \1eq{WInopole} by an amount proportional to the
BSE amplitude for the dynamical pole formation, namely $\Ctilde(r^2)$. 
Similarly, the displaced analogue of \1eq{AwD} is given by
\be
\widetilde{\cal A}(r^2) = 2\left[\frac{\partial {D}^{-1}(r^2)}{\partial r^2} -  \,\Ctilde(r^2)\right]\,.
\label{GDC}
\ee

\section{Ward identity displacement of the three-gluon vertex}
\label{sec:widis3g}

In this section we demonstrate that the WI displacement of $\Gamma^{\alpha\mu\nu}$ 
is expressed precisely in terms of the function ${\mathbb C}(r^2)$, which is thus found to
play a dual role: it is {\it both} the BS amplitude associated with the pole formation {\it and}
the displacement function of the thee-gluon vertex.

The starting point of our analysis is the STI satisfied by the vertex $\fatg_{\alpha\mu \nu}(q,r,p)$, 
\be
q^\alpha {\rm{I}}\!\Gamma_{\alpha \mu \nu}(q,r,p) = F(q^2)
\left[\Delta^{-1}(p^2) P_\nu^\sigma(p) H_{\sigma\mu}(p,q,r) - \Delta^{-1}(r^2) P_\mu^\sigma(r) H_{\sigma\nu}(r,q,p)\right]\,,
\label{STI}
\ee
where $H^{abc}_{\nu\mu}(q,p,r) = -gf^{abc}H_{\nu\mu}(q,p,r)$ is the {\it ghost-gluon kernel}~\cite{Aguilar:2018csq}.
Note that $H_{\sigma\mu}(p,q,r)$ and $H_{\sigma\nu}(r,q,p)$ contain massless poles
in the $r_\mu$ and $p_\nu$ channels, respectively, which are completely eliminated by the transverse projections in \1eq{eq:stidef}.
In what follows we will employ the special relation~\cite{Ibanez:2012zk,Aguilar:2021uwa} 
\be
H_{\nu\mu}(p,q,r) = {\widetilde Z}_1 g_{\nu\mu} + q^\rho K_{\nu\mu\rho}(p,q,r)\,,
\label{HtoK}
\ee
which is particular to the Landau gauge. 
${\widetilde Z}_1$ is the same constant introduced in \1eq{eq:msm}, and the kernel $K$ does not contain poles as \mbox{$q\to 0$}.

It is clear from \2eqs{fullgh}{eq:Vgen} that 
\be
{P}_{\mu'}^{\mu}(r){P}_{\nu'}^{\nu}(p) \left[q^\alpha \fatg_{\alpha \mu \nu}(q,r,p)\right] =  {P}_{\mu'}^{\mu}(r){P}_{\nu'}^{\nu}(p) 
[q^\alpha \Gamma_{\alpha \mu \nu}(q,r,p) +  C_{\mu\nu}(q,r,p)]\,,
\label{eq:stilhs} 
\ee
while, from the STI of \1eq{STI}
\be
{P}_{\mu'}^{\mu}(r){P}_{\nu'}^{\nu}(p) \left[q^\alpha \fatg_{\alpha \mu \nu}(q,r,p)\right]
= {P}_{\mu'}^{\mu}(r){P}_{\nu'}^{\nu}(p)\,F(q^2)\, {R}_{\nu\mu}(p,q,r)\,,
\label{eq:stirhs} 
\ee
where 
\be
{R}_{\nu\mu}(p,q,r) := \Delta^{-1}(p^2) H_{\nu\mu}(p,q,r) - \Delta^{-1}(r^2) H_{\mu\nu}(r,q,p)\,.
\label{eq:defR}
\ee
Then, equating the right-hand sides of \2eqs{eq:stilhs}{eq:stirhs} we obtain
\be
q^\alpha \left[{P}_{\mu'}^{\mu}(r){P}_{\nu'}^{\nu}(p) \Gamma_{\alpha \mu \nu}(q,r,p)\right] =
{P}_{\mu'}^{\mu}(r){P}_{\nu'}^{\nu}(p)   \left[ \,F(q^2)\, {R}_{\nu\mu}(p,q,r) - C_{\mu\nu}(q,r,p) \right]\,.
\label{eq:stidef}
\ee
Next, we carry out the Taylor expansion of
both sides of \1eq{eq:stidef} around $q=0$, keeping terms that are at most linear in $q$. 

The computation of the l.h.s. of \1eq{eq:stidef} is immediate, yielding   
\be
[{\rm l.h.s}] = q^\alpha {\cal T}_{\mu'\nu'}^{\mu\nu}(r) \Gamma_{\alpha \mu \nu}(0,r,-r) \,, \qquad {\cal T}_{\mu'\nu'}^{\mu\nu}(r):={P}_{\mu'}^{\mu}(r){P}_{\nu'}^{\nu}(-r)\,. 
\label{lhs3gold}
\ee
Given that $\Gamma_{\alpha\mu\nu}(0,r,-r)$ depends on a single momentum ($r$), its  
general tensorial decomposition is given by\footnote{The factor of 2 is motivated by the tree-level result
  of \1eq{G0kk}, such that ${\cal A}_1^{(0)}(r^2)= 1$.}  
\be
\Gamma_{\alpha\mu\nu}(0,r,-r) = 2 {\cal A}_1(r^2) \,r_\alpha g_{\mu\nu} + {\cal A}_2(r^2) (r_\mu g_{\nu\alpha} + r_\nu g_{\mu\alpha}) 
+ {\cal A}_3(r^2)\, r_\alpha r_\mu r_\nu \,.
\label{Gtens}
\ee
The form factors ${\cal A}_i(r^2)$ do not contain poles, but are not regular functions; in particular,
${\cal A}_1(r^2)$ diverges logarithmically as $r \to 0$, due to the 
``unprotected'' logarithms that originate  from the massless ghost loops in the diagrammatic expansion of the vertex~\cite{Aguilar:2013vaa,Athenodorou:2016oyh}.

It is then elementary to derive from \1eq{Gtens} that
\be
{\cal T}_{\mu'\nu'}^{\mu\nu}(r) \Gamma_{\alpha\mu\nu}(0,r,-r) = {\cal A}_1(r^2) \lambda_{\mu'\nu'\alpha}(r)\,, \qquad \lambda_{\mu\nu\alpha}(r) := 2r_\alpha P_{\mu\nu}(r)\,,
\label{TGamma}
\ee
and therefore \1eq{lhs3gold} becomes 
\be
[{\rm l.h.s}] = {\cal A}_1(r^2) \,q^\alpha \lambda_{\mu'\nu'\alpha}(r) \,.
\label{lhs3g}
\ee

The computation of the r.h.s. of \1eq{eq:stidef} is slightly more laborious;
in what follows we will highlight some of the technical issues involved~\cite{Aguilar:2021uwa} .

({\it i}) The action of the projectors ${P}_{\mu'}^{\mu}(r){P}_{\nu'}^{\nu}(p)$ on $C_{\mu\nu}(q,r,p)$ triggers
\1eq{eq:PPG2}, and, to lowest order in $q$, only the term $C_1(q,r,p) g_{\mu\nu}$ survives.

({\it ii}) Since it follows immediately from \1eq{eq:defR} that ${R}_{\nu\mu}(-r,0,r) = 0$,
the vanishing of the zeroth order contribution imposes the condition 
\be
C_1(0,r,-r) = 0 \,,
\label{C1_0b}
\ee
in exact analogy to \1eq{Cant}. Note that we have arrived once again at the result of \1eq{C1_0}, but through an entirely different path:
while \1eq{C1_0} is enforced by the Bose symmetry of the three-gluon vertex, \1eq{C1_0b} is a direct consequence of the STI that this vertex satisfies. 

({\it iii})
The Taylor expansion involves the differentiation of the ghost-gluon kernel. In particular, to lowest order in $q$,
we encounter the partial derivatives
\be
\left[\frac{\partial H_{\nu\mu}(p,q,r)}{\partial q^\alpha } \right]_{q=0} \!\!\!\!\!\!=  K_{\nu\mu\alpha}(-r,0,r)\,, \qquad
\left[\frac{\partial H_{\mu\nu}(r,q,p)}{\partial q^\alpha } \right]_{q=0}  \!\!\!\!\!\!= K_{\mu\nu\alpha}(r,0,-r)\,,
\label{Kdef1}
\ee
where \1eq{HtoK} has been used.

({\it iv}) We next employ the tensorial decomposition~\cite{Aguilar:2020yni}, 
\be 
K_{\mu\nu\alpha}(r,0,-r) = - \frac{\w(r^2)}{r^2} g_{\mu\nu}r_\alpha + \cdots \,, 
\label{HKtens}
\ee
where the ellipsis denotes terms that
get annihilated upon contraction with the projector ${\cal T}_{\mu'\nu'}^{\mu\nu}(r)$. \1eq{HKtens}, in conjunction with the elementary relation 
${\cal T}_{\mu'\nu'}^{\mu\nu}(r)K_{\nu\mu\alpha}(-r,0,r) = - {\cal T}_{\mu'\nu'}^{\mu\nu}(r)K_{\mu\nu\alpha}(r,0,-r)$, enables us to finally
express the partial derivatives of \1eq{Kdef1} in terms of the function $\w(r^2)$. 

Taking points ({\it i})-({\it iv}) into account, we can show that the r.h.s. of \1eq{eq:stidef} becomes
\be
   [{\rm r.h.s}] = q^\alpha \lambda_{\mu'\nu'\alpha}(r)\left[
F(0)\left\{\widetilde{Z}_1 [\Delta^{-1}(r^2)]^\prime + \frac{\w(r^2)}{r^2} \Delta^{-1}(r^2)\right\}- \Cfat(r^2)\right] \,,
\label{eq:PPGamma} 
\ee
where the ``prime'' denotes differentiation with respect to $r^2$.
%%%%%%%%%%%%%%%%%%%%%%%%%%%%%%%%%%%%%%%%%%%%%%%%%%%%%%%%%%%%%%%%%%%%%%%%%%%%%
 
The final step is to equate the terms linear in $q$ that appear in \2eqs{lhs3g}{eq:PPGamma}, and thus to obtain
the WI 
\be
 {\cal A}_1(r^2)= F(0)\left\{\widetilde{Z}_1 [\Delta^{-1}(r^2)]^\prime + \frac{\w(r^2)}{r^2} \Delta^{-1}(r^2)\right\}- \Cfat(r^2)\,.
\label{WIdis3g} 
\ee
Thus, the inclusion of the term  $V_{\alpha\mu\nu}(q,r,p)$ in the three-gluon vertex
leads ultimately to the displacement of the WI satisfied by
the pole-free part $\Gamma_{\alpha \mu \nu}(q,r,p)$, by an amount given by the function $\Cfat(r^2)$. 
Evidently, if $\Cfat(r^2)=0$ one recovers the WI in the absence of the Schwinger mechanism.

\section{The displacement function from lattice inputs}
\label{sec:wilat}

In this section we determine the functional form of ${\mathbb C}(r^2)$ 
from  the ``mismatch'' between the quantities entering 
on the two hand-sides of the WI of $\Gamma^{\alpha\mu\nu}$, using inputs obtained almost exclusively from lattice simulations. 
The crucial conceptual advantage
of such a determination is that the lattice is inherently ``blind'' to  
field theoretic constructs such as the Schwinger mechanism; the results are 
obtained through the model-independent functional averaging over gauge-field configurations.
Thus, the emergence of a nontrivial signal would strongly indicate that the Schwinger mechanism,
with the precise field theoretic realization described here, is indeed operational in the gauge sector of QCD.

We first establish a pivotal connection between the form factor ${\cal A}_1(r^2)$ and a special 
projection of the three-gluon vertex, which has been studied extensively in lattice simulations~\mbox{\cite{Parrinello:1994wd,Alles:1996ka,Parrinello:1997wm,Boucaud:1998bq,Cucchieri:2006tf,Cucchieri:2008qm,Duarte:2016ieu,Sternbeck:2017ntv,Vujinovic:2018nqc,Boucaud:2018xup,Aguilar:2019uob, Aguilar:2021lke}}.  Specifically, 
after appropriate amputation of the external legs, the lattice quantity $\Ls(r^2)$ is given by 
\bea
\Ls(r^2) &=&  \frac{\gz^{\alpha\mu \nu}(q,r,p)
P_{\alpha\alpha'}(q)P_{\mu\mu'}(r)P_{\nu\nu'}(p) \fatg^{\alpha'\mu'\nu'}(q,r,p)}
{\rule[0cm]{0cm}{0.45cm}\; {\gz^{\alpha\mu\nu}(q,r,p) P_{\alpha\alpha'}(q)P_{\mu\mu'}(r)P_{\nu\nu'}(p) \gz^{\alpha'\mu'\nu'}(q,r,p)}}
\rule[0cm]{0cm}{0.5cm} \Bigg|_{\substack{\!\!q\to 0 \\ p\to -r}} \,,
\label{asymlat}
\eea
where the suffix ``sg'' stands for ``soft gluon''. 

Clearly, by virtue of \1eq{eq:transvp},  
the term $V^{\alpha'\mu'\nu'}(q,r,p)$, associated with the massless poles, drops out from \1eq{asymlat} in its entirety,
amounting to the effective replacement \mbox{$\fatg^{\alpha'\mu'\nu'}(q,r,p) \to \Gamma^{\alpha'\mu'\nu'}(q,r,p)$}.

%%%%%%%%%%%%%%%%%%%%%%%%%%%%%%%%%%
%Fig. 7
%%%%%%%%%%%%%%%%%%%%%%%%%%%%%%%%%%
\begin{figure}[t!]
\begin{minipage}[b]{0.47\linewidth}
\includegraphics[scale=0.23]{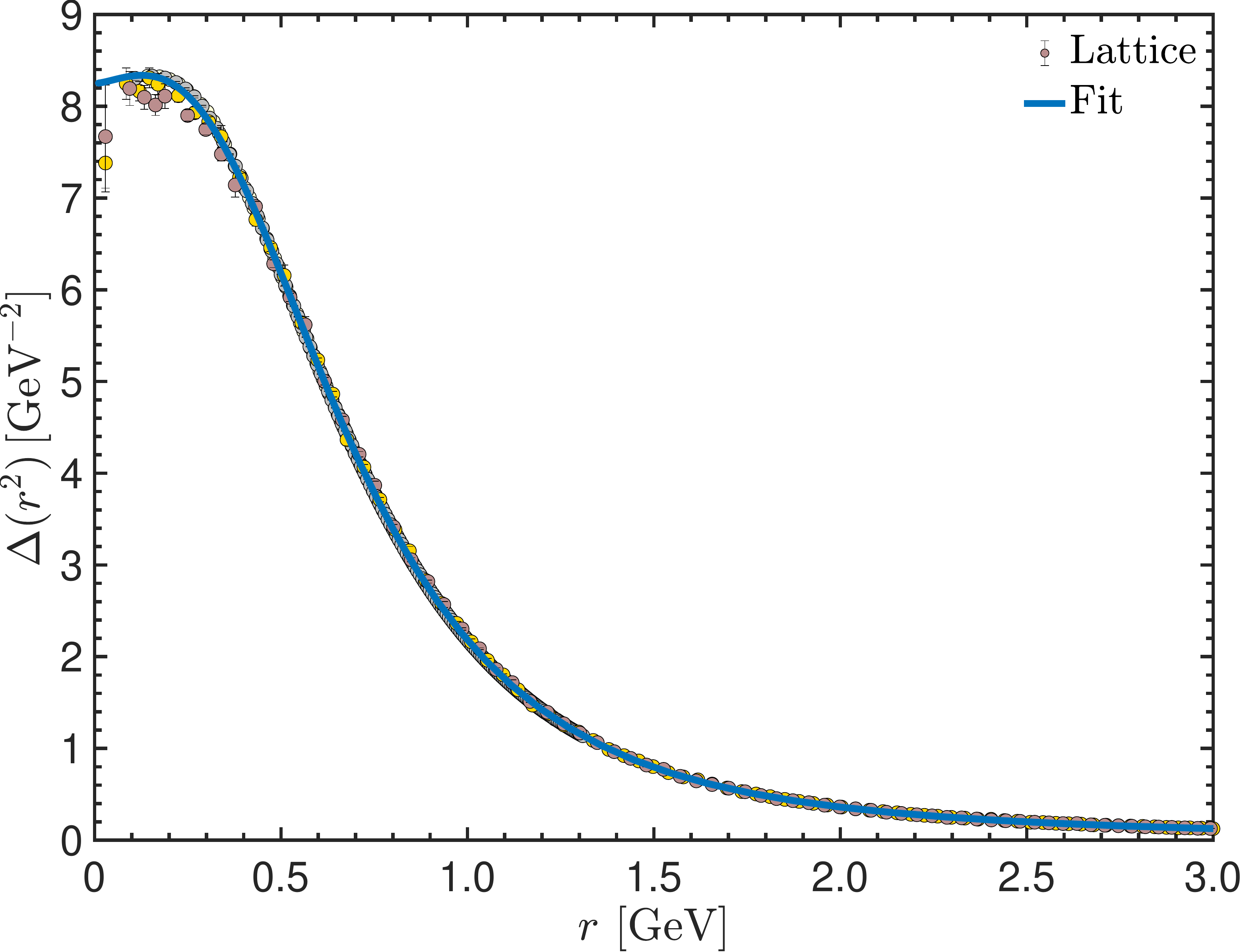} 
\end{minipage}
\begin{minipage}[b]{0.47\linewidth}
\includegraphics[scale=0.23]{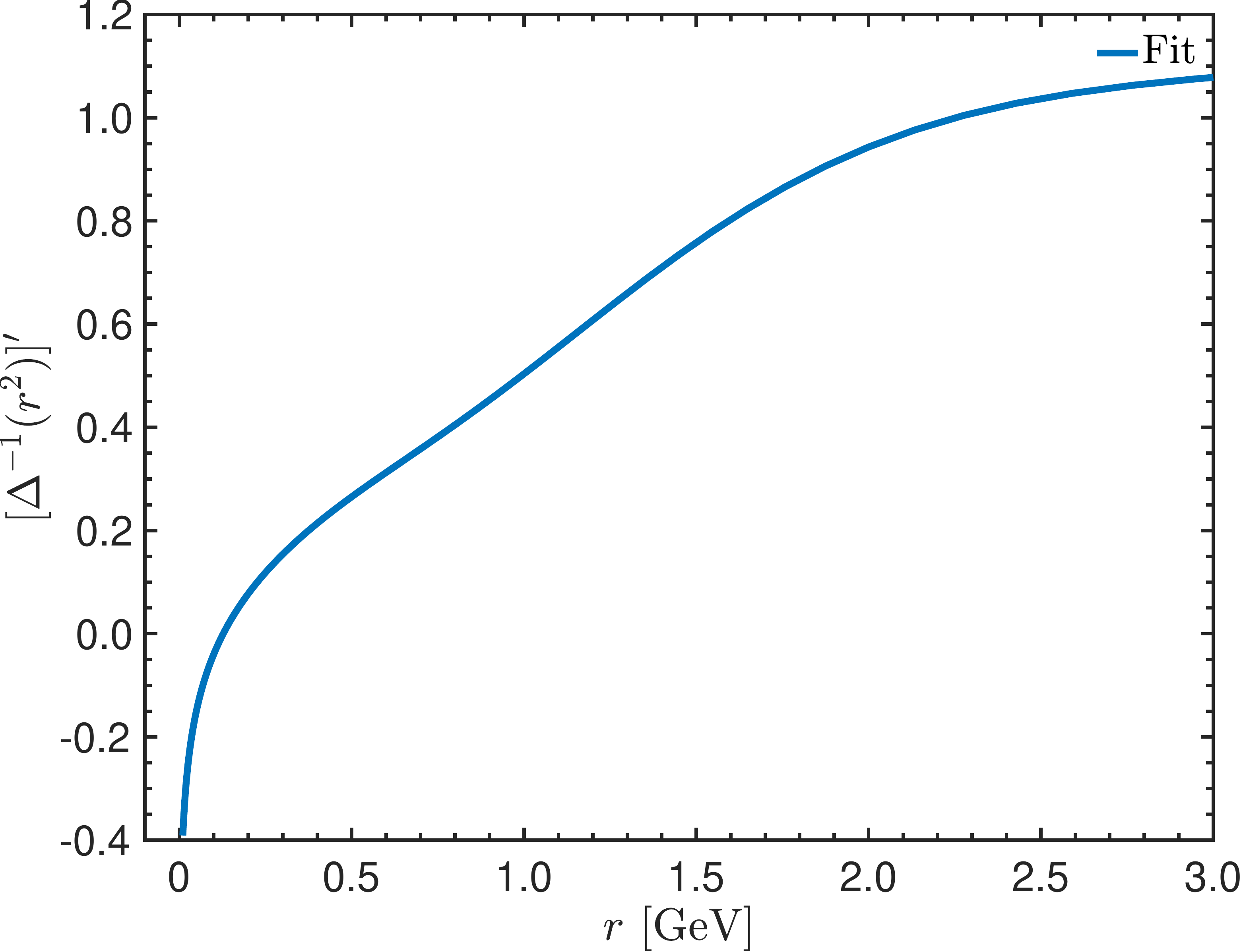}
\end{minipage}
\begin{minipage}[b]{0.47\linewidth}
\vspace{.5cm}  
\includegraphics[scale=0.23]{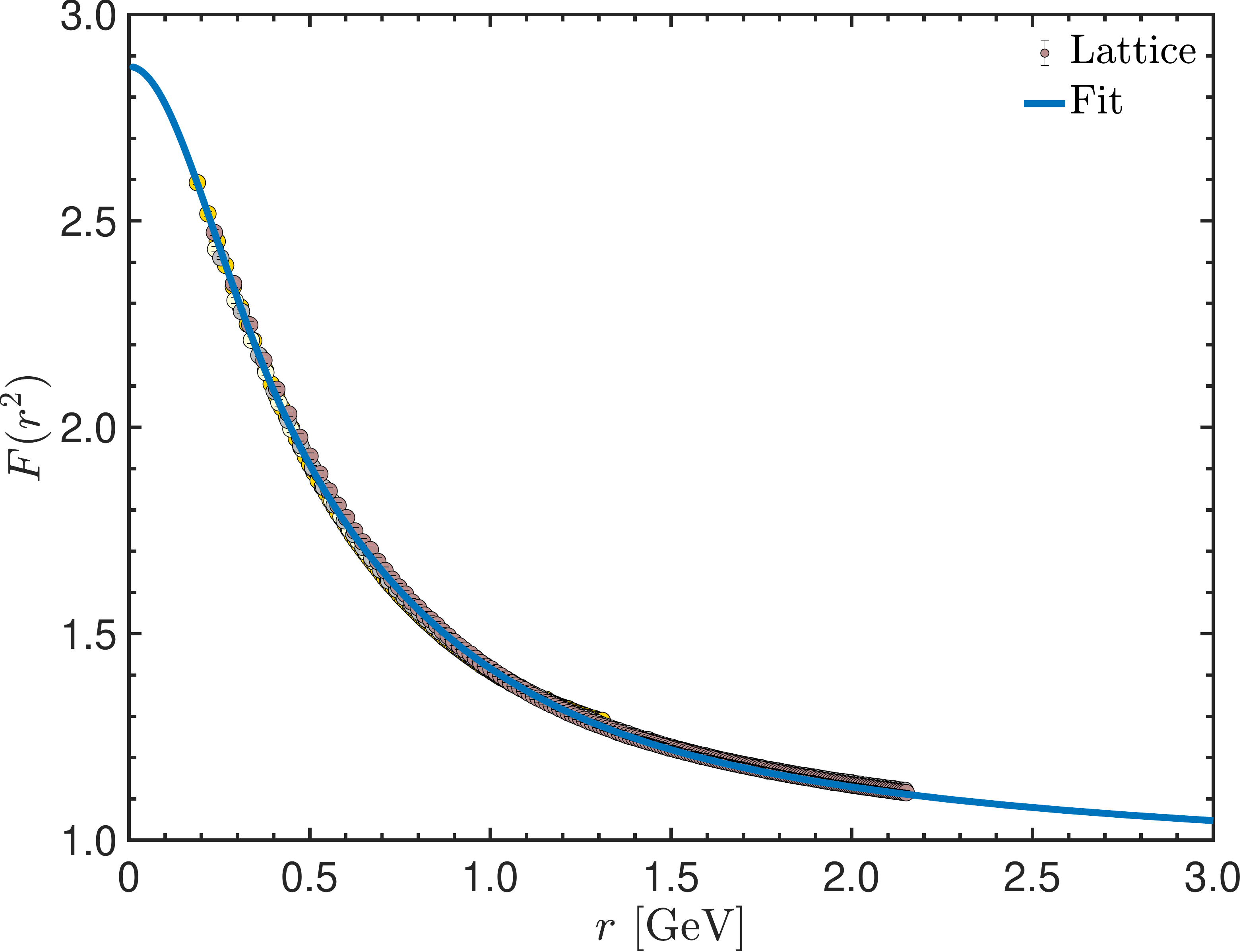}
\end{minipage}
\begin{minipage}[b]{0.47\linewidth}
\includegraphics[scale=0.23]{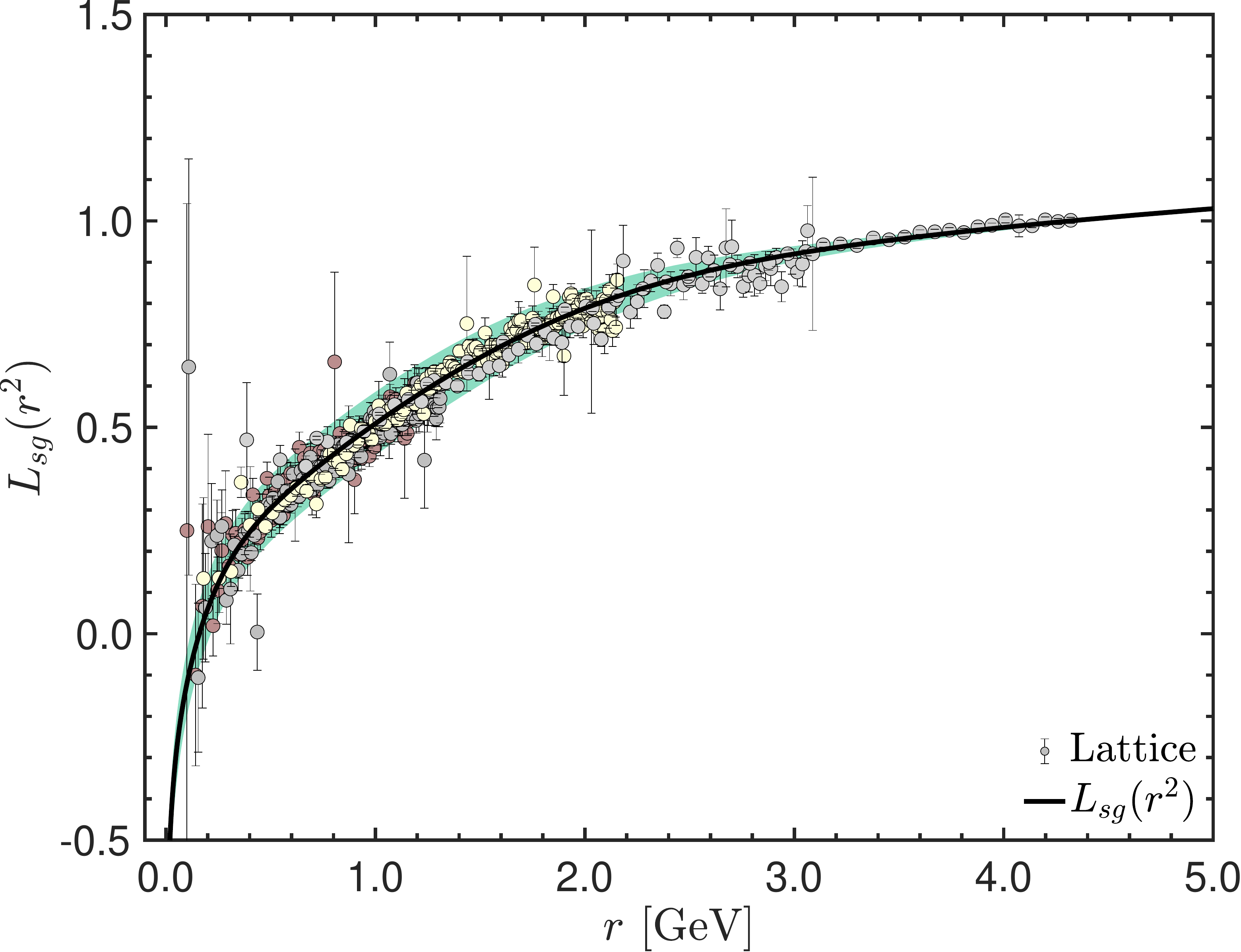}
\end{minipage}
\caption{{\it Upper panel}: The gluon propagator (left) and the first derivative of its inverse (right).
  {\it Lower panel}: The ghost dressing function (left) and the soft gluon form factor $\Ls(r^2)$ of the three-gluon vertex (right).
  All items are taken from~\cite{Aguilar:2021okw}, and have been cured from volume and discretization artifacts.
  Note that $\Ls(r^2)$ is markedly below unity in the infrared, displaying the characteristic zero crossing and the attendant logarithmic
  divergence at the origin~\mbox{\cite{Aguilar:2013vaa,Athenodorou:2016oyh,Boucaud:2017obn,Blum:2015lsa,Corell:2018yil,Aguilar:2019jsj}}.
}
\label{fig:lQCD}
\end{figure}
%%%%%%%%%%%%%%%%%%%%%%%%%%%%%%%%%%%%%%%

Then, the numerator, ${\cal N}$, and denominator, ${\cal D}$,
of the fraction on the r.h.s. of \1eq{asymlat}, after employing
\1eq{Gtens}, become
\be
{\cal N} = 4 (d-1) [r^2 - (r\cdot q)^2/q^2] {\cal A}_1(r^2)\,, \qquad  {\cal D} = 4 (d-1) [r^2 - (r\cdot q)^2/q^2] \,.
\label{NandD}
\ee
At this point, the path-dependent contribution contained in the square bracket drops out when forming the ratio ${\cal N}/{\cal D}$, and \1eq{asymlat} yields the important relation~\cite{Aguilar:2021okw}
\be
\Ls(r^2) =  {\cal A}_1(r^2) \,.
\label{LisB}
\ee
In conclusion, the form factor ${\cal A}_1(r^2)$ appearing in \1eq{WIdis3g} is precisely the one measured on the lattice
in the soft-gluon kinematics; \1eq{LisB} is to be employed in \1eq{WIdis3g}, in order to substitute ${\cal A}_1(r^2)$ by $\Ls(r^2)$.

After this last operation, we pass  
the result of \1eq{WIdis3g} from Minkowski to Euclidean space, following the
standard conversion rules. Specifically, we set $r^2 = - r_{\s {\rm E}}^2$, with $r^2_{\s {\rm E}} > 0$ the positive square of an Euclidean four-vector, and use 
\bea
 \Delta_{\s {\rm E}}(r^2_{\s {\rm E}}) =& - \Delta(-r^2_{\s {\rm E}}) \,, \qquad F_{\s {\rm E}}(r^2_{\s {\rm E}}) =& F(-r^2_{\s {\rm E}}) \,,  \nonumber\\
 \Ls^{\s {\rm E}}(r^2_{\s {\rm E}}) =& \Ls(-r^2_{\s {\rm E}}) \,, \qquad \Cfat_{\s {\rm E}}(r^2_{\s {\rm E}}) =& - \Cfat(-r^2_{\s {\rm E}}) \,. 
\eea
Then, solving for $\Cfat(r^2)$,  we get (suppressing the indices ``${\rm E}$'')
\be
\Cfat(r^2) = \Ls(r^2) - F(0)\left\{\frac{\w(r^2)}{r^2}\Delta^{-1}(r^2) + \widetilde{Z}_1 [\Delta^{-1}(r^2)]^\prime \right\} \,.
\label{centeuc}
\ee

\begin{figure}[t!]
%\hspace{-1.5cm}
\begin{minipage}[b]{0.45\linewidth}
\centering
%\hspace{-1.5cm}
\includegraphics[scale=0.228]{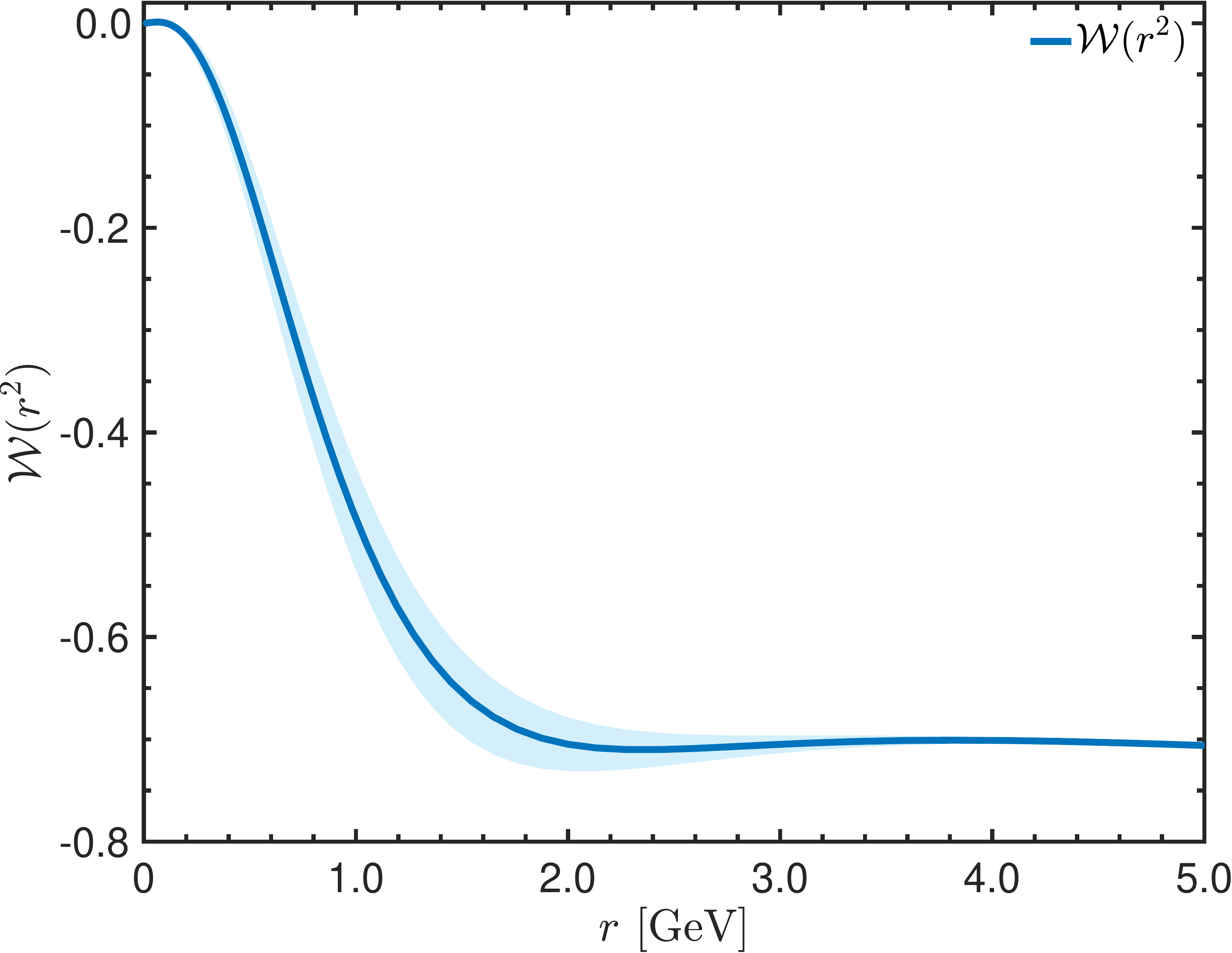}
\end{minipage}
\hspace{0.3cm}
\begin{minipage}[b]{0.45\linewidth}
\includegraphics[scale=0.31]{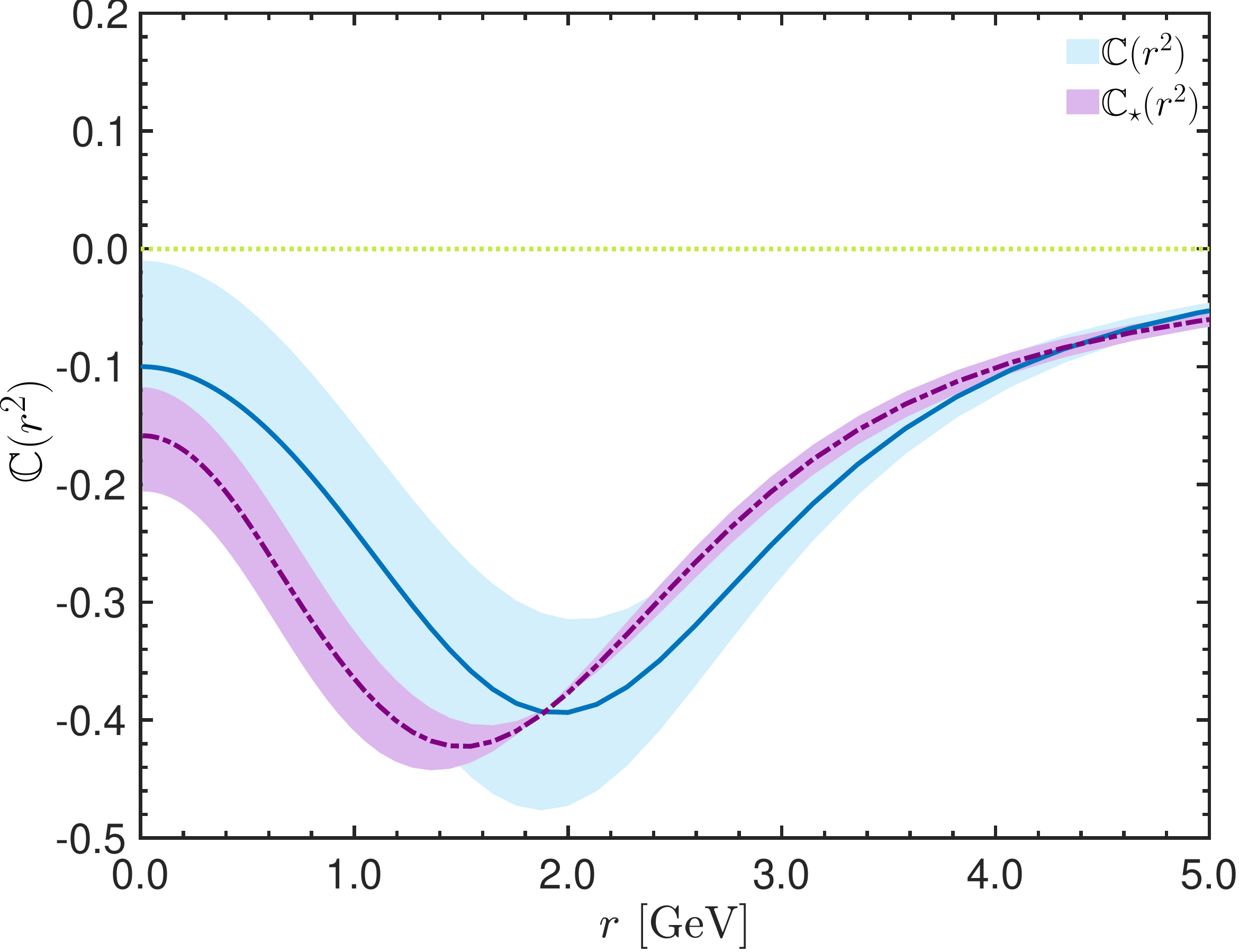} 
\end{minipage}
\caption{{\it Left panel}: The function $\w(r^2)$, computed from the one-loop dressed SDE that governs the ghost-gluon kernel.
  {\it Right panel}: The displacement function $\Cfat(r^2)$ obtained from \1eq{centeuc} (blue continuous curve), compared to the same quantity
obtained from the BSE in \1eq{BSE_hom} (purple dot-dashed curve).}
\label{fig:wcc}
\end{figure}

For the determination of $\Cfat(r^2)$, 
we use lattice inputs for all the quantities that appear on the r.h.s. of \1eq{centeuc}, with the
exception of the function $\w(r^2)$, which will be computed from the SDE satisfied by the ghost-gluon kernel.
The lattice inputs are shown in \fig{fig:lQCD}; all curves are renormalized at $\mu = 4.3$ MeV. 
The computation of $\w(r^2)$ is rather technical, and has been given in detail in~\cite{Aguilar:2021uwa}, Appendix B; the result is shown
in the left panel of \fig{fig:wcc}. 

When all aforementioned quantities are inserted into the r.h.s. of \1eq{centeuc}, a nontrivial result emerges for $\Cfat(r^2)$,
which is shown in the right panel of \fig{fig:wcc}.
The blue error band assigned to  $\Cfat(r^2)$ represents the total propagation of the individual errors
associated with all the inputs entering in \1eq{centeuc}.  
Quite interestingly, the result obtained is not only markedly different from the case $\Cfat(r^2)=0$  
(green dotted horizontal line in the right panel of \fig{fig:wcc}), 
but it bares a striking resemblance to the $\Cfat(r^2)$ obtained from the  
BSE solution. 
In fact, the marked similarity between the two curves provides strong evidence in support of the veracity
of the approximations employed in deriving these results, and 
corroborates the SDE treatment that yields
the result for $\w(r^2)$ shown in the left panel of \fig{fig:wcc}.

%%%%%%%%%%%%%%%%%%%%%%%%%%%%%%%%%%
%Fig. 8
%%%%%%%%%%%%%%%%%%%%%%%%%%%%%%%%%%
%%%%%%%%%%%%%%%%%%%%%%

\section{Ward identity displacement and seagull identity}
\label{sec:int}

When deriving the mass formula of \1eq{eq:msm} we dealt exclusively 
with the $q^{\mu}q^{\nu}$ component of gluon propagator, given that the pole terms $V$ contribute only to this particular tensorial structure.
Of course, the transversality of the self-energy, as captured by \1eq{pitr}, clearly states that the complete treatment of the $g^{\mu\nu}$
component must yield precisely the same answer; nonetheless, the detailed demonstration of this fact in the present context
is highly non-trivial. In particular, the WI displacement turns out to be crucial for the appearance of a term
$g^{\mu\nu}\Delta^{-1}(0)$, as can be 
best exposed within the formalism emerging from the fusion of the pinch technique (PT)~\cite{Cornwall:1981zr,Cornwall:1989gv,Pilaftsis:1996fh,Binosi:2002ft,Binosi:2009qm} and the BFM,
known as ``PT-BFM scheme''. In this section we will briefly outline the key elements of this general construction;
for further details, the reader is referred to~\cite{Aguilar:2006gr,Binosi:2007pi}.

\begin{figure}[t]
  \includegraphics[width=1.0\textwidth]{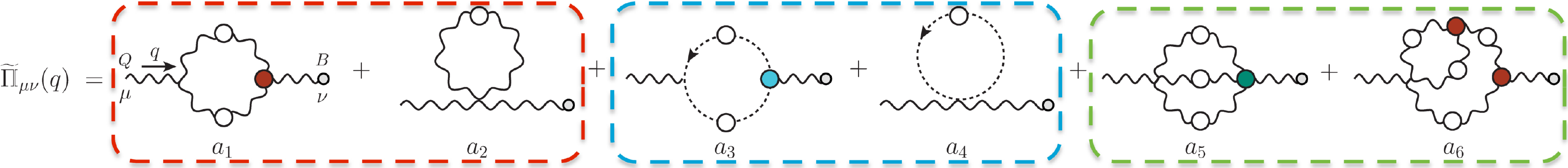}\\
  \caption{The diagrammatic representation of the self-energy $\widetilde\Pi_{\mu\nu}(q)$; 
    the grey circles at the end of the gluon lines indicate a background gluon.
  The corresponding Feynman rules are given in the Appendix B of~\cite{Binosi:2009qm}.}
\label{fig:BQSDE}
\end{figure}

({\it i}) Within the PT-BFM framework, the starting point of our analysis 
is the propagator $\widetilde\Delta_{\mu\nu}(q)$ connecting a quantum gluon, $Q^a_\mu(q)$, with a background one, $B^a_\mu(-q)$; 
the corresponding self-energy, $\widetilde\Pi_{\mu\nu}(q)$, is diagrammatically shown in \fig{fig:BQSDE}. 
One of the most striking properties of $\widetilde{\Pi}_{\mu\nu}(q)$ is its ``block-wise'' transversality~\cite{Aguilar:2006gr,Binosi:2007pi,Binosi:2008qk}: each of the three subsets of diagrams shown in \fig{fig:BQSDE} is individually transverse,  \ie 
\be
q_{\nu} \widetilde{\Pi}_i^{\mu\nu}(q)= 0 \qquad i=1,2,3.
\label{blockwise}
\ee
This result is a direct consequence of the Abelian STIs satisfied by the fully dressed vertices, denoted by $\widetilde{\fatg}$, 
entering in these diagrams, namely
\bea
q^\mu \widetilde{\fatg}_{\mu\alpha\beta}(q,r,p) &=& \Delta_{\alpha\beta}^{-1}(r) - \Delta_{\alpha\beta}^{-1}(p)\,,
\nonumber \\
q^\mu \widetilde{\fatg}^{mnrs}_{\mu\alpha\beta\gamma}(q,r,p,t) &=& f^{mse}f^{ern} {\fatg}_{\alpha\beta\gamma}(r,p,q+t) + f^{mne}f^{esr}{\fatg}_{\beta\gamma\alpha}(p,t,q+r)
\nonumber \\
&+& f^{mre}f^{ens} {\fatg}_{\gamma\alpha\beta}(t,r,q+p)\,,
\label{AbWIthree}
\eea
together with \1eq{STI1sm}.

%%%%%%%%%%%%%%%%%%%
({\it ii}) $\Delta(q^2)$ and $\widetilde\Delta(q^2)$ are related by the exact identity
\begin{align}
\Delta(q^2) = [1 + G(q^2)] \widetilde\Delta(q^2)\,,  
\label{propBQI}
\end{align}
where $G(q^2)$ is the $g_{\mu\nu}$ component of a special two-point function~\cite{Grassi:1999tp,Binosi:2002ez,Binosi:2013cea}.
\1eq{propBQI} allows one to recast the SDE governing $\Delta(q^2)$ in the alternative form 
\be
\Delta^{-1}(q^2)P_{\mu\nu}(q) = \frac{q^2P_{\mu\nu}(q)  + i \widetilde{\Pi}_{\mu\nu}(q)}{1 + G(q^2)} \,, 
\label{sdebq}
\ee
which has the advantage that its diagrammatic expansion contains vertices that satisfy Abelian STIs. Note finally that, in the
Landau gauge only, the powerful identity \mbox{$F^{-1}(0)=1+G(0)$}~\cite{Aguilar:2009pp} expresses the function $G$ at the origin in terms of the
saturation value of the  ghost dressing function.

({\it iii}) The corresponding vertices develop massless poles, following the exact same 
pattern indicated in \2eqs{fullgh}{eq:Vgen}. We can set generically, 
\be
\widetilde{\fatg} = \widetilde{\Gamma} + \widetilde{V}\,,
\label{gvbfm}
\ee 
and the tensorial structures of the vertices $\widetilde{V}$ are those given in \1eq{eq:Vgen}, but with the corresponding
form factors carrying a ``tilde'', \eg 
\be
\widetilde{V}_\alpha(q,r,p) =\frac{q_\alpha}{q^2}\widetilde{C}(q,r,p).
\ee

({\it iv}) In order to isolate the $g^{\mu\nu}$ component, we simply set $q=0$ in the parts of the diagrams that contain the pole-free
vertices,  $\widetilde{\Gamma}$; the implementation of this limit, in turn, triggers the corresponding WIs.  
The block-wise transversality property of \1eq{blockwise} enables one to meaningfully consider this limit within each block, 
in the sense that there is no communication between blocks that enforces cancellations, as happens in the
conventional formulation within the ordinary covariant gauges.
We can therefore illustrate the basic point by means of the block that is operationally
simpler, namely $\widetilde{\Pi}_2^{\mu\nu}(q)$, composed by diagrams $a_3$ and $a_4$ of \fig{fig:BQSDE}. 

({\it v}) A crucial ingredient in this demonstration is the {\it seagull identity}~\cite{Aguilar:2009ke,Aguilar:2016vin},  which states that 
\be 
\int\!\!  d^d k  \,k^2\frac{\partial f(k^2)}{\partial k^2} + \frac{d}{2}\int\!\!  d^d k \,f(k^2) = 0 \,,
\label{sea}
\ee
for functions $f(k^2)$ that satisfy Wilson's criterion~\cite{Wilson:1972cf}; the cases of physical interest are
\mbox{$f(k^2) = \Delta(k^2), D(k^2)$}. This identity is particularly powerful, because, in conjunctions with the WIs 
of the PT-BFM formalism, it enforces the nonperturbative
masslessness of the gluon in the absence of the Schwinger mechanism.  

%%%%%%%%%%%%%%%%%%%%%%%%%%%%%%%%%%%%%%%%%%%

({\it vi}) 
We now want to determine the value of the $g^{\mu\nu}$ component of $\widetilde{\Pi}_2^{\mu\nu}(q)$ at $q=0$.
We have that the ($q$-independent) contribution from $a^{\mu\nu}_4$ is proportional to $g^{\mu\nu}$, while $a_3^{\mu\nu}(q)$ contains both
$g^{\mu\nu}$ and $q^{\mu}q^{\nu}$ components; however, in the limit $q\to 0$ the latter vanish, 
precisely due to the absence of a pole in $q^2$. Let us denote by ${\cal B}(q^2)$ the total contribution proportional to $g^{\mu\nu}$
originating from both diagrams; using the Feynman rules of the BFM~\cite{Binosi:2009qm},
it is rather straightforward to show that, as $q\to0$,  
\be
i{\cal B}(0)  = \frac{2\lambda}{d} F(0)\left[ \int_k   k_\mu D^2(k^2)\widetilde{\Gamma}^\mu(0,-k,k) - d \int_k D(k^2) \right] \,.
\label{Pi2}
\ee
When the Schwinger mechanism is turned off, the WI of \1eq{WInopole} may be recast in the form 
\be 
\widetilde{\Gamma}^\mu(0,-k,k) = - 2 k^\mu D^{-2}(k^2) \frac{\partial D(k^2)}{\partial k^2}\,,
\label{der}
\ee
and so \1eq{Pi2} becomes 
\be
i{\cal B}(0)  = -\frac{4\lambda}{d}F(0)
\underbrace{\left[ \int_k  k^2 \frac{\partial D(k^2)}{\partial k^2} +   \frac{d}{2}\int_k D(k^2) \right]}_{\rm seagull\,\, identity} = 0 \,.
\label{Pi2s}
\ee
%
%%%%%%%%%%%%%%%%%%%%%%%%%%%%%%%%%%
%Fig. 9
%%%%%%%%%%%%%%%%%%%%%%%%%%%%%%%%%%
%%%%%%%%%%%%%%%%%%%%%%%%%%%%%%%%%

When the Schwinger mechanism is activated, the displaced WI of \1eq{WIdis} must be employed, such that  
\be 
\widetilde{\Gamma}^\mu(0,-k,k) = - 2 k^\mu \left[D^{-2}(k^2) \frac{\partial D(k^2)}{\partial k^2} + \widetilde{\cal C}(k^2)\right]\,.
\label{derdis}
\ee
Upon insertion of \1eq{derdis} into  \1eq{Pi2}, the first term triggers the seagull identity as before and vanishes,
while the second furnishes a nonvanishing finite result
\be
i{\cal B}(0) = \frac{4\lambda}{d} F(0) \int_k k^2  D^2(k^2)\widetilde{\cal C}(k^2)\,.
\label{mcalt}
\ee
({\it vii}) Using the exact relation~\cite{Aguilar:2021uwa} 
\be
\C(k^2) = F(0)\,\Ctilde(k^2) \,,
\label{cvsc}
\ee
which is derived from the ``background-quantum identity'' that relates $\widetilde\fatg_{\alpha}(q,r,p)$
and  $\fatg_\alpha(q,r,p)$~\cite{Binosi:2002ez,Binosi:2009qm}, \1eq{mcalt} becomes
\be
i{\cal B}(0) = \frac{4\lambda}{d}  \int_k k^2  D^2(k^2)\C(k^2) \,.
\label{mcalt2}
\ee
Setting $d=4$, introducing the renormalization constant $\widetilde{Z}_1$ and the ghost dressing function $F$,
then passing to Euclidean space and employing spherical coordinates, 
it is straightforward to confirm that the expression in \1eq{mcalt2} is identical to 
that given by the second term in \1eq{eq:msm}.
Completely analogous procedures may be applied to 
the remaining two blocks, $\widetilde{\Pi}_1^{\mu\nu}(q)$ and $\widetilde{\Pi}_3^{\mu\nu}(q)$,
by exploiting the Abelian STIs of \1eq{AbWIthree}~\cite{Binosi:2012sj}. 

In summary, the WI displacement of the vertices 
{\it evades} the seagull identity, and endows the $g^{\mu\nu}$ component of the gluon propagator with the exact amount of mass
required by its transverse nature.

\section{Relating the gluon mass with the transition amplitude}
\label{sec:theI} 

It is particularly instructive to zoom into the detailed composition of the vertex $V_{\alpha\mu\nu}(q,r,p)$, by essentially unfolding the black 
circles in \fig{fig:poles} and exposing the diagrammatic structure of the transition amplitude $I_{\alpha}(q)$,
introduced in the paragraph following \1eq{bare3g}. This analysis
unravels interesting diagrammatic properties, and  
allows us to derive a simple relation between the transition amplitude
and the gluon mass.

%%%%%%%%%%%%%%%%%%%%%%%%%%%%%%%%%%
%Fig. 10
%%%%%%%%%%%%%%%%%%%%%%%%%%%%%%%%%%
\begin{figure}[h!]
\includegraphics[width=1.0\textwidth]{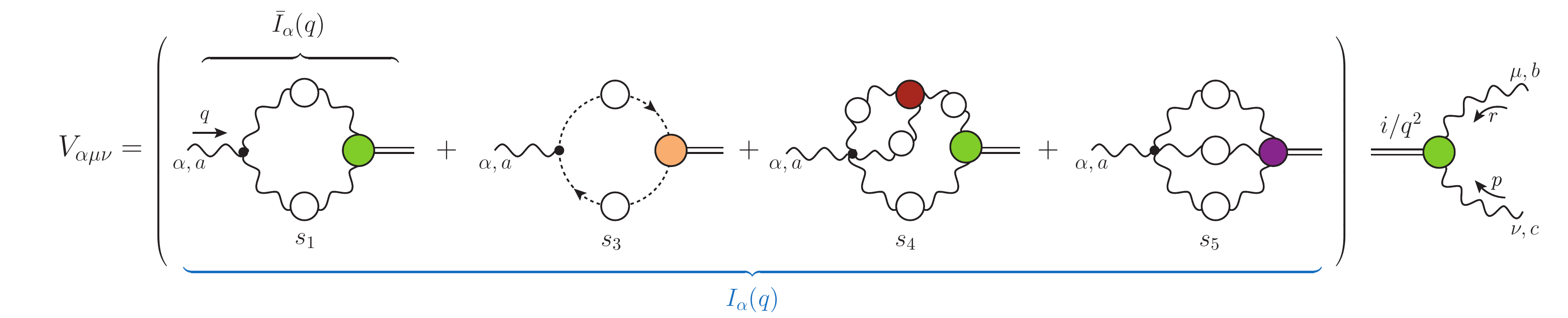} 
\caption{The diagrammatic representation of the vertex $V_{\alpha\mu\nu}(q,r,p)$ in terms of the transition amplitude
  $I_{\alpha}(q)$, the propagator of the massless excitation, and the vertex function.
  Note that the diagrams $s_i$ are in one-to-one correspondence with the $d_i$ of \fig{fig:SDEs}, except for the
  seagull graph $d_2$, which has no analogue $s_2$.}
\label{fig:theI}
\end{figure}
%%%%%%%%%%%%%%%%%%%%%%%%%%%%%%%

The basic elements composing the vertex $V_{\alpha\mu\nu}(q,r,p)$, shown in \fig{fig:theI}, are:
({\it i}) the transition amplitude $I_{\alpha}(q)$, which connects a gluon to the massless 
composite excitation, ({\it ii}) the propagator of the latter, and ({\it iii}) the vertex function $B_{\mu\nu}(q,r,p)$,
connecting the massless excitation to a pair of gluons.
Since color indices are suppressed in \fig{fig:theI}, we emphasize that the propagator of 
the colored massless excitations is given by the expression $i\delta^{ab}/q^2$, \ie it carries  
color, as it should.
Similarly, the vertex
$B_{\mu\nu}(q,r,p)$ is multiplied by the structure constants $f^{abc}$, providing precisely the color term     
that has been factored out from $V_{\alpha\mu\nu}(q,r,p)$ in \1eq{eq:Vgen}.
Thus, we have
\be
V_{\alpha\mu\nu}(q,r,p) = I_{\alpha}(q) \left(\frac{i}{q^2}\right) B_{\mu\nu}(q,r,p) \,, \qquad  I_{\alpha}(q) = q_{\alpha} I(q^2)\,,
\label{eq:Valt}
\ee
with  
\be 
B_{\mu\nu}(q,r,p) = B_1 \, g_{\mu\nu}  + B_2\, r_\mu r_\nu  + B_3 \, p_\mu p_\nu  +  B_4 \, r_\mu p_\nu  + B_5 \,  p_\mu r_\nu  \,.
\label{eq:Bdec}
\ee
Given that the $V_{\alpha\mu\nu}(q,r,p)$ appearing in \2eqs{eq:Vgen}{eq:Valt} represents the same vertex, 
we deduce immediately that
\be
C_1 (q,r,p) = i I(q^2) B_1(q,r,p)\,.
\label{cb1}
\ee
Evidently, since $C_1(0,r,-r)=0$ [see \2eqs{C1_0}{C1_0b}], one obtains from \1eq{cb1} that $B_1(0,r,-r)=0$, so that  
\be
\lim_{q \to 0} {B}_1(q,r,p) =  2 (q\cdot r) {\mathbb B}(r^2) + \cdots \,, \qquad {\mathbb B}(r^2) := \left[ \frac{\partial {B}_1(q,r,p)}{\partial p^2} \right]_{q = 0}\,,
\label{cb2}
\ee
and therefore, from  \1eq{cb1} we have 
\be
\Cfat(r^2) = i I(0) {\mathbb B}(r^2) \,.
\label{cb3}
\ee

In order to illustrate the origin of a key relation between $I(0)$ and $m^2$, let us simplify the discussion
by assuming that graph $s_1$ in \fig{fig:theI} represents the only contribution to $I^{\alpha}(q)$,
to be denoted by ${\bar I}^{\alpha}(q)$. Clearly, the equivalent approximation at the level of
the analysis presented in Sec.~\ref{sec:mass}
would be to assume that only graph $d_1$ in \fig{fig:SDEs} contributes to the gluon mass. 

It is straightforward to deduce from \fig{fig:theI} that ${\bar I}^{\alpha}(q)$ is given by 
\be
{\bar I}^{\alpha}(q) =  \frac{C_A}{2} 
\int_k \Gamma_0^{\alpha\beta\lambda}(-q,-k,k+q) \Delta_{\lambda\mu}(k) \Delta_{\beta\nu}(k+q)  B^{\mu\nu}(q,k,-k-q)\,,
\label{I1}
\ee
where $\frac{1}{2}$ is the corresponding symmetry factor. Since, by Lorentz invariance,
\mbox{${\bar I}^{\alpha}(q) = q^{\alpha} {\bar I}(q^2)$}, we have that \mbox{${\bar I}(q^2) = q_{\alpha}{\bar I}^{\alpha}(q)/q^2$}.
Therefore, from \1eq{I1} we obtain 
\be
{\bar I}(q^2) =  \frac{C_A}{2 \,q^2} 
\int_k \{q_{\alpha}\Gamma_0^{\alpha\beta\lambda}(-q,-k,k+q)\} \Delta_{\lambda\mu}(k) \Delta_{\beta\nu}(k+q)  B^{\mu\nu}(-q,-k,k+q)\,.
\label{I2}
\ee
Next, employing the expression for $\gz^{\alpha\mu\nu}(q,r,p)$ given in \1eq{bare3g}, we have that 
\be
q_{\alpha}\Gamma_0^{\alpha\beta\lambda}(q,k,-k-q) = -(q^2 + 2 q\cdot k)g^{\beta\lambda} + \cdots
\label{I3}
\ee
where the ellipsis indicates terms that get annihilated upon contraction with the Landau-gauge propagators
$\Delta_{\lambda\mu}(k)$ and $\Delta_{\beta\nu}(k+q)$ in the integrand of \1eq{I2}. 

In order to determine from \1eq{I2} the expression for ${\bar I}(0)$, note that,
by virtue of \1eq{cb2}, 
only the term  $(2 q\cdot k)$ in \1eq{I3} contributes to ${\bar I}(0)$,
yielding the combined contribution $(2 q\cdot k)^2 \,{\mathbb B}(k^2)$. Thus, 
using that $P_{\mu}^{\mu}(k) = 3$, 
\bea
{\bar I}(0) &=& -6 C_A \frac{q^{\mu}q^{\nu}}{q^2} \int_k k_{\mu}k_{\nu}  \Delta^2(k^2) {\mathbb B}(k^2)
\nonumber\\
&=& -\frac{3 C_A}{2} \int_k k^2  \Delta^2(k^2) {\mathbb B}(k^2)\,.
\label{I4}  
\eea

\begin{figure}[t]
\includegraphics[width=0.95\textwidth]{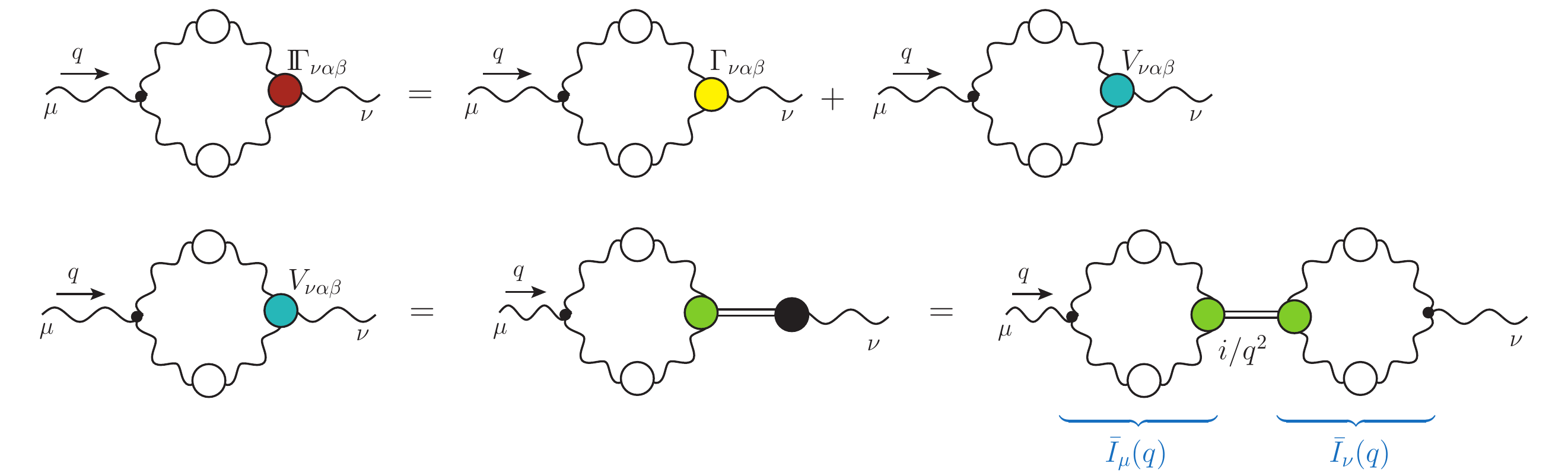}
\caption{The diagrammatic representation of the sequence that leads to \1eq{mI0}. {\it First row}: The
diagram $d_1$ of \fig{fig:SDEs}, once the replacement $\fatg_{\alpha\mu\nu}(q,r,p) = \g_{\alpha\mu\nu}(q,r,p) + V_{\alpha\mu\nu}(q,r,p)$
of \1eq{fullgh} has  been implemented. {\it Second row}: The diagrammatic steps that leads to the replication of the ${\bar I}_{\alpha}(q)$,
and eventually to \1eq{mI0}.}
\label{fig:squared}
\end{figure}

Returning to  \1eq{mfin}, and substituting in it \1eq{cb3}, it is clear that ($\lambda := i g^2 C_{\rm A}/2$)
\be
i {\widehat d}_{1}(0) = g^2 {\bar I}(0) \underbrace{\left\{-\frac{3 C_A}{2} \int_k k^2 \Delta^2(k^2) {\mathbb B}(k^2)\right\}}_{\bar I(0)}\,.
\label{I5}
\ee
As mentioned above, at this level of approximation, $i {\widehat d}_{1}(0)$ is the only
contribution to the gluon mass, to be denoted by  ${\overline m}^2$; so, \1eq{I5} becomes   
\be
{\overline m}^2 = g^2 {\bar I}^2(0) \,.
\label{mI0} 
\ee
Thus, the pattern that emerges from the study of this particular example may be summarized as follows: 
when the vertex $V_{\alpha\mu\nu}$ is given by \1eq{eq:Valt} with $I_{\alpha}(q) \to {\bar I}_{\alpha}(q)$,  
its insertion in the corresponding propagator graph $d_1$ 
leads to the replication of the ${\bar I}_{\alpha}(q)$, as shown schematically in Fig.~\ref{fig:squared}. 

It turns out that this property may be generalized to include the entire $I_{\alpha}(q)$,
composed by the graphs $s_1$, $s_3$, $s_4$, and $s_5$ in \fig{fig:theI}, provided that the
propagator graphs $d_1$, $d_3$, $d_4$, and $d_5$ of \fig{fig:SDEs} are correspondingly included;
for details, see~\cite{Ibanez:2012zk}.
The final result is precisely the generalization of \1eq{mI0}, namely
\be
m^2 = g^2 I^2(0)\,.
\label{mI0f} 
\ee

%%%%%%%%%%%%%%%%%%%%%%%%%%%%%%%%%%
%Fig. 11
%%%%%%%%%%%%%%%%%%%%%%%%%%%%%%%%%%
%%%%%%%%%%%%%%%%%%%%%%%%%%%

We emphasize that, as far as the numerical determination of the gluon mass is concerned, 
\1eq{mI0f} contains the same information as \1eq{eq:msm}, once the diagrammatic expansion of $I_{\alpha}(q)$
has been implemented. Nonetheless, the formulation presented above  
exposes an elaborate diagrammatic pattern,
and is particularly useful for the analysis presented in the next section, mainly due to the role played by 
the vertex function $B_{\mu\nu}(q,r,p)$.

\section{Absence of pole divergences in the $S$-matrix}
\label{sec:sdec}

As has been emphasized in the early literature on the subject, 
one of the main properties of the massless composite excitations that trigger the Schwinger mechanism
is that they do not induce divergences in on-shell amplitudes, see, \eg~\cite{Jackiw:1973tr,Jackiw:1973ha}.
In the case of the Yang-Mills theories that we study, 
the elimination of potentially divergent terms  
hinges on the longitudinality of the vertices $V$, as captured by \1eq{eq:Vgen}, in conjunction with the special limit 
given by \1eq{cb2}.
In this section we demonstrate with a specific example
how all terms containing massless poles are either annihilated in their entirety,
or, in the kinematic limit where poles might in principle cause divergences, they actually give finite contributions, i.e., they correspond to {\it evitable} singularities. 

\begin{figure}[t]
\includegraphics[width=0.95\textwidth]{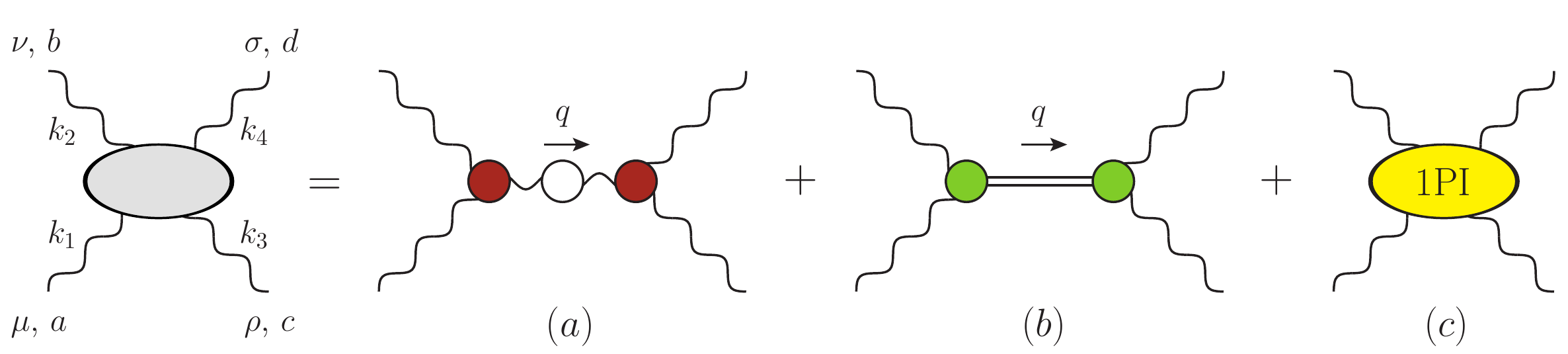}
\caption{The four-gluon scattering amplitude and the three types of diagrams contributing to it. The abbreviation ``1PI'' stands
for ``one-particle irreducible''. Note that diagram ($b$) is composed entirely by nonperturbative structures.}
\label{fig:Smatrix}
\end{figure}

Let us consider the elastic scattering process $g_{\mu}^a(k_1) g_{\nu}^b(k_2) \to g_{\rho}^c(k_3) g_{\sigma}^d(k_4)$, 
depicted in \fig{fig:Smatrix},
where the $g_{\alpha}^a(k_i)$ denotes an ``on-shell'' gluon
of momentum $k_i$, and $q= k_2-k_1 = k_4-k_3$ is the corresponding momentum transfer, with $q^2=t$ is the relevant Mandelstam variable.
The scattering amplitude consists of the three distinct terms denoted by ($a$), ($b$), and ($c$) in \fig{fig:Smatrix}.  
We notice that, unlike ($a$) and ($c$), diagram ($b$) has no perturbative analogue, since all components that comprise it are
generated through nonperturbative effects; observe, in particular, the appearance of the vertex function $B_{\mu\nu}(q,r,p)$, introduced in the previous section.

As we will show at the end of this section, the elimination of all pole divergences does not rely
on any particular properties associated with the ``on-shellness'' of the gluons; this feature is especially welcome,
given that gluons do not appear as asymptotic states.
Nonetheless, it is instructive to first examine how the cancellations proceed when the gluons are assumed to be on shell,
because this will allow us to identify the crucial properties that must be fulfilled in the off-shell case.

To that end, let us assume that, due to the on-shellness of the gluons, 
each external leg is contracted by the corresponding transverse polarization vector, $\epsilon^{\mu}(k)$, for which
\be
k^{\mu} \epsilon_{\mu}(k) = 0 \,. 
\label{polv}
\ee

We start our discussion with diagram $(a)$, given by
\be
(a)=  \epsilon^{\mu}(k_1)\epsilon^{\nu}(k_2)\fatg^{\alpha\mu\nu}(q,k_1,k_2) \Delta(q) P^{\alpha\beta}(q)  \fatg^{\beta\rho\sigma}(q,k_3,k_4)
\epsilon^{\rho}(k_3)\epsilon^{\sigma}(k_4)\,.
\label{sma}
\ee
Noting that, due to the longitudinality of $V_{\alpha\mu\nu}(q,r,p)$, 
\be
V_{\alpha\mu\nu}(q,r,p) P^{\alpha\beta}(q) \epsilon^{\mu}(r)\epsilon^{\nu}(p) =0 \,,
\label{tee}
\ee
it is clear that the terms $V$ drop out from \1eq{sma}, and the two fully dressed vertices $\fatg$ are replaced by their pole-free counterparts, $\Gamma$.
As a result, the contribution from graph ($a$) is finite.

Turning to diagram ($b$), we have that

\be
(b) =  B_{\mu\nu} (q,k_1,k_2) \left[\frac{i}{q^2}\right] B_{\rho\sigma} (q,k_3,k_4) \,. 
\label{smb}
\ee

As the limit $q\to 0$ is taken, \1eq{cb2} is triggered, such that 
\bea
\lim_{q \to 0}\, (b) &=& \{ 2 (q\cdot k_1)\, {\mathbb B}(k_1^2)\} \left[\frac{i}{q^2}\right]  \{ 2 (q\cdot k_3) \,{\mathbb B}(k_3^2)\}
\nonumber\\
 &=& 4 i \lvert k_1 \rvert \lvert k_3 \rvert  \cos \theta_1 \cos \theta_3 \, {\mathbb B}(k_1^2) \,{\mathbb B}(k_3^2) \,,
\eea
where $\theta_1$ and $\theta_3$ are the angles formed between $q$ and $k_1$ and $k_3$, respectively.
The above contribution is clearly finite. 

%%%%%%%%%%%%%%%%%%%%%%%%%%%%%%%%%%
%Fig. 12
%%%%%%%%%%%%%%%%%%%%%%%%%%%%%%%%%%
%%%%%%%%%%%%%%%%%%%%%%%%%%%%%%%%%%

Finally, note that none of the possible $V$-type vertices 
survives in the diagrams contributing to ($c$), precisely due to their longitudinal nature. 
Indeed, if a vertex is fully inserted in a diagram, \ie when none of its momenta is any of the $k_i$,
then it is contracted by three Landau-gauge gluon propagators, and \1eq{eq:transvp} is automatically triggered.  
If, on the other hand, some of the legs of the vertex carry a momentum $k_i$,
say $k_1$, then the part of $V$ that does not get cancelled by the transverse gluon propagators will necessary
be proportional to $k_1^{\mu}/k^2$; it is therefore annihilated upon contraction with the 
$\epsilon_{\mu}(k_1)$, by virtue of \1eq{polv}. 

Thus, it is evident from the above considerations that, in the
limit  $q\to 0$, the terms associated with the massless composite excitations
furnish only finite contributions to the amplitude.

It is important to recognize that, in the above demonstration, the only element introduced due to the assumed 
on-shellness of the scattered gluons is the contraction of the amplitude
by the corresponding polarization vectors, satisfying \1eq{polv}. 
In particular, note that at no point have we assumed anything special about the values of $k_i^2$, \ie neither that $k_i^2=0$
nor that $k_i^2=m^2$.  

\begin{figure}[t]
\includegraphics[width=0.45\textwidth]{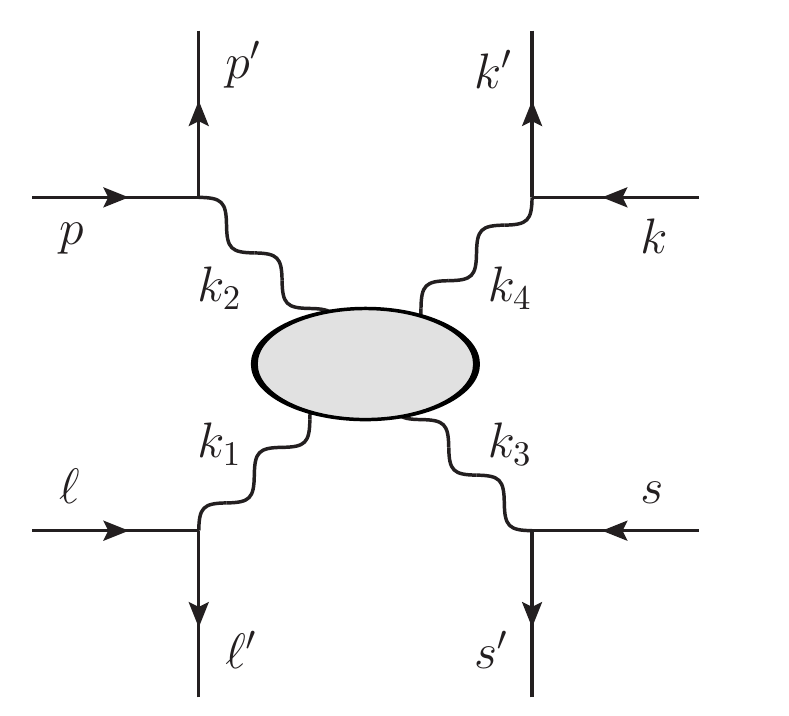}
\caption{The four-gluon amplitude regarded as a completely off-shell subprocess, embedded into a multi-quark scattering process.}
\label{fig:Squarks}
\end{figure}

As a result, it is possible to relax the on-shellness condition completely, and consider the above amplitude 
as an off-shell sub-process, embedded into a more complicated scattering process, as depicted in \fig{fig:Squarks}.
Indeed, the demonstration presented above goes through unaltered, because the on-shell gluons are replaced by
Landau-gauge propagators, which trigger precisely the same relations that the polarization vectors did in the on-shell case.

In conclusion, the above analysis, albeit restricted to a special example, strongly supports the notion that 
the terms associated with the massless excitations do not induce any divergences in the
QCD scattering amplitudes.

%%%%%%%%%%%%%%%%%%%%%%%%%%%%%%%%%%
%Fig. 13
%%%%%%%%%%%%%%%%%%%%%%%%%%%%%%%%%%
%%%%%%%%%%%%%%%%%%%%%%%%%%%%%%

\section{Conclusions}
\label{sec:conc} 

The apparent simplicity of the QCD Lagrangian conceals an enormous wealth of dynamical patterns, giving rise to 
a vast array of complex ``emergent phenomena''~\cite{Roberts:2016mhh}. As has been argued in a series of recent works~\cite{Roberts:2020hiw,Roberts:2021xnz,Binosi:2022djx},  
the pivotal notion that unifies all these phenomena
is the emergence of a hadronic mass, which leaves its imprint on a wide range of physical observables. 
The generation of a mass scale due to the self-interactions of the gluons represents arguably the most fundamental
expression of such an emergence. 
In the present work we have highlighted certain salient aspects   
of the research activity dedicated to this subject,  
within a framework that is based predominantly on the SDEs of the theory, but capitalizes 
extensively on a number of results obtained from large-volume lattice simulations.

The central idea underlying the approach summarized in this article is the implementation of the celebrated Schwinger mechanism
in the context of nonperturbative QCD.
The activation of this mechanism hinges on the 
formation of massless longitudinal poles in the fundamental vertices of the theory.
These poles are {\it composite and carry color}, and play a dual role: they provide the required structures in the gluon vacuum polarization, 
and affect nontrivially the way that the STIs of these vertices are realized. 
This dual nature of the poles is perfectly encoded in the function $\Cfat(r^2)$, which
describes the distinct displacement to the WI satisfied by the pole-free part of the three-gluon vertex, and, at the
same time, is the bound-state wave function that governs the dynamical formation 
of the massless poles by the merging of two gluons, through a characteristic BSE.
This duality, in turn, unveils a profound connection between dynamics (BSEs) and symmetry (STIs),
which becomes particularly patent within the PT-BFM framework.

In fact, it is especially tempting to interpret these massless poles as composite would-be
{\it Nambu-Goldstone} bosons,
given that they appear to fulfill precisely all crucial functions typically ascribed to the latter,
namely ({\it i}) they are ``absorbed'' by the gluons to make them massive, ({\it ii}) maintain the STIs  
of the theory intact once the gluons have been endowed with mass, ({\it iii})
are longitudinally coupled, and ({\it iv}) do not introduce divergences in physical observables, \eg 
the $S$-matrix, as was shown in Sec.~\ref{sec:sdec}.
It would be clearly interesting to pursue this point further; a most promising starting point for such an
investigation is offered by the PT-based analysis first presented in~\cite{Papavassiliou:1989zd}.   

The displacement of the WI, quantified by the function $\Cfat(r^2)$, is exclusive to the special realization of the  Schwinger mechanism
reviewed here. The use   
of lattice results for the main ingredients entering in the pertinent WI reveals the  
presence of a robust model-independent signal for $\Cfat(r^2)$, which is in agreement with the result obtained for it from the
solution of the BSE, under certain simplifying assumptions. 
These findings corroborate the operation of the Schwinger mechanism in QCD, and set the stage for further novel developments.

\section*{Acknowledgments}
\label{sec:acknowledgments}
The author thanks A.C.~Aguilar, D.~Binosi, M.N.~Ferreira, D.~Ib\'a\~nez, J.~Pawlowski, C.D.~Roberts, and
J.~Rodr\'iguez-Quintero for several collaborations, and Craig Roberts for prompting the completion of this work. 
This research is supported 
by the grant PID2020-113334GB-I00/AEI/10.13039/501100011033 of the  Spanish AEI-MICINN,
and the  grant  Prometeo/2019/087 of the Generalitat Valenciana.

\newpage
%\bibliography{bibliography_new}

\begin{thebibliography}{156}%
\makeatletter
\providecommand \@ifxundefined [1]{%
 \@ifx{#1\undefined}
}%
\providecommand \@ifnum [1]{%
 \ifnum #1\expandafter \@firstoftwo
 \else \expandafter \@secondoftwo
 \fi
}%
\providecommand \@ifx [1]{%
 \ifx #1\expandafter \@firstoftwo
 \else \expandafter \@secondoftwo
 \fi
}%
\providecommand \natexlab [1]{#1}%
\providecommand \enquote  [1]{``#1''}%
\providecommand \bibnamefont  [1]{#1}%
\providecommand \bibfnamefont [1]{#1}%
\providecommand \citenamefont [1]{#1}%
\providecommand \href@noop [0]{\@secondoftwo}%
\providecommand \href [0]{\begingroup \@sanitize@url \@href}%
\providecommand \@href[1]{\@@startlink{#1}\@@href}%
\providecommand \@@href[1]{\endgroup#1\@@endlink}%
\providecommand \@sanitize@url [0]{\catcode `\\12\catcode `\$12\catcode
  `\&12\catcode `\#12\catcode `\^12\catcode `\_12\catcode `\%12\relax}%
\providecommand \@@startlink[1]{}%
\providecommand \@@endlink[0]{}%
\providecommand \url  [0]{\begingroup\@sanitize@url \@url }%
\providecommand \@url [1]{\endgroup\@href {#1}{\urlprefix }}%
\providecommand \urlprefix  [0]{URL }%
\providecommand \Eprint [0]{\href }%
\providecommand \doibase [0]{http://dx.doi.org/}%
\providecommand \selectlanguage [0]{\@gobble}%
\providecommand \bibinfo  [0]{\@secondoftwo}%
\providecommand \bibfield  [0]{\@secondoftwo}%
\providecommand \translation [1]{[#1]}%
\providecommand \BibitemOpen [0]{}%
\providecommand \bibitemStop [0]{}%
\providecommand \bibitemNoStop [0]{.\EOS\space}%
\providecommand \EOS [0]{\spacefactor3000\relax}%
\providecommand \BibitemShut  [1]{\csname bibitem#1\endcsname}%
\let\auto@bib@innerbib\@empty
%</preamble>
\bibitem [{\citenamefont {Marciano}\ and\ \citenamefont
  {Pagels}(1978)}]{Marciano:1977su}%
  \BibitemOpen
  \bibfield  {author} {\bibinfo {author} {\bibfnamefont {W.~J.}\ \bibnamefont
  {Marciano}}\ and\ \bibinfo {author} {\bibfnamefont {H.}~\bibnamefont
  {Pagels}},\ }\href {\doibase 10.1016/0370-1573(78)90208-9} {\bibfield
  {journal} {\bibinfo  {journal} {Phys. Rept.}\ }\textbf {\bibinfo {volume}
  {36}},\ \bibinfo {pages} {137} (\bibinfo {year} {1978})}\BibitemShut
  {NoStop}%
\bibitem [{\citenamefont {Collins}(1986)}]{Collins:1984xc}%
  \BibitemOpen
  \bibfield  {author} {\bibinfo {author} {\bibfnamefont {J.~C.}\ \bibnamefont
  {Collins}},\ }\href@noop{} {\emph {\bibinfo
  {title} {{Renormalization}: {An Introduction to Renormalization, The
  Renormalization Group, and the Operator Product Expansion}}}},\ \bibinfo
  {series} {Cambridge Monographs on Mathematical Physics}, Vol.~\bibinfo
  {volume} {26}\ (\bibinfo  {publisher} {Cambridge University Press},\ \bibinfo
  {address} {Cambridge},\ \bibinfo {year} {1986})\BibitemShut {NoStop}%
\bibitem [{\citenamefont {Cloet}\ and\ \citenamefont
  {Roberts}(2014)}]{Cloet:2013jya}%
  \BibitemOpen
  \bibfield  {author} {\bibinfo {author} {\bibfnamefont {I.~C.}\ \bibnamefont
  {Cloet}}\ and\ \bibinfo {author} {\bibfnamefont {C.~D.}\ \bibnamefont
  {Roberts}},\ }\href {\doibase 10.1016/j.ppnp.2014.02.001} {\bibfield
  {journal} {\bibinfo  {journal} {Prog. Part. Nucl. Phys.}\ }\textbf {\bibinfo
  {volume} {77}},\ \bibinfo {pages} {1} (\bibinfo {year} {2014})}\BibitemShut
  {NoStop}%
%%CITATION = ARXIV:1310.2651;%%
\bibitem [{\citenamefont {Aguilar}\ \emph
  {et~al.}(2016{\natexlab{a}})\citenamefont {Aguilar}, \citenamefont {Binosi},\
  and\ \citenamefont {Papavassiliou}}]{Aguilar:2015bud}%
  \BibitemOpen
  \bibfield  {author} {\bibinfo {author} {\bibfnamefont {A.~C.}\ \bibnamefont
  {Aguilar}}, \bibinfo {author} {\bibfnamefont {D.}~\bibnamefont {Binosi}}, \
  and\ \bibinfo {author} {\bibfnamefont {J.}~\bibnamefont {Papavassiliou}},\
  }\href {\doibase 10.1007/s11467-015-0517-6} {\bibfield  {journal} {\bibinfo
  {journal} {Front. Phys.(Beijing)}\ }\textbf {\bibinfo {volume} {11}},\
  \bibinfo {pages} {111203} (\bibinfo {year} {2016}{\natexlab{a}})}\BibitemShut
  {NoStop}%
%%CITATION = ARXIV:1511.08361;%%
\bibitem [{\citenamefont {Roberts}(2020)}]{Roberts:2020hiw}%
  \BibitemOpen
  \bibfield  {author} {\bibinfo {author} {\bibfnamefont {C.~D.}\ \bibnamefont
  {Roberts}},\ }\href {\doibase 10.3390/sym12091468} {\bibfield  {journal}
  {\bibinfo  {journal} {Symmetry}\ }\textbf {\bibinfo {volume} {12}},\ \bibinfo
  {pages} {1468} (\bibinfo {year} {2020})}\BibitemShut {NoStop}%
\bibitem [{\citenamefont {Roberts}(2021)}]{Roberts:2021xnz}%
  \BibitemOpen
  \bibfield  {author} {\bibinfo {author} {\bibfnamefont {C.~D.}\ \bibnamefont
  {Roberts}},\ }\href {\doibase 10.1007/s43673-021-00005-4} {\bibfield
  {journal} {\bibinfo  {journal} {AAPPS Bull.}\ }\textbf {\bibinfo {volume}
  {31}},\ \bibinfo {pages} {6} (\bibinfo {year} {2021})}\BibitemShut {NoStop}%
\bibitem [{\citenamefont {Horak}\ \emph {et~al.}(2022)\citenamefont {Horak},
  \citenamefont {Ihssen}, \citenamefont {Papavassiliou}, \citenamefont
  {Pawlowski}, \citenamefont {Weber},\ and\ \citenamefont
  {Wetterich}}]{Horak:2022aqx}%
  \BibitemOpen
  \bibfield  {author} {\bibinfo {author} {\bibfnamefont {J.}~\bibnamefont
  {Horak}}, \bibinfo {author} {\bibfnamefont {F.}~\bibnamefont {Ihssen}},
  \bibinfo {author} {\bibfnamefont {J.}~\bibnamefont {Papavassiliou}}, \bibinfo
  {author} {\bibfnamefont {J.~M.}\ \bibnamefont {Pawlowski}}, \bibinfo {author}
  {\bibfnamefont {A.}~\bibnamefont {Weber}}, \ and\ \bibinfo {author}
  {\bibfnamefont {C.}~\bibnamefont {Wetterich}},\ }\href{https://arxiv.org/abs/2201.09747} 
 {\bibfield  {journal} {\bibinfo
  {journal} {[arXiv:2201.09747 [hep-ph]]}\ }}\BibitemShut {NoStop}%
  \bibitem [{\citenamefont {Cornwall}(1979)}]{Cornwall:1979hz}%
  \BibitemOpen
  \bibfield  {author} {\bibinfo {author} {\bibfnamefont {J.~M.}\ \bibnamefont
  {Cornwall}},\ }\href {\doibase 10.1016/0550-3213(79)90111-1} {\bibfield
  {journal} {\bibinfo  {journal} {Nucl. Phys. B}\ }\textbf {\bibinfo {volume}
  {157}},\ \bibinfo {pages} {392} (\bibinfo {year} {1979})}\BibitemShut
  {NoStop}%
\bibitem [{\citenamefont {Parisi}\ and\ \citenamefont
  {Petronzio}(1980)}]{Parisi:1980jy}%
  \BibitemOpen
  \bibfield  {author} {\bibinfo {author} {\bibfnamefont {G.}~\bibnamefont
  {Parisi}}\ and\ \bibinfo {author} {\bibfnamefont {R.}~\bibnamefont
  {Petronzio}},\ }\href {\doibase 10.1016/0370-2693(80)90822-9} {\bibfield
  {journal} {\bibinfo  {journal} {Phys. Lett. B}\ }\textbf {\bibinfo {volume}
  {94}},\ \bibinfo {pages} {51} (\bibinfo {year} {1980})}\BibitemShut {NoStop}%
\bibitem [{\citenamefont {Cornwall}(1982)}]{Cornwall:1981zr}%
  \BibitemOpen
  \bibfield  {author} {\bibinfo {author} {\bibfnamefont {J.~M.}\ \bibnamefont
  {Cornwall}},\ }\href {\doibase 10.1103/PhysRevD.26.1453} {\bibfield
  {journal} {\bibinfo  {journal} {Phys. Rev. D}\ }\textbf {\bibinfo {volume}
  {26}},\ \bibinfo {pages} {1453} (\bibinfo {year} {1982})}\BibitemShut
  {NoStop}%
\bibitem [{\citenamefont {Bernard}(1982)}]{Bernard:1981pg}%
  \BibitemOpen
  \bibfield  {author} {\bibinfo {author} {\bibfnamefont {C.~W.}\ \bibnamefont
  {Bernard}},\ }\href {\doibase 10.1016/0370-2693(82)91228-X} {\bibfield
  {journal} {\bibinfo  {journal} {Phys. Lett. B}\ }\textbf {\bibinfo {volume}
  {108}},\ \bibinfo {pages} {431} (\bibinfo {year} {1982})}\BibitemShut
  {NoStop}%
\bibitem [{\citenamefont {Bernard}(1983)}]{Bernard:1982my}%
  \BibitemOpen
  \bibfield  {author} {\bibinfo {author} {\bibfnamefont {C.~W.}\ \bibnamefont
  {Bernard}},\ }\href {\doibase 10.1016/0550-3213(83)90645-4} {\bibfield
  {journal} {\bibinfo  {journal} {Nucl. Phys. B}\ }\textbf {\bibinfo {volume}
  {219}},\ \bibinfo {pages} {341} (\bibinfo {year} {1983})}\BibitemShut
  {NoStop}%
\bibitem [{\citenamefont {Donoghue}(1984)}]{Donoghue:1983fy}%
  \BibitemOpen
  \bibfield  {author} {\bibinfo {author} {\bibfnamefont {J.~F.}\ \bibnamefont
  {Donoghue}},\ }\href {\doibase 10.1103/PhysRevD.29.2559} {\bibfield
  {journal} {\bibinfo  {journal} {Phys. Rev. D}\ }\textbf {\bibinfo {volume}
  {29}},\ \bibinfo {pages} {2559} (\bibinfo {year} {1984})}\BibitemShut
  {NoStop}%
\bibitem [{\citenamefont {Mandula}\ and\ \citenamefont
  {Ogilvie}(1987)}]{Mandula:1987rh}%
  \BibitemOpen
  \bibfield  {author} {\bibinfo {author} {\bibfnamefont {J.}~\bibnamefont
  {Mandula}}\ and\ \bibinfo {author} {\bibfnamefont {M.}~\bibnamefont
  {Ogilvie}},\ }\href {\doibase 10.1016/0370-2693(87)91541-3} {\bibfield
  {journal} {\bibinfo  {journal} {Phys. Lett. B}\ }\textbf {\bibinfo {volume}
  {185}},\ \bibinfo {pages} {127} (\bibinfo {year} {1987})}\BibitemShut
  {NoStop}%
\bibitem [{\citenamefont {Cornwall}\ and\ \citenamefont
  {Papavassiliou}(1989)}]{Cornwall:1989gv}%
  \BibitemOpen
  \bibfield  {author} {\bibinfo {author} {\bibfnamefont {J.~M.}\ \bibnamefont
  {Cornwall}}\ and\ \bibinfo {author} {\bibfnamefont {J.}~\bibnamefont
  {Papavassiliou}},\ }\href {\doibase 10.1103/PhysRevD.40.3474} {\bibfield
  {journal} {\bibinfo  {journal} {Phys. Rev. D}\ }\textbf {\bibinfo {volume}
  {40}},\ \bibinfo {pages} {3474} (\bibinfo {year} {1989})}\BibitemShut
  {NoStop}%
\bibitem [{\citenamefont {Wilson}\ \emph {et~al.}(1994)\citenamefont {Wilson},
  \citenamefont {Walhout}, \citenamefont {Harindranath}, \citenamefont {Zhang},
  \citenamefont {Perry},\ and\ \citenamefont {Glazek}}]{Wilson:1994fk}%
  \BibitemOpen
  \bibfield  {author} {\bibinfo {author} {\bibfnamefont {K.~G.}\ \bibnamefont
  {Wilson}}, \bibinfo {author} {\bibfnamefont {T.~S.}\ \bibnamefont {Walhout}},
  \bibinfo {author} {\bibfnamefont {A.}~\bibnamefont {Harindranath}}, \bibinfo
  {author} {\bibfnamefont {W.-M.}\ \bibnamefont {Zhang}}, \bibinfo {author}
  {\bibfnamefont {R.~J.}\ \bibnamefont {Perry}}, \ and\ \bibinfo {author}
  {\bibfnamefont {S.~D.}\ \bibnamefont {Glazek}},\ }\href {\doibase
  10.1103/PhysRevD.49.6720} {\bibfield  {journal} {\bibinfo  {journal} {Phys.
  Rev.}\ }\textbf {\bibinfo {volume} {D49}},\ \bibinfo {pages} {6720} (\bibinfo
  {year} {1994})}\BibitemShut {NoStop}%
%%CITATION = HEP-TH/9401153;%%
\bibitem [{\citenamefont {Lavelle}(1991)}]{Lavelle:1991ve}%
  \BibitemOpen
  \bibfield  {author} {\bibinfo {author} {\bibfnamefont {M.}~\bibnamefont
  {Lavelle}},\ }\href {\doibase 10.1103/PhysRevD.44.R26} {\bibfield  {journal}
  {\bibinfo  {journal} {Phys. Rev. D}\ }\textbf {\bibinfo {volume} {44}},\
  \bibinfo {pages} {26} (\bibinfo {year} {1991})}\BibitemShut {NoStop}%
\bibitem [{\citenamefont {Halzen}\ \emph {et~al.}(1993)\citenamefont {Halzen},
  \citenamefont {Krein},\ and\ \citenamefont {Natale}}]{Halzen:1992vd}%
  \BibitemOpen
  \bibfield  {author} {\bibinfo {author} {\bibfnamefont {F.}~\bibnamefont
  {Halzen}}, \bibinfo {author} {\bibfnamefont {G.~I.}\ \bibnamefont {Krein}}, \
  and\ \bibinfo {author} {\bibfnamefont {A.~A.}\ \bibnamefont {Natale}},\
  }\href {\doibase 10.1103/PhysRevD.47.295} {\bibfield  {journal} {\bibinfo
  {journal} {Phys. Rev. D}\ }\textbf {\bibinfo {volume} {47}},\ \bibinfo
  {pages} {295} (\bibinfo {year} {1993})}\BibitemShut {NoStop}%
\bibitem [{\citenamefont {Mihara}\ and\ \citenamefont
  {Natale}(2000)}]{Mihara:2000wf}%
  \BibitemOpen
  \bibfield  {author} {\bibinfo {author} {\bibfnamefont {A.}~\bibnamefont
  {Mihara}}\ and\ \bibinfo {author} {\bibfnamefont {A.~A.}\ \bibnamefont
  {Natale}},\ }\href {\doibase 10.1016/S0370-2693(00)00546-3} {\bibfield
  {journal} {\bibinfo  {journal} {Phys. Lett. B}\ }\textbf {\bibinfo {volume}
  {482}},\ \bibinfo {pages} {378} (\bibinfo {year} {2000})}\BibitemShut
  {NoStop}%
\bibitem [{\citenamefont {Philipsen}(2002)}]{Philipsen:2001ip}%
  \BibitemOpen
  \bibfield  {author} {\bibinfo {author} {\bibfnamefont {O.}~\bibnamefont
  {Philipsen}},\ }\href {\doibase 10.1016/S0550-3213(02)00089-5} {\bibfield
  {journal} {\bibinfo  {journal} {Nucl. Phys.}\ }\textbf {\bibinfo {volume}
  {B628}},\ \bibinfo {pages} {167} (\bibinfo {year} {2002})}\BibitemShut
  {NoStop}%
%%CITATION = HEP-LAT/0112047;%%
\bibitem [{\citenamefont {Kondo}(2001)}]{Kondo:2001nq}%
  \BibitemOpen
  \bibfield  {author} {\bibinfo {author} {\bibfnamefont {K.-I.}\ \bibnamefont
  {Kondo}},\ }\href {\doibase 10.1016/S0370-2693(01)00817-6} {\bibfield
  {journal} {\bibinfo  {journal} {Phys. Lett. B}\ }\textbf {\bibinfo {volume}
  {514}},\ \bibinfo {pages} {335} (\bibinfo {year} {2001})}\BibitemShut
  {NoStop}%
\bibitem [{\citenamefont {Aguilar}\ \emph {et~al.}(2003)\citenamefont
  {Aguilar}, \citenamefont {Natale},\ and\ \citenamefont {Rodrigues~da
  Silva}}]{Aguilar:2002tc}%
  \BibitemOpen
  \bibfield  {author} {\bibinfo {author} {\bibfnamefont {A.~C.}\ \bibnamefont
  {Aguilar}}, \bibinfo {author} {\bibfnamefont {A.~A.}\ \bibnamefont {Natale}},
  \ and\ \bibinfo {author} {\bibfnamefont {P.~S.}\ \bibnamefont {Rodrigues~da
  Silva}},\ }\href {\doibase 10.1103/PhysRevLett.90.152001} {\bibfield
  {journal} {\bibinfo  {journal} {Phys. Rev. Lett.}\ }\textbf {\bibinfo
  {volume} {90}},\ \bibinfo {pages} {152001} (\bibinfo {year}
  {2003})}\BibitemShut {NoStop}%
\bibitem [{\citenamefont {Athenodorou}\ and\ \citenamefont
  {Teper}(2020)}]{Athenodorou:2020ani}%
  \BibitemOpen
  \bibfield  {author} {\bibinfo {author} {\bibfnamefont {A.}~\bibnamefont
  {Athenodorou}}\ and\ \bibinfo {author} {\bibfnamefont {M.}~\bibnamefont
  {Teper}},\ }\href {\doibase 10.1007/JHEP11(2020)172} {\bibfield  {journal}
  {\bibinfo  {journal} {JHEP}\ }\textbf {\bibinfo {volume} {11}},\ \bibinfo
  {pages} {172} (\bibinfo {year} {2020})}\BibitemShut {NoStop}%
\bibitem [{\citenamefont {Athenodorou}\ and\ \citenamefont
  {Teper}(2021)}]{Athenodorou:2021qvs}%
  \BibitemOpen
  \bibfield  {author} {\bibinfo {author} {\bibfnamefont {A.}~\bibnamefont
  {Athenodorou}}\ and\ \bibinfo {author} {\bibfnamefont {M.}~\bibnamefont
  {Teper}},\ }\href {\doibase 10.1007/JHEP12(2021)082} {\bibfield  {journal}
  {\bibinfo  {journal} {JHEP}\ }\textbf {\bibinfo {volume} {12}},\ \bibinfo
  {pages} {082} (\bibinfo {year} {2021})}\BibitemShut {NoStop}%
\bibitem [{\citenamefont {Collins}\ \emph {et~al.}(1977)\citenamefont
  {Collins}, \citenamefont {Duncan},\ and\ \citenamefont
  {Joglekar}}]{Collins:1976yq}%
  \BibitemOpen
  \bibfield  {author} {\bibinfo {author} {\bibfnamefont {J.~C.}\ \bibnamefont
  {Collins}}, \bibinfo {author} {\bibfnamefont {A.}~\bibnamefont {Duncan}}, \
  and\ \bibinfo {author} {\bibfnamefont {S.~D.}\ \bibnamefont {Joglekar}},\
  }\href {\doibase 10.1103/PhysRevD.16.438} {\bibfield  {journal} {\bibinfo
  {journal} {Phys. Rev. D}\ }\textbf {\bibinfo {volume} {16}},\ \bibinfo
  {pages} {438} (\bibinfo {year} {1977})}\BibitemShut {NoStop}%
\bibitem [{\citenamefont {Brodsky}\ and\ \citenamefont
  {Shrock}(2008)}]{Brodsky:2008be}%
  \BibitemOpen
  \bibfield  {author} {\bibinfo {author} {\bibfnamefont {S.~J.}\ \bibnamefont
  {Brodsky}}\ and\ \bibinfo {author} {\bibfnamefont {R.}~\bibnamefont
  {Shrock}},\ }\href {\doibase 10.1016/j.physletb.2008.06.054} {\bibfield
  {journal} {\bibinfo  {journal} {Phys. Lett.}\ }\textbf {\bibinfo {volume}
  {B666}},\ \bibinfo {pages} {95} (\bibinfo {year} {2008})}\BibitemShut
  {NoStop}%
%%CITATION = 0806.1535;%%
\bibitem [{\citenamefont {Braun}\ \emph {et~al.}(2010)\citenamefont {Braun},
  \citenamefont {Gies},\ and\ \citenamefont {Pawlowski}}]{Braun:2007bx}%
  \BibitemOpen
  \bibfield  {author} {\bibinfo {author} {\bibfnamefont {J.}~\bibnamefont
  {Braun}}, \bibinfo {author} {\bibfnamefont {H.}~\bibnamefont {Gies}}, \ and\
  \bibinfo {author} {\bibfnamefont {J.~M.}\ \bibnamefont {Pawlowski}},\ }\href
  {\doibase 10.1016/j.physletb.2010.01.009} {\bibfield  {journal} {\bibinfo
  {journal} {Phys. Lett.}\ }\textbf {\bibinfo {volume} {B684}},\ \bibinfo
  {pages} {262} (\bibinfo {year} {2010})}\BibitemShut {NoStop}%
%%CITATION = ARXIV:0708.2413;%%
\bibitem [{\citenamefont {Binosi}\ \emph {et~al.}(2015)\citenamefont {Binosi},
  \citenamefont {Chang}, \citenamefont {Papavassiliou},\ and\ \citenamefont
  {Roberts}}]{Binosi:2014aea}%
  \BibitemOpen
  \bibfield  {author} {\bibinfo {author} {\bibfnamefont {D.}~\bibnamefont
  {Binosi}}, \bibinfo {author} {\bibfnamefont {L.}~\bibnamefont {Chang}},
  \bibinfo {author} {\bibfnamefont {J.}~\bibnamefont {Papavassiliou}}, \ and\
  \bibinfo {author} {\bibfnamefont {C.~D.}\ \bibnamefont {Roberts}},\ }\href
  {\doibase 10.1016/j.physletb.2015.01.031} {\bibfield  {journal} {\bibinfo
  {journal} {Phys. Lett.}\ }\textbf {\bibinfo {volume} {B742}},\ \bibinfo
  {pages} {183} (\bibinfo {year} {2015})}\BibitemShut {NoStop}%
%%CITATION = ARXIV:1412.4782;%%
\bibitem [{\citenamefont {Gao}\ \emph {et~al.}(2018)\citenamefont {Gao},
  \citenamefont {Qin}, \citenamefont {Roberts},\ and\ \citenamefont
  {Rodriguez-Quintero}}]{Gao:2017uox}%
  \BibitemOpen
  \bibfield  {author} {\bibinfo {author} {\bibfnamefont {F.}~\bibnamefont
  {Gao}}, \bibinfo {author} {\bibfnamefont {S.-X.}\ \bibnamefont {Qin}},
  \bibinfo {author} {\bibfnamefont {C.~D.}\ \bibnamefont {Roberts}}, \ and\
  \bibinfo {author} {\bibfnamefont {J.}~\bibnamefont {Rodriguez-Quintero}},\
  }\href {\doibase 10.1103/PhysRevD.97.034010} {\bibfield  {journal} {\bibinfo
  {journal} {Phys. Rev.}\ }\textbf {\bibinfo {volume} {D97}},\ \bibinfo {pages}
  {034010} (\bibinfo {year} {2018})}\BibitemShut {NoStop}%
%%CITATION = ARXIV:1706.04681;%%
\bibitem [{\citenamefont {Binosi}\ and\ \citenamefont
  {Papavassiliou}(2003)}]{Binosi:2002vk}%
  \BibitemOpen
  \bibfield  {author} {\bibinfo {author} {\bibfnamefont {D.}~\bibnamefont
  {Binosi}}\ and\ \bibinfo {author} {\bibfnamefont {J.}~\bibnamefont
  {Papavassiliou}},\ }\href {\doibase 10.1016/S0920-5632(03)01862-0} {\bibfield
   {journal} {\bibinfo  {journal} {Nucl. Phys. Proc. Suppl.}\ }\textbf
  {\bibinfo {volume} {121}},\ \bibinfo {pages} {281} (\bibinfo {year}
  {2003})}\BibitemShut {NoStop}%
\bibitem [{\citenamefont {Aguilar}\ \emph
  {et~al.}(2009{\natexlab{a}})\citenamefont {Aguilar}, \citenamefont {Binosi},
  \citenamefont {Papavassiliou},\ and\ \citenamefont
  {Rodriguez-Quintero}}]{Aguilar:2009nf}%
  \BibitemOpen
  \bibfield  {author} {\bibinfo {author} {\bibfnamefont {A.~C.}\ \bibnamefont
  {Aguilar}}, \bibinfo {author} {\bibfnamefont {D.}~\bibnamefont {Binosi}},
  \bibinfo {author} {\bibfnamefont {J.}~\bibnamefont {Papavassiliou}}, \ and\
  \bibinfo {author} {\bibfnamefont {J.}~\bibnamefont {Rodriguez-Quintero}},\
  }\href {\doibase 10.1103/PhysRevD.80.085018} {\bibfield  {journal} {\bibinfo
  {journal} {Phys. Rev.}\ }\textbf {\bibinfo {volume} {D80}},\ \bibinfo {pages}
  {085018} (\bibinfo {year} {2009}{\natexlab{a}})}\BibitemShut {NoStop}%
\bibitem [{\citenamefont {Binosi}\ \emph {et~al.}(2017)\citenamefont {Binosi},
  \citenamefont {Mezrag}, \citenamefont {Papavassiliou}, \citenamefont
  {Roberts},\ and\ \citenamefont {Rodriguez-Quintero}}]{Binosi:2016nme}%
  \BibitemOpen
  \bibfield  {author} {\bibinfo {author} {\bibfnamefont {D.}~\bibnamefont
  {Binosi}}, \bibinfo {author} {\bibfnamefont {C.}~\bibnamefont {Mezrag}},
  \bibinfo {author} {\bibfnamefont {J.}~\bibnamefont {Papavassiliou}}, \bibinfo
  {author} {\bibfnamefont {C.~D.}\ \bibnamefont {Roberts}}, \ and\ \bibinfo
  {author} {\bibfnamefont {J.}~\bibnamefont {Rodriguez-Quintero}},\ }\href
  {\doibase 10.1103/PhysRevD.96.054026} {\bibfield  {journal} {\bibinfo
  {journal} {Phys. Rev.}\ }\textbf {\bibinfo {volume} {D96}},\ \bibinfo {pages}
  {054026} (\bibinfo {year} {2017})}\BibitemShut {NoStop}%
%%CITATION = ARXIV:1612.04835;%%
\bibitem [{\citenamefont {Cui}\ \emph {et~al.}(2020)\citenamefont {Cui},
  \citenamefont {Zhang}, \citenamefont {Binosi}, \citenamefont {de~Soto},
  \citenamefont {Mezrag}, \citenamefont {Papavassiliou}, \citenamefont
  {Roberts}, \citenamefont {Rodríguez-Quintero}, \citenamefont {Segovia},\
  and\ \citenamefont {Zafeiropoulos}}]{Cui:2019dwv}%
  \BibitemOpen
  \bibfield  {author} {\bibinfo {author} {\bibfnamefont {Z.-F.}\ \bibnamefont
  {Cui}}, \bibinfo {author} {\bibfnamefont {J.-L.}\ \bibnamefont {Zhang}},
  \bibinfo {author} {\bibfnamefont {D.}~\bibnamefont {Binosi}}, \bibinfo
  {author} {\bibfnamefont {F.}~\bibnamefont {de~Soto}}, \bibinfo {author}
  {\bibfnamefont {C.}~\bibnamefont {Mezrag}}, \bibinfo {author} {\bibfnamefont
  {J.}~\bibnamefont {Papavassiliou}}, \bibinfo {author} {\bibfnamefont {C.~D.}\
  \bibnamefont {Roberts}}, \bibinfo {author} {\bibfnamefont {J.}~\bibnamefont
  {Rodríguez-Quintero}}, \bibinfo {author} {\bibfnamefont {J.}~\bibnamefont
  {Segovia}}, \ and\ \bibinfo {author} {\bibfnamefont {S.}~\bibnamefont
  {Zafeiropoulos}},\ }\href {\doibase 10.1088/1674-1137/44/8/083102} {\bibfield
   {journal} {\bibinfo  {journal} {Chin. Phys. C}\ }\textbf {\bibinfo {volume}
  {44}},\ \bibinfo {pages} {083102} (\bibinfo {year} {2020})}\BibitemShut
  {NoStop}%
\bibitem [{\citenamefont {Cucchieri}\ and\ \citenamefont
  {Mendes}(2007)}]{Cucchieri:2007md}%
  \BibitemOpen
  \bibfield  {author} {\bibinfo {author} {\bibfnamefont {A.}~\bibnamefont
  {Cucchieri}}\ and\ \bibinfo {author} {\bibfnamefont {T.}~\bibnamefont
  {Mendes}},\ }\href {\doibase 10.22323/1.042.0297} {\bibfield  {journal}
  {\bibinfo  {journal} {PoS}\ }\textbf {\bibinfo {volume} {LATTICE2007}},\
  \bibinfo {pages} {297} (\bibinfo {year} {2007})}\BibitemShut {NoStop}%
\bibitem [{\citenamefont {Cucchieri}\ and\ \citenamefont
  {Mendes}(2008)}]{Cucchieri:2007rg}%
  \BibitemOpen
  \bibfield  {author} {\bibinfo {author} {\bibfnamefont {A.}~\bibnamefont
  {Cucchieri}}\ and\ \bibinfo {author} {\bibfnamefont {T.}~\bibnamefont
  {Mendes}},\ }\href {\doibase 10.1103/PhysRevLett.100.241601} {\bibfield
  {journal} {\bibinfo  {journal} {Phys. Rev. Lett.}\ }\textbf {\bibinfo
  {volume} {100}},\ \bibinfo {pages} {241601} (\bibinfo {year}
  {2008})}\BibitemShut {NoStop}%
%%CITATION = ARXIV:0712.3517;%%
\bibitem [{\citenamefont {Bogolubsky}\ \emph {et~al.}(2007)\citenamefont
  {Bogolubsky}, \citenamefont {Ilgenfritz}, \citenamefont {Muller-Preussker},\
  and\ \citenamefont {Sternbeck}}]{Bogolubsky:2007ud}%
  \BibitemOpen
  \bibfield  {author} {\bibinfo {author} {\bibfnamefont {I.}~\bibnamefont
  {Bogolubsky}}, \bibinfo {author} {\bibfnamefont {E.}~\bibnamefont
  {Ilgenfritz}}, \bibinfo {author} {\bibfnamefont {M.}~\bibnamefont
  {Muller-Preussker}}, \ and\ \bibinfo {author} {\bibfnamefont
  {A.}~\bibnamefont {Sternbeck}},\ }\href {\doibase 10.22323/1.042.0290}
  {\bibfield  {journal} {\bibinfo  {journal} {PoS}\ }\textbf {\bibinfo {volume}
  {LATTICE2007}},\ \bibinfo {pages} {290} (\bibinfo {year} {2007})}\BibitemShut
  {NoStop}%
\bibitem [{\citenamefont {Bogolubsky}\ \emph {et~al.}(2009)\citenamefont
  {Bogolubsky}, \citenamefont {Ilgenfritz}, \citenamefont {Muller-Preussker},\
  and\ \citenamefont {Sternbeck}}]{Bogolubsky:2009dc}%
  \BibitemOpen
  \bibfield  {author} {\bibinfo {author} {\bibfnamefont {I.}~\bibnamefont
  {Bogolubsky}}, \bibinfo {author} {\bibfnamefont {E.}~\bibnamefont
  {Ilgenfritz}}, \bibinfo {author} {\bibfnamefont {M.}~\bibnamefont
  {Muller-Preussker}}, \ and\ \bibinfo {author} {\bibfnamefont
  {A.}~\bibnamefont {Sternbeck}},\ }\href {\doibase
  10.1016/j.physletb.2009.04.076} {\bibfield  {journal} {\bibinfo  {journal}
  {Phys. Lett.}\ }\textbf {\bibinfo {volume} {B676}},\ \bibinfo {pages} {69}
  (\bibinfo {year} {2009})}\BibitemShut {NoStop}%
%%CITATION = ARXIV:0901.0736;%%
\bibitem [{\citenamefont {Oliveira}\ and\ \citenamefont
  {Silva}(2009)}]{Oliveira:2009eh}%
  \BibitemOpen
  \bibfield  {author} {\bibinfo {author} {\bibfnamefont {O.}~\bibnamefont
  {Oliveira}}\ and\ \bibinfo {author} {\bibfnamefont {P.}~\bibnamefont
  {Silva}},\ }\href {\doibase 10.22323/1.091.0226} {\bibfield  {journal}
  {\bibinfo  {journal} {PoS}\ }\textbf {\bibinfo {volume} {LAT2009}},\ \bibinfo
  {pages} {226} (\bibinfo {year} {2009})}\BibitemShut {NoStop}%
\bibitem [{\citenamefont {Oliveira}\ and\ \citenamefont
  {Bicudo}(2011)}]{Oliveira:2010xc}%
  \BibitemOpen
  \bibfield  {author} {\bibinfo {author} {\bibfnamefont {O.}~\bibnamefont
  {Oliveira}}\ and\ \bibinfo {author} {\bibfnamefont {P.}~\bibnamefont
  {Bicudo}},\ }\href {\doibase 10.1088/0954-3899/38/4/045003} {\bibfield
  {journal} {\bibinfo  {journal} {J. Phys. G}\ }\textbf {\bibinfo {volume}
  {G38}},\ \bibinfo {pages} {045003} (\bibinfo {year} {2011})}\BibitemShut
  {NoStop}%
%%CITATION = ARXIV:1002.4151;%%
\bibitem [{\citenamefont {Cucchieri}\ and\ \citenamefont
  {Mendes}(2010)}]{Cucchieri:2009zt}%
  \BibitemOpen
  \bibfield  {author} {\bibinfo {author} {\bibfnamefont {A.}~\bibnamefont
  {Cucchieri}}\ and\ \bibinfo {author} {\bibfnamefont {T.}~\bibnamefont
  {Mendes}},\ }\href {\doibase 10.1103/PhysRevD.81.016005} {\bibfield
  {journal} {\bibinfo  {journal} {Phys. Rev.}\ }\textbf {\bibinfo {volume}
  {D81}},\ \bibinfo {pages} {016005} (\bibinfo {year} {2010})}\BibitemShut
  {NoStop}%
%%CITATION = ARXIV:0904.4033;%%
\bibitem [{\citenamefont {Aguilar}\ \emph {et~al.}(2008)\citenamefont
  {Aguilar}, \citenamefont {Binosi},\ and\ \citenamefont
  {Papavassiliou}}]{Aguilar:2008xm}%
  \BibitemOpen
  \bibfield  {author} {\bibinfo {author} {\bibfnamefont {A.~C.}\ \bibnamefont
  {Aguilar}}, \bibinfo {author} {\bibfnamefont {D.}~\bibnamefont {Binosi}}, \
  and\ \bibinfo {author} {\bibfnamefont {J.}~\bibnamefont {Papavassiliou}},\
  }\href {\doibase 10.1103/PhysRevD.78.025010} {\bibfield  {journal} {\bibinfo
  {journal} {Phys. Rev.}\ }\textbf {\bibinfo {volume} {D78}},\ \bibinfo {pages}
  {025010} (\bibinfo {year} {2008})}\BibitemShut {NoStop}%
\bibitem [{\citenamefont {Fischer}\ \emph {et~al.}(2009)\citenamefont
  {Fischer}, \citenamefont {Maas},\ and\ \citenamefont
  {Pawlowski}}]{Fischer:2008uz}%
  \BibitemOpen
  \bibfield  {author} {\bibinfo {author} {\bibfnamefont {C.~S.}\ \bibnamefont
  {Fischer}}, \bibinfo {author} {\bibfnamefont {A.}~\bibnamefont {Maas}}, \
  and\ \bibinfo {author} {\bibfnamefont {J.~M.}\ \bibnamefont {Pawlowski}},\
  }\href {\doibase 10.1016/j.aop.2009.07.009} {\bibfield  {journal} {\bibinfo
  {journal} {Annals Phys.}\ }\textbf {\bibinfo {volume} {324}},\ \bibinfo
  {pages} {2408} (\bibinfo {year} {2009})}\BibitemShut {NoStop}%
%%CITATION = ARXIV:0810.1987;%%
\bibitem [{\citenamefont {Binosi}\ \emph {et~al.}(2012)\citenamefont {Binosi},
  \citenamefont {Iba\~nez},\ and\ \citenamefont
  {Papavassiliou}}]{Binosi:2012sj}%
  \BibitemOpen
  \bibfield  {author} {\bibinfo {author} {\bibfnamefont {D.}~\bibnamefont
  {Binosi}}, \bibinfo {author} {\bibfnamefont {D.}~\bibnamefont {Iba\~nez}}, \
  and\ \bibinfo {author} {\bibfnamefont {J.}~\bibnamefont {Papavassiliou}},\
  }\href {\doibase 10.1103/PhysRevD.86.085033} {\bibfield  {journal} {\bibinfo
  {journal} {Phys. Rev.}\ }\textbf {\bibinfo {volume} {D86}},\ \bibinfo {pages}
  {085033} (\bibinfo {year} {2012})}\BibitemShut {NoStop}%
%%CITATION = ARXIV:1208.1451;%%
\bibitem [{\citenamefont {Serreau}\ and\ \citenamefont
  {Tissier}(2012)}]{Serreau:2012cg}%
  \BibitemOpen
  \bibfield  {author} {\bibinfo {author} {\bibfnamefont {J.}~\bibnamefont
  {Serreau}}\ and\ \bibinfo {author} {\bibfnamefont {M.}~\bibnamefont
  {Tissier}},\ }\href {\doibase 10.1016/j.physletb.2012.04.041} {\bibfield
  {journal} {\bibinfo  {journal} {Phys. Lett.}\ }\textbf {\bibinfo {volume}
  {B712}},\ \bibinfo {pages} {97} (\bibinfo {year} {2012})}\BibitemShut
  {NoStop}%
%%CITATION = ARXIV:1202.3432;%%
\bibitem [{\citenamefont {Tissier}\ and\ \citenamefont
  {Wschebor}(2010)}]{Tissier:2010ts}%
  \BibitemOpen
  \bibfield  {author} {\bibinfo {author} {\bibfnamefont {M.}~\bibnamefont
  {Tissier}}\ and\ \bibinfo {author} {\bibfnamefont {N.}~\bibnamefont
  {Wschebor}},\ }\href {\doibase 10.1103/PhysRevD.82.101701} {\bibfield
  {journal} {\bibinfo  {journal} {Phys. Rev. D}\ }\textbf {\bibinfo {volume}
  {82}},\ \bibinfo {pages} {101701} (\bibinfo {year} {2010})}\BibitemShut
  {NoStop}%
\bibitem [{\citenamefont {Aguilar}\ \emph
  {et~al.}(2016{\natexlab{b}})\citenamefont {Aguilar}, \citenamefont {Binosi},
  \citenamefont {Figueiredo},\ and\ \citenamefont
  {Papavassiliou}}]{Aguilar:2016vin}%
  \BibitemOpen
  \bibfield  {author} {\bibinfo {author} {\bibfnamefont {A.~C.}\ \bibnamefont
  {Aguilar}}, \bibinfo {author} {\bibfnamefont {D.}~\bibnamefont {Binosi}},
  \bibinfo {author} {\bibfnamefont {C.~T.}\ \bibnamefont {Figueiredo}}, \ and\
  \bibinfo {author} {\bibfnamefont {J.}~\bibnamefont {Papavassiliou}},\ }\href
  {\doibase 10.1103/PhysRevD.94.045002} {\bibfield  {journal} {\bibinfo
  {journal} {Phys. Rev.}\ }\textbf {\bibinfo {volume} {D94}},\ \bibinfo {pages}
  {045002} (\bibinfo {year} {2016}{\natexlab{b}})}\BibitemShut {NoStop}%
%%CITATION = ARXIV:1604.08456;%%
\bibitem [{\citenamefont {Pel\'aez}\ \emph {et~al.}(2014)\citenamefont
  {Pel\'aez}, \citenamefont {Tissier},\ and\ \citenamefont
  {Wschebor}}]{Pelaez:2014mxa}%
  \BibitemOpen
  \bibfield  {author} {\bibinfo {author} {\bibfnamefont {M.}~\bibnamefont
  {Pel\'aez}}, \bibinfo {author} {\bibfnamefont {M.}~\bibnamefont {Tissier}}, \
  and\ \bibinfo {author} {\bibfnamefont {N.}~\bibnamefont {Wschebor}},\ }\href
  {\doibase 10.1103/PhysRevD.90.065031} {\bibfield  {journal} {\bibinfo
  {journal} {Phys. Rev. D}\ }\textbf {\bibinfo {volume} {90}},\ \bibinfo
  {pages} {065031} (\bibinfo {year} {2014})}\BibitemShut {NoStop}%
\bibitem [{\citenamefont {Dudal}\ \emph {et~al.}(2008)\citenamefont {Dudal},
  \citenamefont {Gracey}, \citenamefont {Sorella}, \citenamefont
  {Vandersickel},\ and\ \citenamefont {Verschelde}}]{Dudal:2008sp}%
  \BibitemOpen
  \bibfield  {author} {\bibinfo {author} {\bibfnamefont {D.}~\bibnamefont
  {Dudal}}, \bibinfo {author} {\bibfnamefont {J.~A.}\ \bibnamefont {Gracey}},
  \bibinfo {author} {\bibfnamefont {S.~P.}\ \bibnamefont {Sorella}}, \bibinfo
  {author} {\bibfnamefont {N.}~\bibnamefont {Vandersickel}}, \ and\ \bibinfo
  {author} {\bibfnamefont {H.}~\bibnamefont {Verschelde}},\ }\href {\doibase
  10.1103/PhysRevD.78.065047} {\bibfield  {journal} {\bibinfo  {journal} {Phys.
  Rev.}\ }\textbf {\bibinfo {volume} {D78}},\ \bibinfo {pages} {065047}
  (\bibinfo {year} {2008})}\BibitemShut {NoStop}%
%%CITATION = 0806.4348;%%
\bibitem [{\citenamefont
  {Rodriguez-Quintero}(2011)}]{RodriguezQuintero:2010wy}%
  \BibitemOpen
  \bibfield  {author} {\bibinfo {author} {\bibfnamefont {J.}~\bibnamefont
  {Rodriguez-Quintero}},\ }\href {\doibase 10.1007/JHEP01(2011)105} {\bibfield
  {journal} {\bibinfo  {journal} {J. High Energy Phys.}\ }\textbf {\bibinfo
  {volume} {01}},\ \bibinfo {pages} {105} (\bibinfo {year} {2011})}\BibitemShut
  {NoStop}%
%%CITATION = ARXIV:1005.4598;%%
\bibitem [{\citenamefont {Pennington}\ and\ \citenamefont
  {Wilson}(2011)}]{Pennington:2011xs}%
  \BibitemOpen
  \bibfield  {author} {\bibinfo {author} {\bibfnamefont {M.~R.}\ \bibnamefont
  {Pennington}}\ and\ \bibinfo {author} {\bibfnamefont {D.~J.}\ \bibnamefont
  {Wilson}},\ }\href {\doibase 10.1103/PhysRevD.84.094028,
  10.1103/PhysRevD.84.119901} {\bibfield  {journal} {\bibinfo  {journal} {Phys.
  Rev.}\ }\textbf {\bibinfo {volume} {D84}},\ \bibinfo {pages} {119901}
  (\bibinfo {year} {2011})}\BibitemShut {NoStop}%
%%CITATION = ARXIV:1109.2117;%%
\bibitem [{\citenamefont {Meyers}\ and\ \citenamefont
  {Swanson}(2014)}]{Meyers:2014iwa}%
  \BibitemOpen
  \bibfield  {author} {\bibinfo {author} {\bibfnamefont {J.}~\bibnamefont
  {Meyers}}\ and\ \bibinfo {author} {\bibfnamefont {E.~S.}\ \bibnamefont
  {Swanson}},\ }\href {\doibase 10.1103/PhysRevD.90.045037} {\bibfield
  {journal} {\bibinfo  {journal} {Phys. Rev.}\ }\textbf {\bibinfo {volume}
  {D90}},\ \bibinfo {pages} {045037} (\bibinfo {year} {2014})}\BibitemShut
  {NoStop}%
%%CITATION = ARXIV:1403.4350;%%
\bibitem [{\citenamefont {Siringo}(2016)}]{Siringo:2015wtx}%
  \BibitemOpen
  \bibfield  {author} {\bibinfo {author} {\bibfnamefont {F.}~\bibnamefont
  {Siringo}},\ }\href {\doibase 10.1016/j.nuclphysb.2016.04.028} {\bibfield
  {journal} {\bibinfo  {journal} {Nucl. Phys.}\ }\textbf {\bibinfo {volume}
  {B907}},\ \bibinfo {pages} {572} (\bibinfo {year} {2016})}\BibitemShut
  {NoStop}%
%%CITATION = ARXIV:1511.01015;%%
\bibitem [{\citenamefont {Cyrol}\ \emph
  {et~al.}(2018{\natexlab{a}})\citenamefont {Cyrol}, \citenamefont {Pawlowski},
  \citenamefont {Rothkopf},\ and\ \citenamefont {Wink}}]{Cyrol:2018xeq}%
  \BibitemOpen
  \bibfield  {author} {\bibinfo {author} {\bibfnamefont {A.~K.}\ \bibnamefont
  {Cyrol}}, \bibinfo {author} {\bibfnamefont {J.~M.}\ \bibnamefont
  {Pawlowski}}, \bibinfo {author} {\bibfnamefont {A.}~\bibnamefont {Rothkopf}},
  \ and\ \bibinfo {author} {\bibfnamefont {N.}~\bibnamefont {Wink}},\ }\href
  {\doibase 10.21468/SciPostPhys.5.6.065} {\bibfield  {journal} {\bibinfo
  {journal} {SciPost Phys.}\ }\textbf {\bibinfo {volume} {5}},\ \bibinfo
  {pages} {065} (\bibinfo {year} {2018}{\natexlab{a}})}\BibitemShut {NoStop}%
%%CITATION = ARXIV:1804.00945;%%
\bibitem [{\citenamefont {Cucchieri}\ \emph {et~al.}(2009)\citenamefont
  {Cucchieri}, \citenamefont {Mendes},\ and\ \citenamefont
  {Santos}}]{Cucchieri:2009kk}%
  \BibitemOpen
  \bibfield  {author} {\bibinfo {author} {\bibfnamefont {A.}~\bibnamefont
  {Cucchieri}}, \bibinfo {author} {\bibfnamefont {T.}~\bibnamefont {Mendes}}, \
  and\ \bibinfo {author} {\bibfnamefont {E.~M.~S.}\ \bibnamefont {Santos}},\
  }\href {\doibase 10.1103/PhysRevLett.103.141602} {\bibfield  {journal}
  {\bibinfo  {journal} {Phys. Rev. Lett.}\ }\textbf {\bibinfo {volume} {103}},\
  \bibinfo {pages} {141602} (\bibinfo {year} {2009})}\BibitemShut {NoStop}%
%%CITATION = ARXIV:0907.4138;%%
\bibitem [{\citenamefont {Cucchieri}\ \emph {et~al.}(2010)\citenamefont
  {Cucchieri}, \citenamefont {Mendes}, \citenamefont {Nakamura},\ and\
  \citenamefont {Santos}}]{Cucchieri:2011pp}%
  \BibitemOpen
  \bibfield  {author} {\bibinfo {author} {\bibfnamefont {A.}~\bibnamefont
  {Cucchieri}}, \bibinfo {author} {\bibfnamefont {T.}~\bibnamefont {Mendes}},
  \bibinfo {author} {\bibfnamefont {G.~M.}\ \bibnamefont {Nakamura}}, \ and\
  \bibinfo {author} {\bibfnamefont {E.~M.~S.}\ \bibnamefont {Santos}},\ }\href
  {\doibase 10.22323/1.117.0026} {\bibfield  {journal} {\bibinfo  {journal}
  {PoS}\ }\textbf {\bibinfo {volume} {FACESQCD}},\ \bibinfo {pages} {026}
  (\bibinfo {year} {2010})}\BibitemShut {NoStop}%
%%CITATION = ARXIV:1102.5233;%%
\bibitem [{\citenamefont {Bicudo}\ \emph {et~al.}(2015)\citenamefont {Bicudo},
  \citenamefont {Binosi}, \citenamefont {Cardoso}, \citenamefont {Oliveira},\
  and\ \citenamefont {Silva}}]{Bicudo:2015rma}%
  \BibitemOpen
  \bibfield  {author} {\bibinfo {author} {\bibfnamefont {P.}~\bibnamefont
  {Bicudo}}, \bibinfo {author} {\bibfnamefont {D.}~\bibnamefont {Binosi}},
  \bibinfo {author} {\bibfnamefont {N.}~\bibnamefont {Cardoso}}, \bibinfo
  {author} {\bibfnamefont {O.}~\bibnamefont {Oliveira}}, \ and\ \bibinfo
  {author} {\bibfnamefont {P.~J.}\ \bibnamefont {Silva}},\ }\href {\doibase
  10.1103/PhysRevD.92.114514} {\bibfield  {journal} {\bibinfo  {journal} {Phys.
  Rev.}\ }\textbf {\bibinfo {volume} {D92}},\ \bibinfo {pages} {114514}
  (\bibinfo {year} {2015})}\BibitemShut {NoStop}%
%%CITATION = ARXIV:1505.05897;%%
\bibitem [{\citenamefont {Epple}\ \emph {et~al.}(2008)\citenamefont {Epple},
  \citenamefont {Reinhardt}, \citenamefont {Schleifenbaum},\ and\ \citenamefont
  {Szczepaniak}}]{Epple:2007ut}%
  \BibitemOpen
  \bibfield  {author} {\bibinfo {author} {\bibfnamefont {D.}~\bibnamefont
  {Epple}}, \bibinfo {author} {\bibfnamefont {H.}~\bibnamefont {Reinhardt}},
  \bibinfo {author} {\bibfnamefont {W.}~\bibnamefont {Schleifenbaum}}, \ and\
  \bibinfo {author} {\bibfnamefont {A.~P.}\ \bibnamefont {Szczepaniak}},\
  }\href {\doibase 10.1103/PhysRevD.77.085007} {\bibfield  {journal} {\bibinfo
  {journal} {Phys. Rev.}\ }\textbf {\bibinfo {volume} {D77}},\ \bibinfo {pages}
  {085007} (\bibinfo {year} {2008})}\BibitemShut {NoStop}%
%%CITATION = ARXIV:0712.3694;%%
\bibitem [{\citenamefont {Campagnari}\ and\ \citenamefont
  {Reinhardt}(2010)}]{Campagnari:2010wc}%
  \BibitemOpen
  \bibfield  {author} {\bibinfo {author} {\bibfnamefont {D.~R.}\ \bibnamefont
  {Campagnari}}\ and\ \bibinfo {author} {\bibfnamefont {H.}~\bibnamefont
  {Reinhardt}},\ }\href {\doibase 10.1103/PhysRevD.82.105021} {\bibfield
  {journal} {\bibinfo  {journal} {Phys. Rev.}\ }\textbf {\bibinfo {volume}
  {D82}},\ \bibinfo {pages} {105021} (\bibinfo {year} {2010})}\BibitemShut
  {NoStop}%
%%CITATION = ARXIV:1009.4599;%%
\bibitem [{\citenamefont {Aguilar}\ \emph {et~al.}(2017)\citenamefont
  {Aguilar}, \citenamefont {Binosi},\ and\ \citenamefont
  {Papavassiliou}}]{Aguilar:2016ock}%
  \BibitemOpen
  \bibfield  {author} {\bibinfo {author} {\bibfnamefont {A.~C.}\ \bibnamefont
  {Aguilar}}, \bibinfo {author} {\bibfnamefont {D.}~\bibnamefont {Binosi}}, \
  and\ \bibinfo {author} {\bibfnamefont {J.}~\bibnamefont {Papavassiliou}},\
  }\href {\doibase 10.1103/PhysRevD.95.034017} {\bibfield  {journal} {\bibinfo
  {journal} {Phys. Rev.}\ }\textbf {\bibinfo {volume} {D95}},\ \bibinfo {pages}
  {034017} (\bibinfo {year} {2017})}\BibitemShut {NoStop}%
%%CITATION = ARXIV:1611.02096;%%
\bibitem [{\citenamefont {Glazek}\ \emph {et~al.}(2017)\citenamefont {Glazek},
  \citenamefont {G\'omez-Rocha}, \citenamefont {More},\ and\ \citenamefont
  {Serafin}}]{Glazek:2017rwe}%
  \BibitemOpen
  \bibfield  {author} {\bibinfo {author} {\bibfnamefont {S.~D.}\ \bibnamefont
  {Glazek}}, \bibinfo {author} {\bibfnamefont {M.}~\bibnamefont
  {G\'omez-Rocha}}, \bibinfo {author} {\bibfnamefont {J.}~\bibnamefont {More}},
  \ and\ \bibinfo {author} {\bibfnamefont {K.}~\bibnamefont {Serafin}},\ }\href
  {\doibase 10.1016/j.physletb.2017.08.018} {\bibfield  {journal} {\bibinfo
  {journal} {Phys. Lett.}\ }\textbf {\bibinfo {volume} {B773}},\ \bibinfo
  {pages} {172} (\bibinfo {year} {2017})}\BibitemShut {NoStop}%
%%CITATION = ARXIV:1705.07629;%%
\bibitem [{\citenamefont {Bowman}\ \emph {et~al.}(2007)\citenamefont {Bowman},
  \citenamefont {Heller}, \citenamefont {Leinweber}, \citenamefont
  {Parappilly}, \citenamefont {Sternbeck}, \citenamefont {von Smekal},
  \citenamefont {Williams},\ and\ \citenamefont {Zhang}}]{Bowman:2007du}%
  \BibitemOpen
  \bibfield  {author} {\bibinfo {author} {\bibfnamefont {P.~O.}\ \bibnamefont
  {Bowman}}, \bibinfo {author} {\bibfnamefont {U.~M.}\ \bibnamefont {Heller}},
  \bibinfo {author} {\bibfnamefont {D.~B.}\ \bibnamefont {Leinweber}}, \bibinfo
  {author} {\bibfnamefont {M.~B.}\ \bibnamefont {Parappilly}}, \bibinfo
  {author} {\bibfnamefont {A.}~\bibnamefont {Sternbeck}}, \bibinfo {author}
  {\bibfnamefont {L.}~\bibnamefont {von Smekal}}, \bibinfo {author}
  {\bibfnamefont {A.~G.}\ \bibnamefont {Williams}}, \ and\ \bibinfo {author}
  {\bibfnamefont {J.-b.}\ \bibnamefont {Zhang}},\ }\href {\doibase
  10.1103/PhysRevD.76.094505} {\bibfield  {journal} {\bibinfo  {journal} {Phys.
  Rev. D}\ }\textbf {\bibinfo {volume} {76}},\ \bibinfo {pages} {094505}
  (\bibinfo {year} {2007})}\BibitemShut {NoStop}%
\bibitem [{\citenamefont {Kamleh}\ \emph {et~al.}(2007)\citenamefont {Kamleh},
  \citenamefont {Bowman}, \citenamefont {Leinweber}, \citenamefont {Williams},\
  and\ \citenamefont {Zhang}}]{Kamleh:2007ud}%
  \BibitemOpen
  \bibfield  {author} {\bibinfo {author} {\bibfnamefont {W.}~\bibnamefont
  {Kamleh}}, \bibinfo {author} {\bibfnamefont {P.~O.}\ \bibnamefont {Bowman}},
  \bibinfo {author} {\bibfnamefont {D.~B.}\ \bibnamefont {Leinweber}}, \bibinfo
  {author} {\bibfnamefont {A.~G.}\ \bibnamefont {Williams}}, \ and\ \bibinfo
  {author} {\bibfnamefont {J.}~\bibnamefont {Zhang}},\ }\href {\doibase
  10.1103/PhysRevD.76.094501} {\bibfield  {journal} {\bibinfo  {journal} {Phys.
  Rev.}\ }\textbf {\bibinfo {volume} {D76}},\ \bibinfo {pages} {094501}
  (\bibinfo {year} {2007})}\BibitemShut {NoStop}%
%%CITATION = ARXIV:0705.4129;%%
\bibitem [{\citenamefont {Ayala}\ \emph {et~al.}(2012)\citenamefont {Ayala},
  \citenamefont {Bashir}, \citenamefont {Binosi}, \citenamefont
  {Cristoforetti},\ and\ \citenamefont {Rodriguez-Quintero}}]{Ayala:2012pb}%
  \BibitemOpen
  \bibfield  {author} {\bibinfo {author} {\bibfnamefont {A.}~\bibnamefont
  {Ayala}}, \bibinfo {author} {\bibfnamefont {A.}~\bibnamefont {Bashir}},
  \bibinfo {author} {\bibfnamefont {D.}~\bibnamefont {Binosi}}, \bibinfo
  {author} {\bibfnamefont {M.}~\bibnamefont {Cristoforetti}}, \ and\ \bibinfo
  {author} {\bibfnamefont {J.}~\bibnamefont {Rodriguez-Quintero}},\ }\href
  {\doibase 10.1103/PhysRevD.86.074512} {\bibfield  {journal} {\bibinfo
  {journal} {Phys. Rev.}\ }\textbf {\bibinfo {volume} {D86}},\ \bibinfo {pages}
  {074512} (\bibinfo {year} {2012})}\BibitemShut {NoStop}%
%%CITATION = ARXIV:1208.0795;%%
\bibitem [{\citenamefont {Aguilar}\ \emph
  {et~al.}(2013{\natexlab{a}})\citenamefont {Aguilar}, \citenamefont {Binosi},\
  and\ \citenamefont {Papavassiliou}}]{Aguilar:2013hoa}%
  \BibitemOpen
  \bibfield  {author} {\bibinfo {author} {\bibfnamefont {A.~C.}\ \bibnamefont
  {Aguilar}}, \bibinfo {author} {\bibfnamefont {D.}~\bibnamefont {Binosi}}, \
  and\ \bibinfo {author} {\bibfnamefont {J.}~\bibnamefont {Papavassiliou}},\
  }\href {\doibase 10.1103/PhysRevD.88.074010} {\bibfield  {journal} {\bibinfo
  {journal} {Phys. Rev.}\ }\textbf {\bibinfo {volume} {D88}},\ \bibinfo {pages}
  {074010} (\bibinfo {year} {2013}{\natexlab{a}})}\BibitemShut {NoStop}%
%%CITATION = ARXIV:1304.5936;%%
\bibitem [{\citenamefont {Aguilar}\ \emph
  {et~al.}(2020{\natexlab{a}})\citenamefont {Aguilar}, \citenamefont {De~Soto},
  \citenamefont {Ferreira}, \citenamefont {Papavassiliou}, \citenamefont
  {Rodríguez-Quintero},\ and\ \citenamefont
  {Zafeiropoulos}}]{Aguilar:2019uob}%
  \BibitemOpen
  \bibfield  {author} {\bibinfo {author} {\bibfnamefont {A.~C.}\ \bibnamefont
  {Aguilar}}, \bibinfo {author} {\bibfnamefont {F.}~\bibnamefont {De~Soto}},
  \bibinfo {author} {\bibfnamefont {M.~N.}\ \bibnamefont {Ferreira}}, \bibinfo
  {author} {\bibfnamefont {J.}~\bibnamefont {Papavassiliou}}, \bibinfo {author}
  {\bibfnamefont {J.}~\bibnamefont {Rodríguez-Quintero}}, \ and\ \bibinfo
  {author} {\bibfnamefont {S.}~\bibnamefont {Zafeiropoulos}},\ }\href {\doibase
  10.1140/epjc/s10052-020-7741-0} {\bibfield  {journal} {\bibinfo  {journal}
  {Eur. Phys. J.}\ }\textbf {\bibinfo {volume} {C80}},\ \bibinfo {pages} {154}
  (\bibinfo {year} {2020}{\natexlab{a}})}\BibitemShut {NoStop}%
%%CITATION = ARXIV:1912.12086;%%
\bibitem [{\citenamefont {Schwinger}(1962{\natexlab{a}})}]{Schwinger:1962tn}%
  \BibitemOpen
  \bibfield  {author} {\bibinfo {author} {\bibfnamefont {J.~S.}\ \bibnamefont
  {Schwinger}},\ }\href {\doibase 10.1103/PhysRev.125.397} {\bibfield
  {journal} {\bibinfo  {journal} {Phys. Rev.}\ }\textbf {\bibinfo {volume}
  {125}},\ \bibinfo {pages} {397} (\bibinfo {year}
  {1962}{\natexlab{a}})}\BibitemShut {NoStop}%
\bibitem [{\citenamefont {Schwinger}(1962{\natexlab{b}})}]{Schwinger:1962tp}%
  \BibitemOpen
  \bibfield  {author} {\bibinfo {author} {\bibfnamefont {J.~S.}\ \bibnamefont
  {Schwinger}},\ }\href {\doibase 10.1103/PhysRev.128.2425} {\bibfield
  {journal} {\bibinfo  {journal} {Phys. Rev.}\ }\textbf {\bibinfo {volume}
  {128}},\ \bibinfo {pages} {2425} (\bibinfo {year}
  {1962}{\natexlab{b}})}\BibitemShut {NoStop}%
\bibitem [{\citenamefont {Roberts}\ and\ \citenamefont
  {Williams}(1994)}]{Roberts:1994dr}%
  \BibitemOpen
  \bibfield  {author} {\bibinfo {author} {\bibfnamefont {C.~D.}\ \bibnamefont
  {Roberts}}\ and\ \bibinfo {author} {\bibfnamefont {A.~G.}\ \bibnamefont
  {Williams}},\ }\href {\doibase 10.1016/0146-6410(94)90049-3} {\bibfield
  {journal} {\bibinfo  {journal} {Prog. Part. Nucl. Phys.}\ }\textbf {\bibinfo
  {volume} {33}},\ \bibinfo {pages} {477} (\bibinfo {year} {1994})}\BibitemShut
  {NoStop}%
\bibitem [{\citenamefont {Alkofer}\ and\ \citenamefont {von
  Smekal}(2001)}]{Alkofer:2000wg}%
  \BibitemOpen
  \bibfield  {author} {\bibinfo {author} {\bibfnamefont {R.}~\bibnamefont
  {Alkofer}}\ and\ \bibinfo {author} {\bibfnamefont {L.}~\bibnamefont {von
  Smekal}},\ }\href {\doibase 10.1016/S0370-1573(01)00010-2} {\bibfield
  {journal} {\bibinfo  {journal} {Phys. Rept.}\ }\textbf {\bibinfo {volume}
  {353}},\ \bibinfo {pages} {281} (\bibinfo {year} {2001})}\BibitemShut
  {NoStop}%
\bibitem [{\citenamefont {Fischer}(2006)}]{Fischer:2006ub}%
  \BibitemOpen
  \bibfield  {author} {\bibinfo {author} {\bibfnamefont {C.~S.}\ \bibnamefont
  {Fischer}},\ }\href {\doibase 10.1088/0954-3899/32/8/R02} {\bibfield
  {journal} {\bibinfo  {journal} {J. Phys. G}\ }\textbf {\bibinfo {volume}
  {32}},\ \bibinfo {pages} {R253} (\bibinfo {year} {2006})}\BibitemShut
  {NoStop}%
\bibitem [{\citenamefont {Roberts}(2008)}]{Roberts:2007ji}%
  \BibitemOpen
  \bibfield  {author} {\bibinfo {author} {\bibfnamefont {C.~D.}\ \bibnamefont
  {Roberts}},\ }\href {\doibase 10.1016/j.ppnp.2007.12.034} {\bibfield
  {journal} {\bibinfo  {journal} {Prog. Part. Nucl. Phys.}\ }\textbf {\bibinfo
  {volume} {61}},\ \bibinfo {pages} {50} (\bibinfo {year} {2008})}\BibitemShut
  {NoStop}%
\bibitem [{\citenamefont {Binosi}\ and\ \citenamefont
  {Papavassiliou}(2009)}]{Binosi:2009qm}%
  \BibitemOpen
  \bibfield  {author} {\bibinfo {author} {\bibfnamefont {D.}~\bibnamefont
  {Binosi}}\ and\ \bibinfo {author} {\bibfnamefont {J.}~\bibnamefont
  {Papavassiliou}},\ }\href {\doibase 10.1016/j.physrep.2009.05.001} {\bibfield
   {journal} {\bibinfo  {journal} {Phys. Rept.}\ }\textbf {\bibinfo {volume}
  {479}},\ \bibinfo {pages} {1} (\bibinfo {year} {2009})}\BibitemShut {NoStop}%
%%CITATION = ARXIV:0909.2536;%%
\bibitem [{\citenamefont {Huber}(2020)}]{Huber:2018ned}%
  \BibitemOpen
  \bibfield  {author} {\bibinfo {author} {\bibfnamefont {M.~Q.}\ \bibnamefont
  {Huber}},\ }\href {\doibase 10.1016/j.physrep.2020.04.004} {\bibfield
  {journal} {\bibinfo  {journal} {Phys. Rept.}\ }\textbf {\bibinfo {volume}
  {879}},\ \bibinfo {pages} {1} (\bibinfo {year} {2020})}\BibitemShut {NoStop}%
\bibitem [{\citenamefont {Pawlowski}\ \emph {et~al.}(2004)\citenamefont
  {Pawlowski}, \citenamefont {Litim}, \citenamefont {Nedelko},\ and\
  \citenamefont {von Smekal}}]{Pawlowski:2003hq}%
  \BibitemOpen
  \bibfield  {author} {\bibinfo {author} {\bibfnamefont {J.~M.}\ \bibnamefont
  {Pawlowski}}, \bibinfo {author} {\bibfnamefont {D.~F.}\ \bibnamefont
  {Litim}}, \bibinfo {author} {\bibfnamefont {S.}~\bibnamefont {Nedelko}}, \
  and\ \bibinfo {author} {\bibfnamefont {L.}~\bibnamefont {von Smekal}},\
  }\href {\doibase 10.1103/PhysRevLett.93.152002} {\bibfield  {journal}
  {\bibinfo  {journal} {Phys. Rev. Lett.}\ }\textbf {\bibinfo {volume} {93}},\
  \bibinfo {pages} {152002} (\bibinfo {year} {2004})}\BibitemShut {NoStop}%
%%CITATION = HEP-TH/0312324;%%
\bibitem [{\citenamefont {Pawlowski}(2007)}]{Pawlowski:2005xe}%
  \BibitemOpen
  \bibfield  {author} {\bibinfo {author} {\bibfnamefont {J.~M.}\ \bibnamefont
  {Pawlowski}},\ }\href {\doibase 10.1016/j.aop.2007.01.007} {\bibfield
  {journal} {\bibinfo  {journal} {Annals Phys.}\ }\textbf {\bibinfo {volume}
  {322}},\ \bibinfo {pages} {2831} (\bibinfo {year} {2007})}\BibitemShut
  {NoStop}%
%%CITATION = HEP-TH/0512261;%%
\bibitem [{\citenamefont {Cyrol}\ \emph
  {et~al.}(2018{\natexlab{b}})\citenamefont {Cyrol}, \citenamefont {Mitter},
  \citenamefont {Pawlowski},\ and\ \citenamefont {Strodthoff}}]{Cyrol:2017ewj}%
  \BibitemOpen
  \bibfield  {author} {\bibinfo {author} {\bibfnamefont {A.~K.}\ \bibnamefont
  {Cyrol}}, \bibinfo {author} {\bibfnamefont {M.}~\bibnamefont {Mitter}},
  \bibinfo {author} {\bibfnamefont {J.~M.}\ \bibnamefont {Pawlowski}}, \ and\
  \bibinfo {author} {\bibfnamefont {N.}~\bibnamefont {Strodthoff}},\ }\href
  {\doibase 10.1103/PhysRevD.97.054006} {\bibfield  {journal} {\bibinfo
  {journal} {Phys. Rev.}\ }\textbf {\bibinfo {volume} {D97}},\ \bibinfo {pages}
  {054006} (\bibinfo {year} {2018}{\natexlab{b}})}\BibitemShut {NoStop}%
%%CITATION = ARXIV:1706.06326;%%
\bibitem [{\citenamefont {Corell}\ \emph {et~al.}(2018)\citenamefont {Corell},
  \citenamefont {Cyrol}, \citenamefont {Mitter}, \citenamefont {Pawlowski},\
  and\ \citenamefont {Strodthoff}}]{Corell:2018yil}%
  \BibitemOpen
  \bibfield  {author} {\bibinfo {author} {\bibfnamefont {L.}~\bibnamefont
  {Corell}}, \bibinfo {author} {\bibfnamefont {A.~K.}\ \bibnamefont {Cyrol}},
  \bibinfo {author} {\bibfnamefont {M.}~\bibnamefont {Mitter}}, \bibinfo
  {author} {\bibfnamefont {J.~M.}\ \bibnamefont {Pawlowski}}, \ and\ \bibinfo
  {author} {\bibfnamefont {N.}~\bibnamefont {Strodthoff}},\ }\href {\doibase
  10.21468/SciPostPhys.5.6.066} {\bibfield  {journal} {\bibinfo  {journal}
  {SciPost Phys.}\ }\textbf {\bibinfo {volume} {5}},\ \bibinfo {pages} {066}
  (\bibinfo {year} {2018})}\BibitemShut {NoStop}%
%%CITATION = ARXIV:1803.10092;%%
\bibitem [{\citenamefont {Blaizot}\ \emph {et~al.}(2021)\citenamefont
  {Blaizot}, \citenamefont {Pawlowski},\ and\ \citenamefont
  {Reinosa}}]{Blaizot:2021ikl}%
  \BibitemOpen
  \bibfield  {author} {\bibinfo {author} {\bibfnamefont {J.-P.}\ \bibnamefont
  {Blaizot}}, \bibinfo {author} {\bibfnamefont {J.~M.}\ \bibnamefont
  {Pawlowski}}, \ and\ \bibinfo {author} {\bibfnamefont {U.}~\bibnamefont
  {Reinosa}},\ }\href {\doibase 10.1016/j.aop.2021.168549} {\bibfield
  {journal} {\bibinfo  {journal} {Annals Phys.}\ }\textbf {\bibinfo {volume}
  {431}},\ \bibinfo {pages} {168549} (\bibinfo {year} {2021})}\BibitemShut
  {NoStop}%
\bibitem [{\citenamefont {Aguilar}\ \emph {et~al.}(2012)\citenamefont
  {Aguilar}, \citenamefont {Ibanez}, \citenamefont {Mathieu},\ and\
  \citenamefont {Papavassiliou}}]{Aguilar:2011xe}%
  \BibitemOpen
  \bibfield  {author} {\bibinfo {author} {\bibfnamefont {A.~C.}\ \bibnamefont
  {Aguilar}}, \bibinfo {author} {\bibfnamefont {D.}~\bibnamefont {Ibanez}},
  \bibinfo {author} {\bibfnamefont {V.}~\bibnamefont {Mathieu}}, \ and\
  \bibinfo {author} {\bibfnamefont {J.}~\bibnamefont {Papavassiliou}},\ }\href
  {\doibase 10.1103/PhysRevD.85.014018} {\bibfield  {journal} {\bibinfo
  {journal} {Phys. Rev.}\ }\textbf {\bibinfo {volume} {D85}},\ \bibinfo {pages}
  {014018} (\bibinfo {year} {2012})}\BibitemShut {NoStop}%
%%CITATION = ARXIV:1110.2633;%%
\bibitem [{\citenamefont {Iba{\~n}ez}\ and\ \citenamefont
  {Papavassiliou}(2013)}]{Ibanez:2012zk}%
  \BibitemOpen
  \bibfield  {author} {\bibinfo {author} {\bibfnamefont {D.}~\bibnamefont
  {Iba{\~n}ez}}\ and\ \bibinfo {author} {\bibfnamefont {J.}~\bibnamefont
  {Papavassiliou}},\ }\href {\doibase 10.1103/PhysRevD.87.034008} {\bibfield
  {journal} {\bibinfo  {journal} {Phys. Rev.}\ }\textbf {\bibinfo {volume}
  {D87}},\ \bibinfo {pages} {034008} (\bibinfo {year} {2013})}\BibitemShut
  {NoStop}%
%%CITATION = ARXIV:1211.5314;%%
\bibitem [{\citenamefont {Binosi}\ and\ \citenamefont
  {Papavassiliou}(2018)}]{Binosi:2017rwj}%
  \BibitemOpen
  \bibfield  {author} {\bibinfo {author} {\bibfnamefont {D.}~\bibnamefont
  {Binosi}}\ and\ \bibinfo {author} {\bibfnamefont {J.}~\bibnamefont
  {Papavassiliou}},\ }\href {\doibase 10.1103/PhysRevD.97.054029} {\bibfield
  {journal} {\bibinfo  {journal} {Phys. Rev.}\ }\textbf {\bibinfo {volume}
  {D97}},\ \bibinfo {pages} {054029} (\bibinfo {year} {2018})}\BibitemShut
  {NoStop}%
%%CITATION = ARXIV:1709.09964;%%
\bibitem [{\citenamefont {Aguilar}\ \emph {et~al.}(2018)\citenamefont
  {Aguilar}, \citenamefont {Binosi}, \citenamefont {Figueiredo},\ and\
  \citenamefont {Papavassiliou}}]{Aguilar:2017dco}%
  \BibitemOpen
  \bibfield  {author} {\bibinfo {author} {\bibfnamefont {A.~C.}\ \bibnamefont
  {Aguilar}}, \bibinfo {author} {\bibfnamefont {D.}~\bibnamefont {Binosi}},
  \bibinfo {author} {\bibfnamefont {C.~T.}\ \bibnamefont {Figueiredo}}, \ and\
  \bibinfo {author} {\bibfnamefont {J.}~\bibnamefont {Papavassiliou}},\ }\href
  {\doibase 10.1140/epjc/s10052-018-5679-2} {\bibfield  {journal} {\bibinfo
  {journal} {Eur. Phys. J.}\ }\textbf {\bibinfo {volume} {C78}},\ \bibinfo
  {pages} {181} (\bibinfo {year} {2018})}\BibitemShut {NoStop}%
%%CITATION = ARXIV:1712.06926;%%
\bibitem [{\citenamefont {Eichmann}\ \emph {et~al.}(2021)\citenamefont
  {Eichmann}, \citenamefont {Pawlowski},\ and\ \citenamefont
  {Silva}}]{Eichmann:2021zuv}%
  \BibitemOpen
  \bibfield  {author} {\bibinfo {author} {\bibfnamefont {G.}~\bibnamefont
  {Eichmann}}, \bibinfo {author} {\bibfnamefont {J.~M.}\ \bibnamefont
  {Pawlowski}}, \ and\ \bibinfo {author} {\bibfnamefont {J.~a.~M.}\
  \bibnamefont {Silva}},\ }\href {\doibase 10.1103/PhysRevD.104.114016}
  {\bibfield  {journal} {\bibinfo  {journal} {Phys. Rev. D}\ }\textbf {\bibinfo
  {volume} {104}},\ \bibinfo {pages} {114016} (\bibinfo {year}
  {2021})}\BibitemShut {NoStop}%
\bibitem [{\citenamefont {Cornwall}\ \emph {et~al.}(2010)\citenamefont
  {Cornwall}, \citenamefont {Papavassiliou},\ and\ \citenamefont
  {Binosi}}]{Cornwall:2010upa}%
  \BibitemOpen
  \bibfield  {author} {\bibinfo {author} {\bibfnamefont {J.~M.}\ \bibnamefont
  {Cornwall}}, \bibinfo {author} {\bibfnamefont {J.}~\bibnamefont
  {Papavassiliou}}, \ and\ \bibinfo {author} {\bibfnamefont {D.}~\bibnamefont
  {Binosi}},\ }\href@noop {} {\emph {\bibinfo {title} {{The Pinch Technique and
  its Applications to Non-Abelian Gauge Theories}}}},\ Vol.~\bibinfo {volume}
  {31}\ (\bibinfo  {publisher} {Cambridge University Press},\ \bibinfo {year}
  {2010})\BibitemShut {NoStop}%
%%CITATION = CMPCE,31,;%%
\bibitem [{\citenamefont {Cucchieri}\ \emph {et~al.}(2006)\citenamefont
  {Cucchieri}, \citenamefont {Maas},\ and\ \citenamefont
  {Mendes}}]{Cucchieri:2006tf}%
  \BibitemOpen
  \bibfield  {author} {\bibinfo {author} {\bibfnamefont {A.}~\bibnamefont
  {Cucchieri}}, \bibinfo {author} {\bibfnamefont {A.}~\bibnamefont {Maas}}, \
  and\ \bibinfo {author} {\bibfnamefont {T.}~\bibnamefont {Mendes}},\ }\href
  {\doibase 10.1103/PhysRevD.74.014503} {\bibfield  {journal} {\bibinfo
  {journal} {Phys. Rev.}\ }\textbf {\bibinfo {volume} {D74}},\ \bibinfo {pages}
  {014503} (\bibinfo {year} {2006})}\BibitemShut {NoStop}%
%%CITATION = HEP-LAT/0605011;%%
\bibitem [{\citenamefont {Cucchieri}\ \emph {et~al.}(2008)\citenamefont
  {Cucchieri}, \citenamefont {Maas},\ and\ \citenamefont
  {Mendes}}]{Cucchieri:2008qm}%
  \BibitemOpen
  \bibfield  {author} {\bibinfo {author} {\bibfnamefont {A.}~\bibnamefont
  {Cucchieri}}, \bibinfo {author} {\bibfnamefont {A.}~\bibnamefont {Maas}}, \
  and\ \bibinfo {author} {\bibfnamefont {T.}~\bibnamefont {Mendes}},\ }\href
  {\doibase 10.1103/PhysRevD.77.094510} {\bibfield  {journal} {\bibinfo
  {journal} {Phys. Rev.}\ }\textbf {\bibinfo {volume} {D77}},\ \bibinfo {pages}
  {094510} (\bibinfo {year} {2008})}\BibitemShut {NoStop}%
%%CITATION = ARXIV:0803.1798;%%
\bibitem [{\citenamefont {Athenodorou}\ \emph {et~al.}(2016)\citenamefont
  {Athenodorou}, \citenamefont {Binosi}, \citenamefont {Boucaud}, \citenamefont
  {De~Soto}, \citenamefont {Papavassiliou}, \citenamefont
  {Rodriguez-Quintero},\ and\ \citenamefont
  {Zafeiropoulos}}]{Athenodorou:2016oyh}%
  \BibitemOpen
  \bibfield  {author} {\bibinfo {author} {\bibfnamefont {A.}~\bibnamefont
  {Athenodorou}}, \bibinfo {author} {\bibfnamefont {D.}~\bibnamefont {Binosi}},
  \bibinfo {author} {\bibfnamefont {P.}~\bibnamefont {Boucaud}}, \bibinfo
  {author} {\bibfnamefont {F.}~\bibnamefont {De~Soto}}, \bibinfo {author}
  {\bibfnamefont {J.}~\bibnamefont {Papavassiliou}}, \bibinfo {author}
  {\bibfnamefont {J.}~\bibnamefont {Rodriguez-Quintero}}, \ and\ \bibinfo
  {author} {\bibfnamefont {S.}~\bibnamefont {Zafeiropoulos}},\ }\href {\doibase
  10.1016/j.physletb.2016.08.065} {\bibfield  {journal} {\bibinfo  {journal}
  {Phys. Lett.}\ }\textbf {\bibinfo {volume} {B761}},\ \bibinfo {pages} {444}
  (\bibinfo {year} {2016})}\BibitemShut {NoStop}%
%%CITATION = ARXIV:1607.01278;%%
\bibitem [{\citenamefont {Duarte}\ \emph {et~al.}(2016)\citenamefont {Duarte},
  \citenamefont {Oliveira},\ and\ \citenamefont {Silva}}]{Duarte:2016ieu}%
  \BibitemOpen
  \bibfield  {author} {\bibinfo {author} {\bibfnamefont {A.~G.}\ \bibnamefont
  {Duarte}}, \bibinfo {author} {\bibfnamefont {O.}~\bibnamefont {Oliveira}}, \
  and\ \bibinfo {author} {\bibfnamefont {P.~J.}\ \bibnamefont {Silva}},\ }\href
  {\doibase 10.1103/PhysRevD.94.074502} {\bibfield  {journal} {\bibinfo
  {journal} {Phys. Rev.}\ }\textbf {\bibinfo {volume} {D94}},\ \bibinfo {pages}
  {074502} (\bibinfo {year} {2016})}\BibitemShut {NoStop}%
%%CITATION = ARXIV:1607.03831;%%
\bibitem [{\citenamefont {Boucaud}\ \emph {et~al.}(2018)\citenamefont
  {Boucaud}, \citenamefont {De~Soto}, \citenamefont {Raya}, \citenamefont
  {Rodr\'{\i}guez-Quintero},\ and\ \citenamefont
  {Zafeiropoulos}}]{Boucaud:2018xup}%
  \BibitemOpen
  \bibfield  {author} {\bibinfo {author} {\bibfnamefont {P.}~\bibnamefont
  {Boucaud}}, \bibinfo {author} {\bibfnamefont {F.}~\bibnamefont {De~Soto}},
  \bibinfo {author} {\bibfnamefont {K.}~\bibnamefont {Raya}}, \bibinfo {author}
  {\bibfnamefont {J.}~\bibnamefont {Rodr\'{\i}guez-Quintero}}, \ and\ \bibinfo
  {author} {\bibfnamefont {S.}~\bibnamefont {Zafeiropoulos}},\ }\href {\doibase
  10.1103/PhysRevD.98.114515} {\bibfield  {journal} {\bibinfo  {journal} {Phys.
  Rev.}\ }\textbf {\bibinfo {volume} {D98}},\ \bibinfo {pages} {114515}
  (\bibinfo {year} {2018})}\BibitemShut {NoStop}%
%%CITATION = ARXIV:1809.05776;%%
\bibitem [{\citenamefont {Taylor}(1971)}]{Taylor:1971ff}%
  \BibitemOpen
  \bibfield  {author} {\bibinfo {author} {\bibfnamefont {J.}~\bibnamefont
  {Taylor}},\ }\href {\doibase 10.1016/0550-3213(71)90297-5} {\bibfield
  {journal} {\bibinfo  {journal} {Nucl. Phys. B}\ }\textbf {\bibinfo {volume}
  {33}},\ \bibinfo {pages} {436} (\bibinfo {year} {1971})}\BibitemShut
  {NoStop}%
\bibitem [{\citenamefont {Slavnov}(1972)}]{Slavnov:1972fg}%
  \BibitemOpen
  \bibfield  {author} {\bibinfo {author} {\bibfnamefont {A.}~\bibnamefont
  {Slavnov}},\ }\href {\doibase 10.1007/BF01090719} {\bibfield  {journal}
  {\bibinfo  {journal} {Theor. Math. Phys.}\ }\textbf {\bibinfo {volume}
  {10}},\ \bibinfo {pages} {99} (\bibinfo {year} {1972})}\BibitemShut {NoStop}%
\bibitem [{\citenamefont {Eichten}\ and\ \citenamefont
  {Feinberg}(1974)}]{Eichten:1974et}%
  \BibitemOpen
  \bibfield  {author} {\bibinfo {author} {\bibfnamefont {E.}~\bibnamefont
  {Eichten}}\ and\ \bibinfo {author} {\bibfnamefont {F.}~\bibnamefont
  {Feinberg}},\ }\href {\doibase 10.1103/PhysRevD.10.3254} {\bibfield
  {journal} {\bibinfo  {journal} {Phys. Rev. D}\ }\textbf {\bibinfo {volume}
  {10}},\ \bibinfo {pages} {3254} (\bibinfo {year} {1974})}\BibitemShut
  {NoStop}%
\bibitem [{\citenamefont {Poggio}\ \emph {et~al.}(1975)\citenamefont {Poggio},
  \citenamefont {Tomboulis},\ and\ \citenamefont {Tye}}]{Poggio:1974qs}%
  \BibitemOpen
  \bibfield  {author} {\bibinfo {author} {\bibfnamefont {E.~C.}\ \bibnamefont
  {Poggio}}, \bibinfo {author} {\bibfnamefont {E.}~\bibnamefont {Tomboulis}}, \
  and\ \bibinfo {author} {\bibfnamefont {S.~H.~H.}\ \bibnamefont {Tye}},\
  }\href {\doibase 10.1103/PhysRevD.11.2839} {\bibfield  {journal} {\bibinfo
  {journal} {Phys. Rev.}\ }\textbf {\bibinfo {volume} {D11}},\ \bibinfo {pages}
  {2839} (\bibinfo {year} {1975})}\BibitemShut {NoStop}%
%%CITATION = PHRVA,D11,2839;%%
\bibitem [{\citenamefont {Smit}(1974)}]{Smit:1974je}%
  \BibitemOpen
  \bibfield  {author} {\bibinfo {author} {\bibfnamefont {J.}~\bibnamefont
  {Smit}},\ }\href {\doibase 10.1103/PhysRevD.10.2473} {\bibfield  {journal}
  {\bibinfo  {journal} {Phys. Rev. D}\ }\textbf {\bibinfo {volume} {10}},\
  \bibinfo {pages} {2473} (\bibinfo {year} {1974})}\BibitemShut {NoStop}%
\bibitem [{\citenamefont {Papavassiliou}(1990)}]{Papavassiliou:1989zd}%
  \BibitemOpen
  \bibfield  {author} {\bibinfo {author} {\bibfnamefont {J.}~\bibnamefont
  {Papavassiliou}},\ }\href {\doibase 10.1103/PhysRevD.41.3179} {\bibfield
  {journal} {\bibinfo  {journal} {Phys. Rev. D}\ }\textbf {\bibinfo {volume}
  {41}},\ \bibinfo {pages} {3179} (\bibinfo {year} {1990})}\BibitemShut
  {NoStop}%
\bibitem [{\citenamefont {Aguilar}\ \emph {et~al.}(2022)\citenamefont
  {Aguilar}, \citenamefont {Ferreira},\ and\ \citenamefont
  {Papavassiliou}}]{Aguilar:2021uwa}%
  \BibitemOpen
  \bibfield  {author} {\bibinfo {author} {\bibfnamefont {A.~C.}\ \bibnamefont
  {Aguilar}}, \bibinfo {author} {\bibfnamefont {M.~N.}\ \bibnamefont
  {Ferreira}}, \ and\ \bibinfo {author} {\bibfnamefont {J.}~\bibnamefont
  {Papavassiliou}},\ }\href {\doibase 10.1103/PhysRevD.105.014030} {\bibfield
  {journal} {\bibinfo  {journal} {Phys. Rev. D}\ }\textbf {\bibinfo {volume}
  {105}},\ \bibinfo {pages} {014030} (\bibinfo {year} {2022})}\BibitemShut
  {NoStop}%
\bibitem [{\citenamefont {Aguilar}\ and\ \citenamefont
  {Papavassiliou}(2006)}]{Aguilar:2006gr}%
  \BibitemOpen
  \bibfield  {author} {\bibinfo {author} {\bibfnamefont {A.~C.}\ \bibnamefont
  {Aguilar}}\ and\ \bibinfo {author} {\bibfnamefont {J.}~\bibnamefont
  {Papavassiliou}},\ }\href {\doibase 10.1088/1126-6708/2006/12/012} {\bibfield
   {journal} {\bibinfo  {journal} {J. High Energy Phys.}\ }\textbf {\bibinfo
  {volume} {12}},\ \bibinfo {pages} {012} (\bibinfo {year} {2006})}\BibitemShut
  {NoStop}%
\bibitem [{\citenamefont {Kondo}(2010)}]{Kondo:2009gc}%
  \BibitemOpen
  \bibfield  {author} {\bibinfo {author} {\bibfnamefont {K.-I.}\ \bibnamefont
  {Kondo}},\ }\href {\doibase 10.1143/PTP.122.1455} {\bibfield  {journal}
  {\bibinfo  {journal} {Prog. Theor. Phys.}\ }\textbf {\bibinfo {volume}
  {122}},\ \bibinfo {pages} {1455} (\bibinfo {year} {2010})}\BibitemShut
  {NoStop}%
\bibitem [{\citenamefont {Dudal}\ \emph {et~al.}(2012)\citenamefont {Dudal},
  \citenamefont {Oliveira},\ and\ \citenamefont
  {Rodriguez-Quintero}}]{Dudal:2012zx}%
  \BibitemOpen
  \bibfield  {author} {\bibinfo {author} {\bibfnamefont {D.}~\bibnamefont
  {Dudal}}, \bibinfo {author} {\bibfnamefont {O.}~\bibnamefont {Oliveira}}, \
  and\ \bibinfo {author} {\bibfnamefont {J.}~\bibnamefont
  {Rodriguez-Quintero}},\ }\href {\doibase 10.1103/PhysRevD.86.105005,
  10.1103/PhysRevD.86.109902} {\bibfield  {journal} {\bibinfo  {journal} {Phys.
  Rev.}\ }\textbf {\bibinfo {volume} {D86}},\ \bibinfo {pages} {105005}
  (\bibinfo {year} {2012})}\BibitemShut {NoStop}%
%%CITATION = ARXIV:1207.5118;%%
\bibitem [{\citenamefont {Aguilar}\ \emph
  {et~al.}(2013{\natexlab{b}})\citenamefont {Aguilar}, \citenamefont
  {Iba{\~n}ez},\ and\ \citenamefont {Papavassiliou}}]{Aguilar:2013xqa}%
  \BibitemOpen
  \bibfield  {author} {\bibinfo {author} {\bibfnamefont {A.~C.}\ \bibnamefont
  {Aguilar}}, \bibinfo {author} {\bibfnamefont {D.}~\bibnamefont {Iba{\~n}ez}},
  \ and\ \bibinfo {author} {\bibfnamefont {J.}~\bibnamefont {Papavassiliou}},\
  }\href {\doibase 10.1103/PhysRevD.87.114020} {\bibfield  {journal} {\bibinfo
  {journal} {Phys. Rev.}\ }\textbf {\bibinfo {volume} {D87}},\ \bibinfo {pages}
  {114020} (\bibinfo {year} {2013}{\natexlab{b}})}\BibitemShut {NoStop}%
%%CITATION = ARXIV:1303.3609;%%
\bibitem [{\citenamefont {Cyrol}\ \emph {et~al.}(2016)\citenamefont {Cyrol},
  \citenamefont {Fister}, \citenamefont {Mitter}, \citenamefont {Pawlowski},\
  and\ \citenamefont {Strodthoff}}]{Cyrol:2016tym}%
  \BibitemOpen
  \bibfield  {author} {\bibinfo {author} {\bibfnamefont {A.~K.}\ \bibnamefont
  {Cyrol}}, \bibinfo {author} {\bibfnamefont {L.}~\bibnamefont {Fister}},
  \bibinfo {author} {\bibfnamefont {M.}~\bibnamefont {Mitter}}, \bibinfo
  {author} {\bibfnamefont {J.~M.}\ \bibnamefont {Pawlowski}}, \ and\ \bibinfo
  {author} {\bibfnamefont {N.}~\bibnamefont {Strodthoff}},\ }\href {\doibase
  10.1103/PhysRevD.94.054005} {\bibfield  {journal} {\bibinfo  {journal} {Phys.
  Rev.}\ }\textbf {\bibinfo {volume} {D94}},\ \bibinfo {pages} {054005}
  (\bibinfo {year} {2016})}\BibitemShut {NoStop}%
%%CITATION = ARXIV:1605.01856;%%
\bibitem [{\citenamefont {Aguilar}\ \emph
  {et~al.}(2021{\natexlab{a}})\citenamefont {Aguilar}, \citenamefont
  {Ambr\'osio}, \citenamefont {De~Soto}, \citenamefont {Ferreira},
  \citenamefont {Oliveira}, \citenamefont {Papavassiliou},\ and\ \citenamefont
  {Rodr\'\i{}guez-Quintero}}]{Aguilar:2021okw}%
  \BibitemOpen
  \bibfield  {author} {\bibinfo {author} {\bibfnamefont {A.~C.}\ \bibnamefont
  {Aguilar}}, \bibinfo {author} {\bibfnamefont {C.~O.}\ \bibnamefont
  {Ambr\'osio}}, \bibinfo {author} {\bibfnamefont {F.}~\bibnamefont {De~Soto}},
  \bibinfo {author} {\bibfnamefont {M.~N.}\ \bibnamefont {Ferreira}}, \bibinfo
  {author} {\bibfnamefont {B.~M.}\ \bibnamefont {Oliveira}}, \bibinfo {author}
  {\bibfnamefont {J.}~\bibnamefont {Papavassiliou}}, \ and\ \bibinfo {author}
  {\bibfnamefont {J.}~\bibnamefont {Rodr\'\i{}guez-Quintero}},\ }\href
  {\doibase 10.1103/PhysRevD.104.054028} {\bibfield  {journal} {\bibinfo
  {journal} {Phys. Rev. D}\ }\textbf {\bibinfo {volume} {104}},\ \bibinfo
  {pages} {054028} (\bibinfo {year} {2021}{\natexlab{a}})}\BibitemShut
  {NoStop}%
\bibitem [{\citenamefont {Schleifenbaum}\ \emph {et~al.}(2005)\citenamefont
  {Schleifenbaum}, \citenamefont {Maas}, \citenamefont {Wambach},\ and\
  \citenamefont {Alkofer}}]{Schleifenbaum:2004id}%
  \BibitemOpen
  \bibfield  {author} {\bibinfo {author} {\bibfnamefont {W.}~\bibnamefont
  {Schleifenbaum}}, \bibinfo {author} {\bibfnamefont {A.}~\bibnamefont {Maas}},
  \bibinfo {author} {\bibfnamefont {J.}~\bibnamefont {Wambach}}, \ and\
  \bibinfo {author} {\bibfnamefont {R.}~\bibnamefont {Alkofer}},\ }\href
  {\doibase 10.1103/PhysRevD.72.014017} {\bibfield  {journal} {\bibinfo
  {journal} {Phys. Rev. D}\ }\textbf {\bibinfo {volume} {72}},\ \bibinfo
  {pages} {014017} (\bibinfo {year} {2005})}\BibitemShut {NoStop}%
\bibitem [{\citenamefont {Huber}\ and\ \citenamefont {von
  Smekal}(2013)}]{Huber:2012kd}%
  \BibitemOpen
  \bibfield  {author} {\bibinfo {author} {\bibfnamefont {M.~Q.}\ \bibnamefont
  {Huber}}\ and\ \bibinfo {author} {\bibfnamefont {L.}~\bibnamefont {von
  Smekal}},\ }\href {\doibase 10.1007/JHEP04(2013)149} {\bibfield  {journal}
  {\bibinfo  {journal} {J. High Energy Phys.}\ }\textbf {\bibinfo {volume}
  {04}},\ \bibinfo {pages} {149} (\bibinfo {year} {2013})}\BibitemShut
  {NoStop}%
%%CITATION = ARXIV:1211.6092;%%
\bibitem [{\citenamefont {Huber}\ \emph {et~al.}(2012)\citenamefont {Huber},
  \citenamefont {Maas},\ and\ \citenamefont {von Smekal}}]{Huber:2012zj}%
  \BibitemOpen
  \bibfield  {author} {\bibinfo {author} {\bibfnamefont {M.~Q.}\ \bibnamefont
  {Huber}}, \bibinfo {author} {\bibfnamefont {A.}~\bibnamefont {Maas}}, \ and\
  \bibinfo {author} {\bibfnamefont {L.}~\bibnamefont {von Smekal}},\ }\href
  {\doibase 10.1007/JHEP11(2012)035} {\bibfield  {journal} {\bibinfo  {journal}
  {J. High Energy Phys.}\ }\textbf {\bibinfo {volume} {11}},\ \bibinfo {pages}
  {035} (\bibinfo {year} {2012})}\BibitemShut {NoStop}%
%%CITATION = ARXIV:1207.0222;%%
\bibitem [{\citenamefont {Blum}\ \emph {et~al.}(2014)\citenamefont {Blum},
  \citenamefont {Huber}, \citenamefont {Mitter},\ and\ \citenamefont {von
  Smekal}}]{Blum:2014gna}%
  \BibitemOpen
  \bibfield  {author} {\bibinfo {author} {\bibfnamefont {A.}~\bibnamefont
  {Blum}}, \bibinfo {author} {\bibfnamefont {M.~Q.}\ \bibnamefont {Huber}},
  \bibinfo {author} {\bibfnamefont {M.}~\bibnamefont {Mitter}}, \ and\ \bibinfo
  {author} {\bibfnamefont {L.}~\bibnamefont {von Smekal}},\ }\href {\doibase
  10.1103/PhysRevD.89.061703} {\bibfield  {journal} {\bibinfo  {journal} {Phys.
  Rev.}\ }\textbf {\bibinfo {volume} {D89}},\ \bibinfo {pages} {061703}
  (\bibinfo {year} {2014})}\BibitemShut {NoStop}%
%%CITATION = ARXIV:1401.0713;%%
\bibitem [{\citenamefont {Eichmann}\ \emph {et~al.}(2014)\citenamefont
  {Eichmann}, \citenamefont {Williams}, \citenamefont {Alkofer},\ and\
  \citenamefont {Vujinovic}}]{Eichmann:2014xya}%
  \BibitemOpen
  \bibfield  {author} {\bibinfo {author} {\bibfnamefont {G.}~\bibnamefont
  {Eichmann}}, \bibinfo {author} {\bibfnamefont {R.}~\bibnamefont {Williams}},
  \bibinfo {author} {\bibfnamefont {R.}~\bibnamefont {Alkofer}}, \ and\
  \bibinfo {author} {\bibfnamefont {M.}~\bibnamefont {Vujinovic}},\ }\href
  {\doibase 10.1103/PhysRevD.89.105014} {\bibfield  {journal} {\bibinfo
  {journal} {Phys. Rev.}\ }\textbf {\bibinfo {volume} {D89}},\ \bibinfo {pages}
  {105014} (\bibinfo {year} {2014})}\BibitemShut {NoStop}%
%%CITATION = ARXIV:1402.1365;%%
\bibitem [{\citenamefont {Williams}\ \emph {et~al.}(2016)\citenamefont
  {Williams}, \citenamefont {Fischer},\ and\ \citenamefont
  {Heupel}}]{Williams:2015cvx}%
  \BibitemOpen
  \bibfield  {author} {\bibinfo {author} {\bibfnamefont {R.}~\bibnamefont
  {Williams}}, \bibinfo {author} {\bibfnamefont {C.~S.}\ \bibnamefont
  {Fischer}}, \ and\ \bibinfo {author} {\bibfnamefont {W.}~\bibnamefont
  {Heupel}},\ }\href {\doibase 10.1103/PhysRevD.93.034026} {\bibfield
  {journal} {\bibinfo  {journal} {Phys. Rev.}\ }\textbf {\bibinfo {volume}
  {D93}},\ \bibinfo {pages} {034026} (\bibinfo {year} {2016})}\BibitemShut
  {NoStop}%
%%CITATION = ARXIV:1512.00455;%%
\bibitem [{\citenamefont {Papavassiliou}\ \emph {et~al.}(2022)\citenamefont
  {Papavassiliou}, \citenamefont {Aguilar},\ and\ \citenamefont
  {Ferreira}}]{Papavassiliou:2022umz}%
  \BibitemOpen
  \bibfield  {author} {\bibinfo {author} {\bibfnamefont {J.}~\bibnamefont
  {Papavassiliou}}, \bibinfo {author} {\bibfnamefont {A.~C.}\ \bibnamefont
  {Aguilar}}, \ and\ \bibinfo {author} {\bibfnamefont {M.~N.}\ \bibnamefont
  {Ferreira}},\ }\ \href{https://rmf.smf.mx/ojs/index.php/rmf-s/article/view/6314/6462} {\bibfield
  {journal} {\bibinfo  {journal} {Supl. Rev. Mex. Fis.}\ }\textbf {\bibinfo {volume}
  {3}},\ \bibinfo {pages} {0308112} (\bibinfo {year} {2022})}\BibitemShut
  {NoStop}% 
  \bibitem [{\citenamefont {Aguilar}\ and\ \citenamefont
  {Papavassiliou}(2010)}]{Aguilar:2009ke}%
  \BibitemOpen
  \bibfield  {author} {\bibinfo {author} {\bibfnamefont {A.~C.}\ \bibnamefont
  {Aguilar}}\ and\ \bibinfo {author} {\bibfnamefont {J.}~\bibnamefont
  {Papavassiliou}},\ }\href {\doibase 10.1103/PhysRevD.81.034003} {\bibfield
  {journal} {\bibinfo  {journal} {Phys. Rev.}\ }\textbf {\bibinfo {volume}
  {D81}},\ \bibinfo {pages} {034003} (\bibinfo {year} {2010})}\BibitemShut
  {NoStop}%
\bibitem [{\citenamefont {Jackiw}\ and\ \citenamefont
  {Johnson}(1973)}]{Jackiw:1973tr}%
  \BibitemOpen
  \bibfield  {author} {\bibinfo {author} {\bibfnamefont {R.}~\bibnamefont
  {Jackiw}}\ and\ \bibinfo {author} {\bibfnamefont {K.}~\bibnamefont
  {Johnson}},\ }\href {\doibase 10.1103/PhysRevD.8.2386} {\bibfield  {journal}
  {\bibinfo  {journal} {Phys. Rev. D}\ }\textbf {\bibinfo {volume} {8}},\
  \bibinfo {pages} {2386} (\bibinfo {year} {1973})}\BibitemShut {NoStop}%
\bibitem [{\citenamefont {Jackiw}(1973)}]{Jackiw:1973ha}%
  \BibitemOpen
  \bibfield  {author} {\bibinfo {author} {\bibfnamefont {R.}~\bibnamefont
  {Jackiw}},\ }\href@noop {} {\bibfield  {journal} {\bibinfo  {journal} {In
  *Erice 1973, Proceedings, Laws Of Hadronic Matter*, New York 1975, 225-251
  and M I T Cambridge - COO-3069-190 (73,REC.AUG 74) 23p}\ } (\bibinfo {year}
  {1973})}\BibitemShut {NoStop}%
\bibitem [{\citenamefont {Aguilar}\ \emph
  {et~al.}(2014{\natexlab{a}})\citenamefont {Aguilar}, \citenamefont {Binosi},
  \citenamefont {Iba{\~n}ez},\ and\ \citenamefont
  {Papavassiliou}}]{Aguilar:2013vaa}%
  \BibitemOpen
  \bibfield  {author} {\bibinfo {author} {\bibfnamefont {A.~C.}\ \bibnamefont
  {Aguilar}}, \bibinfo {author} {\bibfnamefont {D.}~\bibnamefont {Binosi}},
  \bibinfo {author} {\bibfnamefont {D.}~\bibnamefont {Iba{\~n}ez}}, \ and\
  \bibinfo {author} {\bibfnamefont {J.}~\bibnamefont {Papavassiliou}},\ }\href
  {\doibase 10.1103/PhysRevD.89.085008} {\bibfield  {journal} {\bibinfo
  {journal} {Phys. Rev.}\ }\textbf {\bibinfo {volume} {D89}},\ \bibinfo {pages}
  {085008} (\bibinfo {year} {2014}{\natexlab{a}})}\BibitemShut {NoStop}%
%%CITATION = ARXIV:1312.1212;%%
\bibitem [{\citenamefont {Binosi}\ and\ \citenamefont
  {Papavassiliou}(2002{\natexlab{a}})}]{Binosi:2002ez}%
  \BibitemOpen
  \bibfield  {author} {\bibinfo {author} {\bibfnamefont {D.}~\bibnamefont
  {Binosi}}\ and\ \bibinfo {author} {\bibfnamefont {J.}~\bibnamefont
  {Papavassiliou}},\ }\href {\doibase 10.1103/PhysRevD.66.025024} {\bibfield
  {journal} {\bibinfo  {journal} {Phys. Rev.}\ }\textbf {\bibinfo {volume}
  {D66}},\ \bibinfo {pages} {025024} (\bibinfo {year}
  {2002}{\natexlab{a}})}\BibitemShut {NoStop}%
\bibitem [{\citenamefont {Hasenfratz}\ and\ \citenamefont
  {Hasenfratz}(1980)}]{Hasenfratz:1980kn}%
  \BibitemOpen
  \bibfield  {author} {\bibinfo {author} {\bibfnamefont {A.}~\bibnamefont
  {Hasenfratz}}\ and\ \bibinfo {author} {\bibfnamefont {P.}~\bibnamefont
  {Hasenfratz}},\ }\href {\doibase 10.1016/0370-2693(80)90118-5} {\bibfield
  {journal} {\bibinfo  {journal} {Phys. Lett. B}\ }\textbf {\bibinfo {volume}
  {93}},\ \bibinfo {pages} {165} (\bibinfo {year} {1980})}\BibitemShut
  {NoStop}%
\bibitem [{\citenamefont {Alles}\ \emph {et~al.}(1997)\citenamefont {Alles},
  \citenamefont {Henty}, \citenamefont {Panagopoulos}, \citenamefont
  {Parrinello}, \citenamefont {Pittori},\ and\ \citenamefont
  {Richards}}]{Alles:1996ka}%
  \BibitemOpen
  \bibfield  {author} {\bibinfo {author} {\bibfnamefont {B.}~\bibnamefont
  {Alles}}, \bibinfo {author} {\bibfnamefont {D.}~\bibnamefont {Henty}},
  \bibinfo {author} {\bibfnamefont {H.}~\bibnamefont {Panagopoulos}}, \bibinfo
  {author} {\bibfnamefont {C.}~\bibnamefont {Parrinello}}, \bibinfo {author}
  {\bibfnamefont {C.}~\bibnamefont {Pittori}}, \ and\ \bibinfo {author}
  {\bibfnamefont {D.~G.}\ \bibnamefont {Richards}},\ }\href {\doibase
  10.1016/S0550-3213(97)00483-5} {\bibfield  {journal} {\bibinfo  {journal}
  {Nucl. Phys.}\ }\textbf {\bibinfo {volume} {B502}},\ \bibinfo {pages} {325}
  (\bibinfo {year} {1997})}\BibitemShut {NoStop}%
%%CITATION = HEP-LAT/9605033;%%
\bibitem [{\citenamefont {Davydychev}\ \emph {et~al.}(1998)\citenamefont
  {Davydychev}, \citenamefont {Osland},\ and\ \citenamefont
  {Tarasov}}]{Davydychev:1997vh}%
  \BibitemOpen
  \bibfield  {author} {\bibinfo {author} {\bibfnamefont {A.~I.}\ \bibnamefont
  {Davydychev}}, \bibinfo {author} {\bibfnamefont {P.}~\bibnamefont {Osland}},
  \ and\ \bibinfo {author} {\bibfnamefont {O.~V.}\ \bibnamefont {Tarasov}},\
  }\href {\doibase 10.1103/PhysRevD.58.036007} {\bibfield  {journal} {\bibinfo
  {journal} {Phys. Rev. D}\ }\textbf {\bibinfo {volume} {58}},\ \bibinfo
  {pages} {036007} (\bibinfo {year} {1998})}\BibitemShut {NoStop}%
\bibitem [{\citenamefont {Boucaud}\ \emph
  {et~al.}(1998{\natexlab{a}})\citenamefont {Boucaud}, \citenamefont {Leroy},
  \citenamefont {Micheli}, \citenamefont {Pene},\ and\ \citenamefont
  {Roiesnel}}]{Boucaud:1998bq}%
  \BibitemOpen
  \bibfield  {author} {\bibinfo {author} {\bibfnamefont {P.}~\bibnamefont
  {Boucaud}}, \bibinfo {author} {\bibfnamefont {J.~P.}\ \bibnamefont {Leroy}},
  \bibinfo {author} {\bibfnamefont {J.}~\bibnamefont {Micheli}}, \bibinfo
  {author} {\bibfnamefont {O.}~\bibnamefont {Pene}}, \ and\ \bibinfo {author}
  {\bibfnamefont {C.}~\bibnamefont {Roiesnel}},\ }\href {\doibase
  10.1088/1126-6708/1998/10/017} {\bibfield  {journal} {\bibinfo  {journal} {J.
  High Energy Phys.}\ }\textbf {\bibinfo {volume} {10}},\ \bibinfo {pages}
  {017} (\bibinfo {year} {1998}{\natexlab{a}})}\BibitemShut {NoStop}%
%%CITATION = HEP-PH/9810322;%%
\bibitem [{\citenamefont {Boucaud}\ \emph
  {et~al.}(1998{\natexlab{b}})\citenamefont {Boucaud}, \citenamefont {Leroy},
  \citenamefont {Micheli}, \citenamefont {Pene},\ and\ \citenamefont
  {Roiesnel}}]{Boucaud:1998xi}%
  \BibitemOpen
  \bibfield  {author} {\bibinfo {author} {\bibfnamefont {P.}~\bibnamefont
  {Boucaud}}, \bibinfo {author} {\bibfnamefont {J.~P.}\ \bibnamefont {Leroy}},
  \bibinfo {author} {\bibfnamefont {J.}~\bibnamefont {Micheli}}, \bibinfo
  {author} {\bibfnamefont {O.}~\bibnamefont {Pene}}, \ and\ \bibinfo {author}
  {\bibfnamefont {C.}~\bibnamefont {Roiesnel}},\ }\href {\doibase
  10.1088/1126-6708/1998/12/004} {\bibfield  {journal} {\bibinfo  {journal}
  {JHEP}\ }\textbf {\bibinfo {volume} {12}},\ \bibinfo {pages} {004} (\bibinfo
  {year} {1998}{\natexlab{b}})}\BibitemShut {NoStop}%
\bibitem [{\citenamefont {Nakanishi}(1969)}]{Nakanishi:1969ph}%
  \BibitemOpen
  \bibfield  {author} {\bibinfo {author} {\bibfnamefont {N.}~\bibnamefont
  {Nakanishi}},\ }\href {\doibase 10.1143/PTPS.43.1} {\bibfield  {journal}
  {\bibinfo  {journal} {Prog. Theor. Phys. Suppl.}\ }\textbf {\bibinfo {volume}
  {43}},\ \bibinfo {pages} {1} (\bibinfo {year} {1969})}\BibitemShut {NoStop}%
\bibitem [{\citenamefont {Maris}\ and\ \citenamefont
  {Roberts}(1997)}]{Maris:1997tm}%
  \BibitemOpen
  \bibfield  {author} {\bibinfo {author} {\bibfnamefont {P.}~\bibnamefont
  {Maris}}\ and\ \bibinfo {author} {\bibfnamefont {C.~D.}\ \bibnamefont
  {Roberts}},\ }\href {\doibase 10.1103/PhysRevC.56.3369} {\bibfield  {journal}
  {\bibinfo  {journal} {Phys. Rev. C}\ }\textbf {\bibinfo {volume} {56}},\
  \bibinfo {pages} {3369} (\bibinfo {year} {1997})}\BibitemShut {NoStop}%
\bibitem [{\citenamefont {Blank}\ and\ \citenamefont
  {Krassnigg}(2011)}]{Blank:2010bp}%
  \BibitemOpen
  \bibfield  {author} {\bibinfo {author} {\bibfnamefont {M.}~\bibnamefont
  {Blank}}\ and\ \bibinfo {author} {\bibfnamefont {A.}~\bibnamefont
  {Krassnigg}},\ }\href {\doibase 10.1016/j.cpc.2011.03.003} {\bibfield
  {journal} {\bibinfo  {journal} {Comput. Phys. Commun.}\ }\textbf {\bibinfo
  {volume} {182}},\ \bibinfo {pages} {1391} (\bibinfo {year}
  {2011})}\BibitemShut {NoStop}%
\bibitem [{\citenamefont {Aguilar}\ \emph
  {et~al.}(2014{\natexlab{b}})\citenamefont {Aguilar}, \citenamefont {Binosi},\
  and\ \citenamefont {Papavassiliou}}]{Aguilar:2014tka}%
  \BibitemOpen
  \bibfield  {author} {\bibinfo {author} {\bibfnamefont {A.~C.}\ \bibnamefont
  {Aguilar}}, \bibinfo {author} {\bibfnamefont {D.}~\bibnamefont {Binosi}}, \
  and\ \bibinfo {author} {\bibfnamefont {J.}~\bibnamefont {Papavassiliou}},\
  }\href {\doibase 10.1103/PhysRevD.89.085032} {\bibfield  {journal} {\bibinfo
  {journal} {Phys. Rev.}\ }\textbf {\bibinfo {volume} {D89}},\ \bibinfo {pages}
  {085032} (\bibinfo {year} {2014}{\natexlab{b}})}\BibitemShut {NoStop}%
%%CITATION = ARXIV:1401.3631;%%
\bibitem [{\citenamefont {Boucaud}\ \emph {et~al.}(2009)\citenamefont
  {Boucaud}, \citenamefont {De~Soto}, \citenamefont {Leroy}, \citenamefont
  {Le~Yaouanc}, \citenamefont {Micheli} \emph {et~al.}}]{Boucaud:2008gn}%
  \BibitemOpen
  \bibfield  {author} {\bibinfo {author} {\bibfnamefont {P.}~\bibnamefont
  {Boucaud}}, \bibinfo {author} {\bibfnamefont {F.}~\bibnamefont {De~Soto}},
  \bibinfo {author} {\bibfnamefont {J.}~\bibnamefont {Leroy}}, \bibinfo
  {author} {\bibfnamefont {A.}~\bibnamefont {Le~Yaouanc}}, \bibinfo {author}
  {\bibfnamefont {J.}~\bibnamefont {Micheli}},  \emph {et~al.},\ }\href
  {\doibase 10.1103/PhysRevD.79.014508} {\bibfield  {journal} {\bibinfo
  {journal} {Phys. Rev.}\ }\textbf {\bibinfo {volume} {D79}},\ \bibinfo {pages}
  {014508} (\bibinfo {year} {2009})}\BibitemShut {NoStop}%
%%CITATION = ARXIV:0811.2059;%%
\bibitem [{\citenamefont {von Smekal}\ \emph {et~al.}(2009)\citenamefont {von
  Smekal}, \citenamefont {Maltman},\ and\ \citenamefont
  {Sternbeck}}]{vonSmekal:2009ae}%
  \BibitemOpen
  \bibfield  {author} {\bibinfo {author} {\bibfnamefont {L.}~\bibnamefont {von
  Smekal}}, \bibinfo {author} {\bibfnamefont {K.}~\bibnamefont {Maltman}}, \
  and\ \bibinfo {author} {\bibfnamefont {A.}~\bibnamefont {Sternbeck}},\ }\href
  {\doibase 10.1016/j.physletb.2009.10.030} {\bibfield  {journal} {\bibinfo
  {journal} {Phys. Lett. B}\ }\textbf {\bibinfo {volume} {681}},\ \bibinfo
  {pages} {336} (\bibinfo {year} {2009})}\BibitemShut {NoStop}%
\bibitem [{\citenamefont {Boucaud}\ \emph {et~al.}(2011)\citenamefont
  {Boucaud}, \citenamefont {Dudal}, \citenamefont {Leroy}, \citenamefont
  {Pene},\ and\ \citenamefont {Rodriguez-Quintero}}]{Boucaud:2011eh}%
  \BibitemOpen
  \bibfield  {author} {\bibinfo {author} {\bibfnamefont {P.}~\bibnamefont
  {Boucaud}}, \bibinfo {author} {\bibfnamefont {D.}~\bibnamefont {Dudal}},
  \bibinfo {author} {\bibfnamefont {J.}~\bibnamefont {Leroy}}, \bibinfo
  {author} {\bibfnamefont {O.}~\bibnamefont {Pene}}, \ and\ \bibinfo {author}
  {\bibfnamefont {J.}~\bibnamefont {Rodriguez-Quintero}},\ }\href {\doibase
  10.1007/JHEP12(2011)018} {\bibfield  {journal} {\bibinfo  {journal} {J. High
  Energy Phys.}\ }\textbf {\bibinfo {volume} {12}},\ \bibinfo {pages} {018}
  (\bibinfo {year} {2011})}\BibitemShut {NoStop}%
%%CITATION = ARXIV:1109.3803;%%
\bibitem [{\citenamefont {Aguilar}\ \emph
  {et~al.}(2021{\natexlab{b}})\citenamefont {Aguilar}, \citenamefont {De~Soto},
  \citenamefont {Ferreira}, \citenamefont {Papavassiliou},\ and\ \citenamefont
  {Rodr\'\i{}guez-Quintero}}]{Aguilar:2021lke}%
  \BibitemOpen
  \bibfield  {author} {\bibinfo {author} {\bibfnamefont {A.~C.}\ \bibnamefont
  {Aguilar}}, \bibinfo {author} {\bibfnamefont {F.}~\bibnamefont {De~Soto}},
  \bibinfo {author} {\bibfnamefont {M.~N.}\ \bibnamefont {Ferreira}}, \bibinfo
  {author} {\bibfnamefont {J.}~\bibnamefont {Papavassiliou}}, \ and\ \bibinfo
  {author} {\bibfnamefont {J.}~\bibnamefont {Rodr\'\i{}guez-Quintero}},\ }\href
  {\doibase 10.1016/j.physletb.2021.136352} {\bibfield  {journal} {\bibinfo
  {journal} {Phys. Lett. B}\ }\textbf {\bibinfo {volume} {818}},\ \bibinfo
  {pages} {136352} (\bibinfo {year} {2021}{\natexlab{b}})}\BibitemShut
  {NoStop}%
\bibitem [{\citenamefont {Aguilar}\ \emph {et~al.}(2010)\citenamefont
  {Aguilar}, \citenamefont {Binosi},\ and\ \citenamefont
  {Papavassiliou}}]{Aguilar:2010gm}%
  \BibitemOpen
  \bibfield  {author} {\bibinfo {author} {\bibfnamefont {A.~C.}\ \bibnamefont
  {Aguilar}}, \bibinfo {author} {\bibfnamefont {D.}~\bibnamefont {Binosi}}, \
  and\ \bibinfo {author} {\bibfnamefont {J.}~\bibnamefont {Papavassiliou}},\
  }\href {\doibase 10.1007/JHEP07(2010)002} {\bibfield  {journal} {\bibinfo
  {journal} {J. High Energy Phys.}\ }\textbf {\bibinfo {volume} {07}},\
  \bibinfo {pages} {002} (\bibinfo {year} {2010})}\BibitemShut {NoStop}%
\bibitem [{\citenamefont {Gell-Mann}\ and\ \citenamefont
  {Low}(1954)}]{Gell-Mann:1954yli}%
  \BibitemOpen
  \bibfield  {author} {\bibinfo {author} {\bibfnamefont {M.}~\bibnamefont
  {Gell-Mann}}\ and\ \bibinfo {author} {\bibfnamefont {F.~E.}\ \bibnamefont
  {Low}},\ }\href {\doibase 10.1103/PhysRev.95.1300} {\bibfield  {journal}
  {\bibinfo  {journal} {Phys. Rev.}\ }\textbf {\bibinfo {volume} {95}},\
  \bibinfo {pages} {1300} (\bibinfo {year} {1954})}\BibitemShut {NoStop}%
\bibitem [{\citenamefont {DeWitt}(1967)}]{DeWitt:1967ub}%
  \BibitemOpen
  \bibfield  {author} {\bibinfo {author} {\bibfnamefont {B.~S.}\ \bibnamefont
  {DeWitt}},\ }\href {\doibase 10.1103/PhysRev.162.1195} {\bibfield  {journal}
  {\bibinfo  {journal} {Phys. Rev.}\ }\textbf {\bibinfo {volume} {162}},\
  \bibinfo {pages} {1195} (\bibinfo {year} {1967})}\BibitemShut {NoStop}%
\bibitem [{\citenamefont {Honerkamp}(1972)}]{Honerkamp:1972fd}%
  \BibitemOpen
  \bibfield  {author} {\bibinfo {author} {\bibfnamefont {J.}~\bibnamefont
  {Honerkamp}},\ }\href {\doibase 10.1016/0550-3213(72)90063-6} {\bibfield
  {journal} {\bibinfo  {journal} {Nucl. Phys. B}\ }\textbf {\bibinfo {volume}
  {48}},\ \bibinfo {pages} {269} (\bibinfo {year} {1972})}\BibitemShut
  {NoStop}%
\bibitem [{\citenamefont {'t~Hooft}(1971)}]{tHooft:1971qjg}%
  \BibitemOpen
  \bibfield  {author} {\bibinfo {author} {\bibfnamefont {G.}~\bibnamefont
  {'t~Hooft}},\ }\href {\doibase 10.1016/0550-3213(71)90139-8} {\bibfield
  {journal} {\bibinfo  {journal} {Nucl. Phys. B}\ }\textbf {\bibinfo {volume}
  {35}},\ \bibinfo {pages} {167} (\bibinfo {year} {1971})}\BibitemShut
  {NoStop}%
\bibitem [{\citenamefont {Kallosh}(1974)}]{Kallosh:1974yh}%
  \BibitemOpen
  \bibfield  {author} {\bibinfo {author} {\bibfnamefont {R.~E.}\ \bibnamefont
  {Kallosh}},\ }\href {\doibase 10.1016/0550-3213(74)90284-3} {\bibfield
  {journal} {\bibinfo  {journal} {Nucl. Phys. B}\ }\textbf {\bibinfo {volume}
  {78}},\ \bibinfo {pages} {293} (\bibinfo {year} {1974})}\BibitemShut
  {NoStop}%
\bibitem [{\citenamefont {Kluberg-Stern}\ and\ \citenamefont
  {Zuber}(1975)}]{Kluberg-Stern:1974nmx}%
  \BibitemOpen
  \bibfield  {author} {\bibinfo {author} {\bibfnamefont {H.}~\bibnamefont
  {Kluberg-Stern}}\ and\ \bibinfo {author} {\bibfnamefont {J.~B.}\ \bibnamefont
  {Zuber}},\ }\href {\doibase 10.1103/PhysRevD.12.482} {\bibfield  {journal}
  {\bibinfo  {journal} {Phys. Rev. D}\ }\textbf {\bibinfo {volume} {12}},\
  \bibinfo {pages} {482} (\bibinfo {year} {1975})}\BibitemShut {NoStop}%
\bibitem [{\citenamefont {Abbott}(1981)}]{Abbott:1980hw}%
  \BibitemOpen
  \bibfield  {author} {\bibinfo {author} {\bibfnamefont {L.}~\bibnamefont
  {Abbott}},\ }\href {\doibase 10.1016/0550-3213(81)90371-0} {\bibfield
  {journal} {\bibinfo  {journal} {Nucl. Phys. B}\ }\textbf {\bibinfo {volume}
  {185}},\ \bibinfo {pages} {189} (\bibinfo {year} {1981})}\BibitemShut
  {NoStop}%
\bibitem [{\citenamefont {Shore}(1981)}]{Shore:1981mj}%
  \BibitemOpen
  \bibfield  {author} {\bibinfo {author} {\bibfnamefont {G.~M.}\ \bibnamefont
  {Shore}},\ }\href {\doibase 10.1016/0003-4916(81)90198-6} {\bibfield
  {journal} {\bibinfo  {journal} {Annals Phys.}\ }\textbf {\bibinfo {volume}
  {137}},\ \bibinfo {pages} {262} (\bibinfo {year} {1981})}\BibitemShut
  {NoStop}%
\bibitem [{\citenamefont {Abbott}\ \emph {et~al.}(1983)\citenamefont {Abbott},
  \citenamefont {Grisaru},\ and\ \citenamefont {Schaefer}}]{Abbott:1983zw}%
  \BibitemOpen
  \bibfield  {author} {\bibinfo {author} {\bibfnamefont {L.~F.}\ \bibnamefont
  {Abbott}}, \bibinfo {author} {\bibfnamefont {M.~T.}\ \bibnamefont {Grisaru}},
  \ and\ \bibinfo {author} {\bibfnamefont {R.~K.}\ \bibnamefont {Schaefer}},\
  }\href {\doibase 10.1016/0550-3213(83)90337-1} {\bibfield  {journal}
  {\bibinfo  {journal} {Nucl. Phys. B}\ }\textbf {\bibinfo {volume} {229}},\
  \bibinfo {pages} {372} (\bibinfo {year} {1983})}\BibitemShut {NoStop}%
\bibitem [{\citenamefont {Aguilar}\ \emph
  {et~al.}(2019{\natexlab{a}})\citenamefont {Aguilar}, \citenamefont
  {Ferreira}, \citenamefont {Figueiredo},\ and\ \citenamefont
  {Papavassiliou}}]{Aguilar:2018csq}%
  \BibitemOpen
  \bibfield  {author} {\bibinfo {author} {\bibfnamefont {A.~C.}\ \bibnamefont
  {Aguilar}}, \bibinfo {author} {\bibfnamefont {M.~N.}\ \bibnamefont
  {Ferreira}}, \bibinfo {author} {\bibfnamefont {C.~T.}\ \bibnamefont
  {Figueiredo}}, \ and\ \bibinfo {author} {\bibfnamefont {J.}~\bibnamefont
  {Papavassiliou}},\ }\href {\doibase 10.1103/PhysRevD.99.034026} {\bibfield
  {journal} {\bibinfo  {journal} {Phys. Rev.}\ }\textbf {\bibinfo {volume}
  {D99}},\ \bibinfo {pages} {034026} (\bibinfo {year}
  {2019}{\natexlab{a}})}\BibitemShut {NoStop}%
%%CITATION = ARXIV:1811.08961;%%
\bibitem [{\citenamefont {Aguilar}\ \emph
  {et~al.}(2020{\natexlab{b}})\citenamefont {Aguilar}, \citenamefont
  {Ferreira},\ and\ \citenamefont {Papavassiliou}}]{Aguilar:2020yni}%
  \BibitemOpen
  \bibfield  {author} {\bibinfo {author} {\bibfnamefont {A.~C.}\ \bibnamefont
  {Aguilar}}, \bibinfo {author} {\bibfnamefont {M.~N.}\ \bibnamefont
  {Ferreira}}, \ and\ \bibinfo {author} {\bibfnamefont {J.}~\bibnamefont
  {Papavassiliou}},\ }\href {\doibase 10.1140/epjc/s10052-020-08453-2}
  {\bibfield  {journal} {\bibinfo  {journal} {Eur. Phys. J. C}\ }\textbf
  {\bibinfo {volume} {80}},\ \bibinfo {pages} {887} (\bibinfo {year}
  {2020}{\natexlab{b}})}\BibitemShut {NoStop}%
\bibitem [{\citenamefont {Parrinello}(1994)}]{Parrinello:1994wd}%
  \BibitemOpen
  \bibfield  {author} {\bibinfo {author} {\bibfnamefont {C.}~\bibnamefont
  {Parrinello}},\ }\href {\doibase 10.1103/PhysRevD.50.R4247} {\bibfield
  {journal} {\bibinfo  {journal} {Phys. Rev.}\ }\textbf {\bibinfo {volume}
  {D50}},\ \bibinfo {pages} {R4247} (\bibinfo {year} {1994})}\BibitemShut
  {NoStop}%
%%CITATION = HEP-LAT/9405024;%%
\bibitem [{\citenamefont {Parrinello}\ \emph {et~al.}(1998)\citenamefont
  {Parrinello}, \citenamefont {Richards}, \citenamefont {Alles}, \citenamefont
  {Panagopoulos},\ and\ \citenamefont {Pittori}}]{Parrinello:1997wm}%
  \BibitemOpen
  \bibfield  {author} {\bibinfo {author} {\bibfnamefont {C.}~\bibnamefont
  {Parrinello}}, \bibinfo {author} {\bibfnamefont {D.}~\bibnamefont
  {Richards}}, \bibinfo {author} {\bibfnamefont {B.}~\bibnamefont {Alles}},
  \bibinfo {author} {\bibfnamefont {H.}~\bibnamefont {Panagopoulos}}, \ and\
  \bibinfo {author} {\bibfnamefont {C.}~\bibnamefont {Pittori}} (\bibinfo
  {collaboration} {UKQCD}),\ }\href {\doibase 10.1016/S0920-5632(97)00734-2}
  {\bibfield  {journal} {\bibinfo  {journal} {Nucl. Phys. B Proc. Suppl.}\
  }\textbf {\bibinfo {volume} {63}},\ \bibinfo {pages} {245} (\bibinfo {year}
  {1998})}\BibitemShut {NoStop}%
\bibitem [{\citenamefont {Sternbeck}\ \emph {et~al.}(2017)\citenamefont
  {Sternbeck}, \citenamefont {Balduf}, \citenamefont {Kizilersu}, \citenamefont
  {Oliveira}, \citenamefont {Silva}, \citenamefont {Skullerud},\ and\
  \citenamefont {Williams}}]{Sternbeck:2017ntv}%
  \BibitemOpen
  \bibfield  {author} {\bibinfo {author} {\bibfnamefont {A.}~\bibnamefont
  {Sternbeck}}, \bibinfo {author} {\bibfnamefont {P.-H.}\ \bibnamefont
  {Balduf}}, \bibinfo {author} {\bibfnamefont {A.}~\bibnamefont {Kizilersu}},
  \bibinfo {author} {\bibfnamefont {O.}~\bibnamefont {Oliveira}}, \bibinfo
  {author} {\bibfnamefont {P.~J.}\ \bibnamefont {Silva}}, \bibinfo {author}
  {\bibfnamefont {J.-I.}\ \bibnamefont {Skullerud}}, \ and\ \bibinfo {author}
  {\bibfnamefont {A.~G.}\ \bibnamefont {Williams}},\ }\href {\doibase
  10.22323/1.256.0349} {\bibfield  {journal} {\bibinfo  {journal} {PoS}\
  }\textbf {\bibinfo {volume} {LATTICE2016}},\ \bibinfo {pages} {349} (\bibinfo
  {year} {2017})}\BibitemShut {NoStop}%
%%CITATION = ARXIV:1702.00612;%%
\bibitem [{\citenamefont {Vujinovic}\ and\ \citenamefont
  {Mendes}(2019)}]{Vujinovic:2018nqc}%
  \BibitemOpen
  \bibfield  {author} {\bibinfo {author} {\bibfnamefont {M.}~\bibnamefont
  {Vujinovic}}\ and\ \bibinfo {author} {\bibfnamefont {T.}~\bibnamefont
  {Mendes}},\ }\href {\doibase 10.1103/PhysRevD.99.034501} {\bibfield
  {journal} {\bibinfo  {journal} {Phys. Rev.}\ }\textbf {\bibinfo {volume}
  {D99}},\ \bibinfo {pages} {034501} (\bibinfo {year} {2019})}\BibitemShut
  {NoStop}%
%%CITATION = ARXIV:1807.03673;%%
\bibitem [{\citenamefont {Boucaud}\ \emph {et~al.}(2017)\citenamefont
  {Boucaud}, \citenamefont {De~Soto}, \citenamefont {Rodr\'{\i}guez-Quintero},\
  and\ \citenamefont {Zafeiropoulos}}]{Boucaud:2017obn}%
  \BibitemOpen
  \bibfield  {author} {\bibinfo {author} {\bibfnamefont {P.}~\bibnamefont
  {Boucaud}}, \bibinfo {author} {\bibfnamefont {F.}~\bibnamefont {De~Soto}},
  \bibinfo {author} {\bibfnamefont {J.}~\bibnamefont
  {Rodr\'{\i}guez-Quintero}}, \ and\ \bibinfo {author} {\bibfnamefont
  {S.}~\bibnamefont {Zafeiropoulos}},\ }\href {\doibase
  10.1103/PhysRevD.95.114503} {\bibfield  {journal} {\bibinfo  {journal} {Phys.
  Rev.}\ }\textbf {\bibinfo {volume} {D95}},\ \bibinfo {pages} {114503}
  (\bibinfo {year} {2017})}\BibitemShut {NoStop}%
%%CITATION = ARXIV:1701.07390;%%
\bibitem [{\citenamefont {Blum}\ \emph {et~al.}(2015)\citenamefont {Blum},
  \citenamefont {Alkofer}, \citenamefont {Huber},\ and\ \citenamefont
  {Windisch}}]{Blum:2015lsa}%
  \BibitemOpen
  \bibfield  {author} {\bibinfo {author} {\bibfnamefont {A.~L.}\ \bibnamefont
  {Blum}}, \bibinfo {author} {\bibfnamefont {R.}~\bibnamefont {Alkofer}},
  \bibinfo {author} {\bibfnamefont {M.~Q.}\ \bibnamefont {Huber}}, \ and\
  \bibinfo {author} {\bibfnamefont {A.}~\bibnamefont {Windisch}},\ }\href
  {\doibase 10.5506/APhysPolBSupp.8.321} {\bibfield  {journal} {\bibinfo
  {journal} {Acta Phys. Polon. Supp.}\ }\textbf {\bibinfo {volume} {8}},\
  \bibinfo {pages} {321} (\bibinfo {year} {2015})}\BibitemShut {NoStop}%
\bibitem [{\citenamefont {Aguilar}\ \emph
  {et~al.}(2019{\natexlab{b}})\citenamefont {Aguilar}, \citenamefont
  {Ferreira}, \citenamefont {Figueiredo},\ and\ \citenamefont
  {Papavassiliou}}]{Aguilar:2019jsj}%
  \BibitemOpen
  \bibfield  {author} {\bibinfo {author} {\bibfnamefont {A.~C.}\ \bibnamefont
  {Aguilar}}, \bibinfo {author} {\bibfnamefont {M.~N.}\ \bibnamefont
  {Ferreira}}, \bibinfo {author} {\bibfnamefont {C.~T.}\ \bibnamefont
  {Figueiredo}}, \ and\ \bibinfo {author} {\bibfnamefont {J.}~\bibnamefont
  {Papavassiliou}},\ }\href {\doibase 10.1103/PhysRevD.99.094010} {\bibfield
  {journal} {\bibinfo  {journal} {Phys. Rev.}\ }\textbf {\bibinfo {volume}
  {D99}},\ \bibinfo {pages} {094010} (\bibinfo {year}
  {2019}{\natexlab{b}})}\BibitemShut {NoStop}%
%%CITATION = ARXIV:1903.01184;%%
\bibitem [{\citenamefont {Pilaftsis}(1997)}]{Pilaftsis:1996fh}%
  \BibitemOpen
  \bibfield  {author} {\bibinfo {author} {\bibfnamefont {A.}~\bibnamefont
  {Pilaftsis}},\ }\href {\doibase 10.1016/S0550-3213(96)00686-4} {\bibfield
  {journal} {\bibinfo  {journal} {Nucl. Phys. B}\ }\textbf {\bibinfo {volume}
  {487}},\ \bibinfo {pages} {467} (\bibinfo {year} {1997})}\BibitemShut
  {NoStop}%
\bibitem [{\citenamefont {Binosi}\ and\ \citenamefont
  {Papavassiliou}(2002{\natexlab{b}})}]{Binosi:2002ft}%
  \BibitemOpen
  \bibfield  {author} {\bibinfo {author} {\bibfnamefont {D.}~\bibnamefont
  {Binosi}}\ and\ \bibinfo {author} {\bibfnamefont {J.}~\bibnamefont
  {Papavassiliou}},\ }\href {\doibase 10.1103/PhysRevD.66.111901} {\bibfield
  {journal} {\bibinfo  {journal} {Phys. Rev. D}\ }\textbf {\bibinfo {volume}
  {66}},\ \bibinfo {pages} {111901} (\bibinfo {year}
  {2002}{\natexlab{b}})}\BibitemShut {NoStop}%
\bibitem [{\citenamefont {Binosi}\ and\ \citenamefont
  {Papavassiliou}(2008{\natexlab{a}})}]{Binosi:2007pi}%
  \BibitemOpen
  \bibfield  {author} {\bibinfo {author} {\bibfnamefont {D.}~\bibnamefont
  {Binosi}}\ and\ \bibinfo {author} {\bibfnamefont {J.}~\bibnamefont
  {Papavassiliou}},\ }\href {\doibase 10.1103/PhysRevD.77.061702} {\bibfield
  {journal} {\bibinfo  {journal} {Phys. Rev.}\ }\textbf {\bibinfo {volume}
  {D77}},\ \bibinfo {pages} {061702} (\bibinfo {year}
  {2008}{\natexlab{a}})}\BibitemShut {NoStop}%
\bibitem [{\citenamefont {Binosi}\ and\ \citenamefont
  {Papavassiliou}(2008{\natexlab{b}})}]{Binosi:2008qk}%
  \BibitemOpen
  \bibfield  {author} {\bibinfo {author} {\bibfnamefont {D.}~\bibnamefont
  {Binosi}}\ and\ \bibinfo {author} {\bibfnamefont {J.}~\bibnamefont
  {Papavassiliou}},\ }\href {\doibase 10.1088/1126-6708/2008/11/063} {\bibfield
   {journal} {\bibinfo  {journal} {J. High Energy Phys.}\ }\textbf {\bibinfo
  {volume} {11}},\ \bibinfo {pages} {063} (\bibinfo {year}
  {2008}{\natexlab{b}})}\BibitemShut {NoStop}%
\bibitem [{\citenamefont {Grassi}\ \emph {et~al.}(2001)\citenamefont {Grassi},
  \citenamefont {Hurth},\ and\ \citenamefont {Steinhauser}}]{Grassi:1999tp}%
  \BibitemOpen
  \bibfield  {author} {\bibinfo {author} {\bibfnamefont {P.~A.}\ \bibnamefont
  {Grassi}}, \bibinfo {author} {\bibfnamefont {T.}~\bibnamefont {Hurth}}, \
  and\ \bibinfo {author} {\bibfnamefont {M.}~\bibnamefont {Steinhauser}},\
  }\href {\doibase 10.1006/aphy.2001.6117} {\bibfield  {journal} {\bibinfo
  {journal} {Annals Phys.}\ }\textbf {\bibinfo {volume} {288}},\ \bibinfo
  {pages} {197} (\bibinfo {year} {2001})}\BibitemShut {NoStop}%
\bibitem [{\citenamefont {Binosi}\ and\ \citenamefont
  {Quadri}(2013)}]{Binosi:2013cea}%
  \BibitemOpen
  \bibfield  {author} {\bibinfo {author} {\bibfnamefont {D.}~\bibnamefont
  {Binosi}}\ and\ \bibinfo {author} {\bibfnamefont {A.}~\bibnamefont
  {Quadri}},\ }\href {\doibase 10.1103/PhysRevD.88.085036} {\bibfield
  {journal} {\bibinfo  {journal} {Phys. Rev.}\ }\textbf {\bibinfo {volume}
  {D88}},\ \bibinfo {pages} {085036} (\bibinfo {year} {2013})}\BibitemShut
  {NoStop}%
%%CITATION = ARXIV:1309.1021;%%
\bibitem [{\citenamefont {Aguilar}\ \emph
  {et~al.}(2009{\natexlab{b}})\citenamefont {Aguilar}, \citenamefont {Binosi},\
  and\ \citenamefont {Papavassiliou}}]{Aguilar:2009pp}%
  \BibitemOpen
  \bibfield  {author} {\bibinfo {author} {\bibfnamefont {A.~C.}\ \bibnamefont
  {Aguilar}}, \bibinfo {author} {\bibfnamefont {D.}~\bibnamefont {Binosi}}, \
  and\ \bibinfo {author} {\bibfnamefont {J.}~\bibnamefont {Papavassiliou}},\
  }\href {\doibase 10.1088/1126-6708/2009/11/066} {\bibfield  {journal}
  {\bibinfo  {journal} {J. High Energy Phys.}\ }\textbf {\bibinfo {volume}
  {11}},\ \bibinfo {pages} {066} (\bibinfo {year}
  {2009}{\natexlab{b}})}\BibitemShut {NoStop}%
\bibitem [{\citenamefont {Wilson}(1973)}]{Wilson:1972cf}%
  \BibitemOpen
  \bibfield  {author} {\bibinfo {author} {\bibfnamefont {K.~G.}\ \bibnamefont
  {Wilson}},\ }\href {\doibase 10.1103/PhysRevD.7.2911} {\bibfield  {journal}
  {\bibinfo  {journal} {Phys. Rev.}\ }\textbf {\bibinfo {volume} {D7}},\
  \bibinfo {pages} {2911} (\bibinfo {year} {1973})}\BibitemShut {NoStop}%
%%CITATION = PHRVA,D7,2911;%%
\bibitem [{\citenamefont {Roberts}\ and\ \citenamefont
  {Mezrag}(2017)}]{Roberts:2016mhh}%
  \BibitemOpen
  \bibfield  {author} {\bibinfo {author} {\bibfnamefont {C.~D.}\ \bibnamefont
  {Roberts}}\ and\ \bibinfo {author} {\bibfnamefont {C.}~\bibnamefont
  {Mezrag}},\ }\href {\doibase 10.1051/epjconf/201713701017} {\bibfield
  {journal} {\bibinfo  {journal} {EPJ Web Conf.}\ }\textbf {\bibinfo {volume}
  {137}},\ \bibinfo {pages} {01017} (\bibinfo {year} {2017})}\BibitemShut
  {NoStop}%
\bibitem [{\citenamefont {Binosi}(2022)}]{Binosi:2022djx}%
  \BibitemOpen
  \bibfield  {author} {\bibinfo {author} {\bibfnamefont {D.}~\bibnamefont
  {Binosi}},\ }\href {\doibase 10.1007/s00601-022-01740-6} {\bibfield
  {journal} {\bibinfo  {journal} {Few Body Syst.}\ }\textbf {\bibinfo {volume}
  {63}},\ \bibinfo {pages} {42} (\bibinfo {year} {2022})}\BibitemShut {NoStop}%
\end{thebibliography}

%Control: production of eprint (-1) disabled
%

\end{document}